\documentclass[11pt]{article}
\usepackage[utf8]{inputenc} 
\usepackage[english]{babel}
\usepackage[T1]{fontenc}
\usepackage[bbgreekl]{mathbbol}
\usepackage[margin=2cm,vmargin=2cm]{geometry}
\usepackage{caption}
\usepackage{cite}
\usepackage{tikz-cd}
\usepackage{amsmath,amsfonts,amssymb,amsthm}
\usepackage{authblk}
\usepackage{csquotes}
\usepackage{url}
\usepackage{fullpage,setspace}
\usepackage{cancel,verbatim,slashed,todonotes}
\usepackage{xcolor,mdframed,graphicx,color}
\usepackage{dsfont}
\usepackage{mathtools}
\usepackage{multirow,stmaryrd}
\usepackage{array}
\usepackage{makecell}
\usepackage{mathrsfs}
\DeclareSymbolFontAlphabet{\mathbbvar}{bbold}
\DeclareSymbolFontAlphabet{\mathbb}{AMSb}

\newcommand{\Coo}{\mathcal{C}^{\infty}}

\newcommand{\Diff}{\mathrm{Diff}}

\usepackage{slashed}

\newcommand{\di}{\mathrm{d}}

\newcommand{\String}{\mathrm{String}}

\newcommand{\tenhie}{\mathscr{T\!\!H}}
\newcommand{\ind}[1]{\mathcal{#1}}
\newcommand{\bigsp}[1]{\underline{#1}}
\usepackage{upgreek}

\DeclareRobustCommand\longtwoheadrightarrow
    {\relbar\joinrel\twoheadrightarrow}
\DeclareRobustCommand\longhookrightarrow
     {\lhook\joinrel\longrightarrow}
\newcommand{\xtwoheadrightarrow}[2][]{%
  \mathrel{\ooalign{$\xrightarrow[#1\mkern4mu]{#2\mkern4mu}$\cr%
  \hidewidth$\rightarrow\mkern4mu$}}
}
\theoremstyle{definition}

\theoremstyle{definition}

\theoremstyle{definition}

\theoremstyle{definition}

\theoremstyle{definition}

\theoremstyle{definition}

\numberwithin{equation}{subsection}
\tikzcdset{%
    triple line/.code={\tikzset{%
        double equal sign distance, 
        double=\pgfkeysvalueof{/tikz/commutative diagrams/background color}}},
    quadruple line/.code={\tikzset{%
        double equal sign distance, 
        double=\pgfkeysvalueof{/tikz/commutative diagrams/background color}}},
    Rrightarrow/.code={\tikzcdset{triple line}\pgfsetarrows{tikzcd implies cap-tikzcd implies}},
    RRightarrow/.code={\tikzcdset{quadruple line}\pgfsetarrows{tikzcd implies cap-tikzcd implies}}
}

\makeatletter
\providecommand{\leftsquigarrow}{%
  \mathrel{\mathpalette\reflect@squig\relax}%
}
\newcommand{\reflect@squig}[2]{%
  \reflectbox{$\m@th#1\rightsquigarrow$}%
}
\makeatother
\usepackage{hyperref}
\title{
\begin{flushright}
\normalsize{QMUL-PH-20-14}
\end{flushright}
\vspace{1cm}
\bigskip
\bf
The puzzle of global Double Field Theory:\\
open problems and the case for \\a {Higher} Kaluza-Klein perspective}
\author{\sc Luigi Alfonsi}
\affil{\em\normalsize Centre for Research in String Theory,\\\em School of Physics and Astronomy,\\\em Queen Mary University of London,\\\em 327 Mile End Road, London E1 4NS, UK\\\vspace{4mm}\tt \href{mailto:l.alfonsi@qmul.ac.uk}{l.alfonsi@qmul.ac.uk}}
\date{\small July 10, 2020}

\begin{document}
\maketitle

\vspace{0.5cm}
\abstract{
\noindent The history of the geometry of Double Field Theory is the history of string theorists' effort to tame higher geometric structures. 
In this spirit, the first part of this paper will contain a brief overview on the literature of geometry of DFT, focusing on the attempts of a global description.\vspace{0.1cm}

\noindent In \cite{Alf19} we proposed that the global doubled space is not a manifold, but the total space of a bundle gerbe. This would mean that DFT is a field theory on a bundle gerbe, in analogy with ordinary Kaluza-Klein Theory being a field theory on a principal bundle.
\vspace{0.1cm}

\noindent In this paper we make the original construction by \cite{Alf19} significantly more immediate. This is achieved by introducing an \textit{atlas for the bundle gerbe}. This atlas is naturally equipped with $2d$-dimensional local charts, where $d$ is the dimension of physical spacetime. We argue that the local charts of this atlas should be identified with the usual coordinate description of DFT.
\vspace{0.1cm}

\noindent In the last part we will discuss aspects of the global geometry of tensor hierarchies in this bundle gerbe picture. This allows to identify their global non-geometric properties and explain how the picture of non-abelian String-bundles emerges. We interpret the abelian T-fold and the \textit{Poisson-Lie T-fold} as global tensor hierarchies.
}

\newpage
\tableofcontents
\section{Introduction}\label{s1}

\paragraph{Double Field Theory.}
The symmetry known as T-duality is one of the main features of String Theory, in comparison with classical field theories.
Double Field Theory (DFT) is an attempt to make this symmetry manifest: in other words it is a T-duality covariant formulation of Type II supergravity. DFT was officially created in \cite{HulZwi09}, but seminal work includes \cite{Siegel1} and \cite{Siegel2}. See \cite{BerTho14} for a review of the subject and \cite{BerBla20} for a review in the broader context of extended field theories.

\paragraph{The bundle gerbe of Kalb-Ramond field.}
Geometrically the Kalb-Ramond field is interpreted as the connection of a \textit{bundle gerbe} $\mathscr{G}\twoheadrightarrow M$, a geometric object which possesses gauge transformations, but also gauge-of-gauge transformations. Bundle gerbes were originally introduced in \cite{Murray} and the definition of their gauge transformation was defined later in \cite{Murray2}. See \cite{Murray3} for an introductory review. In \cite{Hit99} bundle gerbes were reformulated in terms of \v{C}ech cohomology.
Given a good cover $\{U_\alpha\}$ of the base manifold $M$, the local $2$-forms $B_{\alpha}\in\Omega^2(U_\alpha)$ are patched by local $1$-form gauge transformations $\Lambda_{\alpha\beta}\in\Omega^1(U_\alpha\cap U_\beta)$ which are themselves patched by scalar gauge transformations $G_{\alpha\beta\gamma}\in\Coo(U_\alpha\cap U_\beta \cap U_\gamma)$ satisfying the cocycle condition on four-fold overlaps of patches. In other words the differential local data of the Kalb-Ramond field are patched on overlaps of patches by the conditions
\begin{equation}\label{eq:introgerby}
    \begin{aligned}
    H \,&=\, \mathrm{d}B_{(\alpha)} \\
    B_{(\beta)} - B_{(\alpha)} \,&=\, \mathrm{d}\Lambda_{(\alpha\beta)} \\
    \Lambda_{(\alpha\beta)}+\Lambda_{(\beta\gamma)}+\Lambda_{(\gamma\alpha)} \,&=\, \mathrm{d}G_{(\alpha\beta\gamma)} \\
    G_{(\alpha\beta\gamma)}-G_{(\beta\gamma\delta)}+G_{(\gamma\delta\alpha)}-G_{(\delta\alpha\beta)} \,&\in\, 2\pi\mathbb{Z}
    \end{aligned}
\end{equation}
More recently in \cite{Principal1} bundle gerbes were formalized as a specific case of principal $\infty$-bundle, which is a principal bundle where the ordinary Lie group fiber has been generalized to any $L_\infty$-group. Therefore any Kalb-Ramond field is the connection of a particular principal $\infty$-bundle.

\paragraph{Higher geometry of T-duality.}
Notice that T-duality has been naturally formulated in the context of higher geometry as an isomorphism of bundle gerbes between a string background and its dual in \cite{BunNik13}, \cite{FSS16x}, \cite{FSS17x}, \cite{FSS18}, \cite{FSS18x} and \cite{NikWal18}. Let us consider two $T^n$-bundle spacetimes $M\xrightarrow{\pi}M_0$ and $\widetilde{M}\xrightarrow{\tilde{\pi}}M_0$ over a common base manifold $M_0$. Then the couple of bundle gerbes $\mathscr{G}\xrightarrow{\Pi}M$ and $\widetilde{\mathscr{G}}\xrightarrow{\tilde{\Pi}}\widetilde{M}$, encoding two Kalb-Ramond fields respectively on $M$ and $\widetilde{M}$, are geometric T-dual if the following isomorphism exists  
\begin{equation}
    \begin{tikzcd}[row sep={11ex,between origins}, column sep={11ex,between origins}]
    & \mathscr{G}\times_{M_0} \widetilde{M}\arrow[rr, "\cong"', "\text{T-duality}"]\arrow[dr, "\Pi"']\arrow[dl, "\tilde{\pi}"] & & M\times_{M_0}\widetilde{\mathscr{G}}\arrow[dr, "\pi"']\arrow[dl, "\tilde{\Pi}"] \\
    \mathscr{G}\arrow[dr, "\Pi"'] & & M\times_{M_0}\widetilde{M}\arrow[dr, "\pi"']\arrow[dl, "\tilde{\pi}"] & & [-2.5em]\widetilde{\mathscr{G}}\arrow[dl, "\tilde{\Pi}"] \\
    & M\arrow[dr, "\pi"'] & & \widetilde{M}\arrow[dl, "\tilde{\pi}"] & \\
    & & M_0 & &
    \end{tikzcd}
\end{equation}
This diagram can be interpreted as the finite version of the one in \cite{CavGua11} for the respective Courant algebroids. In this sense T-duality is a geometric property of bundle gerbes.\vspace{0.25cm}

\paragraph{Higher Kaluza-Klein Theory.}
As argued in \cite{Ber19} and \cite{BerBla20}, DFT should be interpreted as a generalization of Kaluza-Klein Theory where it is the Kalb-Ramond field, and not a gauge field, that is unified with the pseudo-Riemannian metric in a bigger space. Since the Kalb-Ramond field is geometrized by a bundle gerbe, in \cite{Alf19} we proposed that DFT should be globally interpreted as a field theory on the total space of a bundle gerbe, just like ordinary Kaluza-Klein Theory lives on the total space of a principal bundle. In the reference we showed how to derive some known doubled spaces such as the ones describing T-folds, and how to interpret T-duality. \vspace{0.25cm}

\noindent In this paper we want to clarify some aspects of Higher Kaluza-Klein geometry by comparing it to previous proposals of DFT geometry. In particular we will deal with the problem of equipping a bundle gerbe with suitable coordinates. Finally we will focus on the concept of \textit{tensor hierarchy} and how this emerges from a bundle gerbe perspective.

\paragraph{Plan of the paper.}
In section \ref{s2} we will illustrate a concise review of the main proposals for a global geometry of DFT, together with a discussion of the main open problems. In section \ref{s3} we will give a brief introduction to the Higher Kaluza-Klein proposal. In section \ref{s4} we will introduce an atlas for the bundle gerbe and we will argue that the $2d$ local coordinates of the charts must be interpreted as the local coordinates of the doubled space of DFT. Finally, in section \ref{s5}, we will consider both an abelian T-fold and a Poisson-Lie T-fold and we will interpret them as particular cases of global tensor hierarchies. Thus we will propose a globalization for tensor hierarchies which relies on the dimensional reduction of the bundle gerbe.

\subsection{Double Field Theory}
As explained by \cite{BerBla20}, Double Field Theory (DFT) should be thought as a generalization of Kaluza-Klein Theory from gauge fields to the Kalb-Ramond field. In this subsection we will give a quick introduction to local DFT.\vspace{0.2cm}

\noindent Let us consider an open simply connected $2d$-dimensional patch $\mathcal{U}$. We can introduce coordinates $(x^\mu,\tilde{x}_\mu):\mathcal{U}\rightarrow \mathbb{R}^{2d}$, which we will call collectively $x^M:=(x^\mu,\tilde{x}_\mu)$. We can define a tensor $\eta=\eta_{MN}\di x^M \otimes \di x^N \in \Omega^1(\mathcal{U})^{\otimes 2}$ with matrix $\eta_{MN}:=\left(\begin{smallmatrix}0&1\\1&0\end{smallmatrix}\right)$. We want now to define a \textit{generalized Lie derivative} which is compatible with the $\eta$-tensor, i.e. such that it satisfies the property $\mathfrak{L}_X\eta=0$ for any vector $X\in\mathfrak{X}(\mathcal{U})$. Thus for any couple of vectors $X,Y\in\mathfrak{X}(\mathcal{U})$ we have the following definition:
\begin{equation}
    \big(\mathfrak{L}_XY\big)^M \;:=\;  X^N\partial_N Y^M - \mathbb{P}^{ML}_{\quad\; NP} \partial_LX^NY^P
\end{equation}
where we defined the tensor
\begin{equation}
    \mathbb{P}^{ML}_{\quad\; NP} \,:=\, \delta^M_P\delta^L_N  - \eta^{ML}\eta_{NP}
\end{equation}
projecting the $GL(2d)$-valued function $\partial_LX^N$ into an $\mathfrak{o}(d,d)$-valued one. We also define the \textit{C-bracket} by the anti-symmetrization of the generalized Lie derivative, i.e. by
\begin{equation}
    \llbracket X,Y \rrbracket_{\mathrm{C}} \;:=\; \frac{1}{2}\big(\mathfrak{L}_XY - \mathfrak{L}_YX\big).
\end{equation}

\noindent Now, if we want to construct an algebra of generalized Lie derivatives (or, in other words an \textit{algebra of infinitesimal generalized diffeomorphisms}), we realize that it cannot be close, i.e. we generally have
\begin{equation}
    \big[\mathfrak{L}_X, \,\mathfrak{L}_Y\big] \;\neq\;\mathfrak{L}_{\llbracket X,Y \rrbracket_{\mathrm{C}}}
\end{equation}
Thus, to assure the closure, we need to impose extra conditions. The \textit{weak} and the \textit{strong constraint} (also known collectively as \textit{section condition}) are respectively the conditions
\begin{equation}
    \eta^{MN}\partial_M\partial_N\phi_i =0, \qquad  \eta^{MN}\partial_M\phi_1\partial_N\phi_2 =0
\end{equation}
for any couple of field or parameter $\phi_1,\phi_2$. The immediate solution to the section condition is obtained by considering only fields and parameters $\phi$ which satisfy the condition $\tilde{\partial}^\mu\phi=0$. Therefore, upon application of the strong constraint, all the fields and parameters will depend on the $d$-dimensional submanifold $U := \mathcal{U}/\!\sim \,\,\subset \mathcal{U}$, where $\sim$ is the relation identifying points with the same physical coordinates $(x^\mu,\tilde{x}_\mu)\sim(x^\mu,\tilde{x}_\mu')$. In particular vectors $X\in\mathfrak{X}(\mathcal{U})$ satisfying the strong constraint can be identified with sections of the generalized tangent bundle $TU\otimes T^\ast U$ of Generalized Geometry. Moreover the C-bracket, when restricted to strong constrained vectors, reduces to the Courant bracket of Generalized Geometry, i.e. we have
\begin{equation}
    \llbracket -,- \rrbracket_{\mathrm{C}}\,\Big|_{\tilde{\partial}^\mu=0} \;=\; [-,-]_{\mathrm{Cou}}
\end{equation}
In this sense, strong constrained DFT locally reduces to ordinary Generalized Geometry. 

\vspace{0.2cm}

\noindent Now we want a tensor parametrizing the coset $O(d,d)/\big(O(1,d-1)\times O(1,d-1)\big)$, thus we will define the \textit{generalized metric} $\mathcal{G}_{MN}\di x^M\otimes \di x^N$ by requiring that it is symmetric and it satisfies the property $\mathcal{G}_{ML}\eta^{LP}\mathcal{G}_{PN}=\eta_{MN}$. The matrix $\mathcal{G}_{MN}$ can be then parametrized by
\begin{equation}
    \mathcal{G}_{MN} \;=\; \begin{pmatrix}g_{\mu\nu}- B_{\mu\lambda}g^{\lambda\rho}B_{\rho\beta} & B_{\mu\lambda}g^{\lambda\nu} \\-g^{\mu\lambda}B_{\lambda\nu} & g^{\mu\nu} \end{pmatrix}.
\end{equation}
where $g:=g_{\mu\nu}\di x^\mu\otimes \di x^\nu$ is a symmetric tensor and $B:=\frac{1}{2}B_{\mu\nu}\di x^\mu\wedge \di x^\nu$ is an anti-symmetric tensor on the submanifold $U$. These are respectively interpreted as a metric and a Kalb-Ramond field on the $d$-dimensional patch $U$. Let us now consider a strong constrained vector of the form $X:=v+\lambda\in\mathfrak{X}(U)\oplus \Omega^1(U)$. The infinitesimal gauge transformation given by generalized Lie derivative $\delta \mathcal{G}_{MN}=\mathfrak{L}_X\mathcal{G}_{MN}$ is equivalent to the following gauge transformations:
\begin{equation}
    \delta g = \mathcal{L}_v g, \qquad \delta B = \mathcal{L}_v B + \di\lambda
\end{equation}
where $\mathcal{L}_v$ is an ordinary Lie derivative of vectors $\mathfrak{X}(U)$. Hence it reproduces the gauge transformations of bosonic Type II supergravity. In other words the infinitesimal \textit{generalized diffeomorphisms} of the $2d$-dimensional patch $\mathcal{U}$ unify the infinitesimal diffeomorphisms of the $d$-dimensional subpatch $U\subset \mathcal{U}$ with the infinitesimal gauge transformations of the Kalb-Ramond field, in analogy with Kaluza-Klein Theory.
\vspace{0.2cm}

\noindent Now the globalization problem of DFT can be summed as follows: since the Kalb-Ramond field $B_{\mu\nu}$ is the locally defined $2$-form of the \textit{connection of a bundle gerbe} (see patching conditions \eqref{eq:introgerby}), how can local DFT patches $\big(\mathcal{U},\,\mathcal{G}_{MN},\,\eta_{MN}\big)$ we just introduced be glued together consistently? We will devolve the rest of the paper to try to answer this question.

\subsection{Double Field Theory on group manifolds}

\noindent \textit{Double Field Theory on group manifolds}, also known as $\mathrm{DFT}_{\mathrm{WZW}}$, gives us a well-defined global version of DFT in the particular case of constant generalized fluxes, which can be interpreted as the structure constants of some $2d$-dimensional Lie algebra. $\mathrm{DFT}_{\mathrm{WZW}}$ was originally introduced by \cite{DFTWZW15} and reformulated in terms of generalized metric by \cite{DFTWZW15x}, while fundamental seminal work can already be found in \cite{Hul09}. It includes, in particular, the familiar abelian cases of the doubled torus $T^{2d}$ and of $\mathbb{R}^{2d}$. \vspace{0.2cm}

\noindent In $\mathrm{DFT}_{\mathrm{WZW}}$ one considers a doubled space which is a $2d$-dimensional Lie group $\mathcal{G}$, whose local coordinates near the identity element are $x^I:\mathcal{G}\rightarrow\mathbb{R}^{2d}$ with $I=1,\dots,2d$. Let also $\mathfrak{g}:=\mathrm{Lie}(\mathcal{G})$ be the Lie algebra of $\mathcal{G}$. Given a set of generators $\{T_{\mathcal{I}}\}_{\mathcal{I}=1,\dots,2d}$ of $\mathfrak{g}$, we will have bracket structure $[T_{\mathcal{I}},T_{\mathcal{J}}]_{\mathfrak{g}}=C_{\mathcal{IJ}}^{\quad\mathcal{K}}\, T_{\mathcal{K}}$, where we called $C_{\mathcal{IJ}}^{\quad\mathcal{K}}$ the structure constants of this algebra. \vspace{0.2cm}

\noindent Since Lie groups are always parallelizable, it is possible to introduce a parallelization given by a frame $\xi^\mathcal{I}:=E^\mathcal{I}_{\;\,J}\di x^J\in\Omega^1(\mathcal{G})$ where $E^\mathcal{I}_{\;\,J}$ are $GL(2d)$-valued matrices on $\mathcal{G}$. Notice that the $\mathfrak{g}$-valued $1$-form $\xi^\mathcal{I}T_\mathcal{I}$ is exactly the left-invariant Maurer-Cartan $1$-form of the Lie group $\mathcal{G}$. By using the inverse transpose of the matrices of the frame $E_\mathcal{I}^{\;\,J}$ we immediately find the dual vector basis $D_\mathcal{I} := E_\mathcal{I}^{\;\,J} \partial_J$, which will satisfy the Lie bracket $[D_\mathcal{I},\,D_\mathcal{J}]_{\mathrm{Lie}}=C_{\mathcal{IJ}}^{\quad\mathcal{K}}D_\mathcal{K}$.\vspace{0.2cm}

\noindent It was proven by \cite{Hass18} that the solution of the section condition $\eta^{\mathcal{IJ}}D_{\mathcal{I}}\otimes D_{\mathcal{J}}(-)=0$ applied on any field or parameter, naturally selects a maximal isotropic submanifold $G\in\mathcal{G}$ of the group manifold. Remarkably any two such submanifolds $G,\widetilde{G}\subset \mathcal{G}$ are Poisson-Lie T-dual. If the group manifold is abelian, this reduces to usual abelian T-duality. \vspace{0.2cm}

\noindent  We can define the $\eta$-tensor by $\eta_{\mathcal{IJ}}\xi^\mathcal{I}\otimes\xi^\mathcal{J}$ where $\eta_{\mathcal{IJ}}= \left(\begin{smallmatrix}0&1\\1&0\end{smallmatrix}\right)$. We can also define the generalized metric as a tensor $\mathcal{G}_{\mathcal{IJ}}\xi^\mathcal{I}\otimes\xi^\mathcal{J}$ on the group manifold, where the matrix $\mathcal{G}_{\mathcal{IJ}}$ is constant. If we locally rewrite the generalized metric $\mathcal{G}_{\mathcal{IJ}}\xi^\mathcal{I}\otimes\xi^\mathcal{J} = \mathcal{G}_{IJ}\di x^I\otimes\di x^J$ in the coordinate basis, we obtain metric $g$ and Kalb-Ramond field $B$ depending on the $d$-dimensional submanifold $G$. Finally, the structure constants $C_{\mathcal{IJ}}^{\quad\mathcal{K}}$ can be naturally identified with the generalized fluxes by $C_{\mathcal{IJK}}\, \xi^\mathcal{I} \wedge \xi^\mathcal{J} \wedge \xi^\mathcal{K}$. See \cite{DFTWZW15x} and \cite{Hass18} for more details.

\section{Review of proposals for DFT geometry}\label{s2}

In this section we will give a brief overview on the main proposals for geometry of DFT. We will underline relations between the approaches and we will discuss some open problems.

\subsection{Non-associative proposal}

The non-associative proposal was presented by \cite{HohZwi12} and further developed by \cite{Hohm13space}. Its aim is to realize the group of gauge transformations of DFT by diffeomorphisms of the doubled space. However, since the C-bracket structure on doubled vectors does not satisfy the Jacobi identity, its exponentiation will not give us a Lie group, but a geometric object which does not satisfy the associativity property.

\paragraph{Non-trivial three-fold overlaps.}
In the proposal by \cite{Hohm13space} the doubled space $\mathcal{M}$ is just a $2d$-dimensional smooth manifold. This means that we can consider a cover $\{\mathcal{U}_\alpha\}$, so that $\bigcup_\alpha \mathcal{U}_\alpha = \mathcal{M}$, and glue the coordinate patches on each two-fold overlap $\mathcal{U}_\alpha \cap \mathcal{U}_\beta$ of the doubled space by diffeomorphisms $x_{(\beta)} = f_{(\alpha\beta)}\big(x_{(\alpha)}\big)$. Vectors of the tangent bundle $T\mathcal{M}$ will be then glued on each $T(\mathcal{U}_\alpha \cap \mathcal{U}_\beta)$ by the $GL(2d)$-valued Jacobian matrix $J_{(\alpha\beta)} := {\partial x_{(\alpha)}}/{\partial x_{(\beta)}}$. However these transformations do not work for doubled vectors from DFT, thus \cite{Hohm13space} proposed that the doubled vectors should transform by the $O(d,d)$-valued matrix given by
\begin{equation}\label{eq:fmatrix}
    \mathcal{F}_{(\alpha\beta)} \;:=\; \frac{1}{2}\Big(J_{(\alpha\beta)}J^{-\mathrm{T}}_{(\alpha\beta)}+J^{-\mathrm{T}}_{(\alpha\beta)}J_{(\alpha\beta)}\Big)
\end{equation}
which indeed preserves the $O(d,d)$-metric $\eta:=\eta_{MN}\di x^M \otimes \di x^N$. Now, if we go to the three-fold overlaps of patches $\mathcal{U}_\alpha \cap \mathcal{U}_\beta \cap \mathcal{U}_\gamma$ we realize that these transition functions do not satisfy the expected cocycle condition. In other words we generally have
\begin{equation}\boxed{\quad\label{eq:noncocycle}
    \mathcal{F}_{(\alpha\beta)}\,\mathcal{F}_{(\beta\gamma)}\,\mathcal{F}_{(\gamma\alpha)} \,\neq\, 1 \quad}
\end{equation}
Notice that for the first time we see something resembling a gerbe-like structure spontaneously emerging in DFT geometry.

\paragraph{Modified exponential map.}
The solution proposed by \cite{Hohm13space} consists, first of all, in a modified exponential map ${\exp}\Theta:\mathfrak{X}(\mathcal{U})\rightarrow \Diff(\mathcal{U})$. This will map any vector $X\in\mathfrak{X}(\mathcal{U})$ in the diffeomorphism given by
\begin{equation}\label{eq:thetaexp}
    x'= e^{\Theta(X)} x \;\quad\; \text{with} \;\quad\; \Theta(X)^M \,:=\, X^M + \underbrace{\sum_i \rho_i \partial^M \chi_i}_{\mathcal{O}(X^3)}
\end{equation} \vspace{-0.46cm}

\noindent where $\rho_i$ and $\chi_i$ are functions on $x$ depending on the vector $X$ in a way which guarantees that $\Theta(X)^M\partial_M = X^M\partial_M$ when applied to any field satisfying the strong constraint. This modified diffeomorphism crucially agrees with the gauge transformation $V'(x)=e^{\mathfrak{L}_X}V(x)$ of DFT, where $\mathfrak{L}_X$ is the generalized Lie derivative defined by the D-bracket.

\paragraph{$\star$-product and non-associativity.}
In ordinary differential geometry the exponential map $\exp:\big(\mathfrak{X}(\mathcal{U}),\,[-,-]\big)\rightarrow \big(\Diff(\mathcal{U}),\,\circ\,\big)$ maps a vector $X\mapsto e^X$ into the diffeomorphism that it generates. The usual exponential map notoriously satisfies the property $e^{X}\circ e^{Y} = e^{Z}$ with $Z\in\mathfrak{X}(\mathcal{U})$ given by the Baker-Campbell-Hausdorff series $Z=X+Y+[X,Y]/2+\dots$ for any couple of vectors $X,Y\in\mathfrak{X}(\mathcal{U})$.
The idea by \cite{Hohm13space} consists in equipping the space of vector fields $\mathfrak{X}(\mathcal{U})$ with another bracket structure $\big(\mathfrak{X}(\mathcal{U}),\,\llbracket -,-\rrbracket_{\mathrm{C}}\big)$, where $\llbracket -,-\rrbracket_{\mathrm{C}}$ is the C-bracket of DFT. Now this algebra can be integrated by using the modified exponential map ${\exp}\Theta$ defined in \eqref{eq:thetaexp} to a quasigroup $\big(\Diff(\mathcal{U}),\,\star\,\big)$ that satisfies
\begin{equation}
    e^{\Theta(X)} \star e^{\Theta(Y)} = e^{\Theta(Z)} \;\quad\; \text{with} \;\quad\; Z=X+Y+\frac{1}{2}\llbracket X,Y\rrbracket_{\mathrm{C}}+\dots
\end{equation}
It is possible to check that this $\star$-product is \textit{not associative}: in other words the inequality
\begin{equation}
    (f\star g) \star h \,\neq\, f\star (g \star h),
\end{equation}
where $f,g,h\in\Diff(\mathcal{U})$ are diffeomorphisms, generally holds. Now let us call the diffeomorphisms $f:=e^{\Theta(X)}$, $g:=e^{\Theta(Y)}$ and $h:=e^{\Theta(Z)}$ obtained by exponentiating three vectors $X,Y,Z\in\mathfrak{X}(\mathcal{U})$. Then the obstruction of the $\star$-product from being associative is controlled by an element $W$ which satisfies the equation
\begin{equation}
    (f \star g)\star h \,=\, W \star \big( f \star (g\star h) \big)
\end{equation}
and which is given by $W=\exp\Theta\!\left(-\frac{1}{6}\mathcal{J}(X,Y,Z)+\dots\right)$, where $\mathcal{J}(-,-,-)$ is the Jacobiator of the C-bracket. Even if it is well-known that the Jacobiator is of the form $\mathcal{J}^M=\partial^M\mathcal{N}$ for a function $\mathcal{N}\in\Coo(\mathcal{U})$, notice that the transformation $W$ is non-trivial. Also if we consider diffeomorphisms on doubled space which satisfy $f_{(\alpha\beta)}\star f_{(\beta\gamma)}=f_{(\alpha\gamma)}$, we re-obtain the desired property $\mathcal{F}_{(\alpha\beta)} \mathcal{F}_{(\beta\gamma)}=\mathcal{F}_{(\alpha\gamma)}$ for doubled vectors.
\vspace{0.25cm}

\noindent We know that the diffeomorphisms group of the doubled space is not homeomorphic to the group $ G_{\mathrm{DFT}}$ of DFT gauge transformations $e^{\mathfrak{L}_X}$. But now, by replacing the composition of diffeomorphisms with the $\star$-product, we can define a homomorphism 
\begin{equation}
    \varphi:\big(\Diff(\mathcal{U}),\,\star\,\big) \,\longrightarrow\, G_{\mathrm{DFT}}
\end{equation}
which therefore satisfies the property
\begin{equation}
    \varphi(f\star g) \,=\, \varphi(f)\,\varphi(g).
\end{equation}
This property determines the $\star$-product up to trivial gauge transformation. In the logic of \cite{Hohm13space} this will allow to geometrically realize DFT gauge transformation as diffeomorphisms of the doubled space. 

\subsection{Proposal with gerbe-like local transformations}

The first paper in the literature explicitly recognizing the higher geometrical property of DFT is \cite{BCM14}. In the reference it is argued that we can overcome many of the difficulties of the non-associative proposal by describing the geometry of DFT modulo local $O(d,d)$-transformations.

\paragraph{Gerbes debut.}
Their proposal starts from the same problem \eqref{eq:noncocycle}, but proposes a different solution. 
We can rewrite the C-bracket of doubled vectors by $\llbracket X,Y \rrbracket_{\mathrm{C}}= [X,Y]_{\mathrm{Lie}}+ \lambda(X,Y)$ where we called $\lambda^M(X,Y) := X^N\partial^MY_N$. This means that we can rewrite the algebra of DFT gauge transformations as $[\mathfrak{L}_X,\mathfrak{L}_Y] = \mathfrak{L}_{[X,Y]_{\mathrm{Lie}}}+\Delta(X,Y)$ where we defined $\Delta(X,Y) := \mathfrak{L}_{\lambda(X,Y)}$. In \cite{BCM14} it is noticed that the extra $\Delta$-transformation appearing in the DFT gauge algebra is \textit{non-translating}, i.e. it involves no translation term if acting on tensors satisfying the strong constraint. Thus the diffeomorphism $e^X$ and the gauge transformation $e^{\mathfrak{L}_X}$ agree up to a local transformation $e^{\Delta}=1+\Delta$.
In fact, if we impose the strong constraint on fields and parameters 
\begin{equation}
    \Delta_M^{\;\;N} \;=\; \begin{pmatrix}0 & 0 \\ \partial_{[\mu} \tilde{\lambda}_{\nu]} & 0 \end{pmatrix}
\end{equation}
where $\tilde{\lambda}_\mu = X^N\partial_\mu Y_N$ depends only on the $d$-dimensional physical subset $U\subset\mathcal{U}$ of our doubled space patch. Hence the local $\Delta$-transformation is just an infinitesimal gauge transformation $\mathfrak{L}_{\tilde{\lambda}} B=\di\tilde{\lambda}$ of the Kalb-Ramond field. 

\paragraph{Further discussion.}
As noticed by \cite{Hull14}, $\Delta$-transformations are integrated on a patch $\mathcal{U}$ to the group $\Omega^1(U)$ of finite gauge transformations of the Kalb-Ramond field, while full gauge transformations generated by a strong constrained doubled vectors are integrated to $\Diff(U)\ltimes\Omega^1(U)\subset \Diff(\mathcal{U})$. Now we notice that the group of DFT gauge transformations effectively becomes the homotopy quotient $G_{\mathrm{DFT}}=\big(\Diff(U)\ltimes\Omega^1(U)\big)/\!/\Omega^1(U)$, thus a $2$-group.
\vspace{-0.25cm}

\noindent The doubled space is still a $2d$-dimensional manifold $\mathcal{M}$ and then its coordinate patches on each two-fold overlap $\mathcal{U}_\alpha \cap \mathcal{U}_\beta$ are still glued by by diffeomorphisms $x_{(\beta)} = f_{(\alpha\beta)}\big(x_{(\alpha)}\big)$. The doubled vectors are still glued by the $O(d,d)$-valued matrix $\mathcal{F}_{(\alpha\beta)}$ defined in \eqref{eq:fmatrix}, like in the non-associative proposal. Now, according to \cite{BCM14}, on three-fold overlaps of patches $\mathcal{U}_\alpha\cap \mathcal{U}_\beta \cap \mathcal{U}_\gamma$ the transition functions of doubled vectors satisfy 
\begin{equation}\label{eq:bbb}
    \mathcal{F}_{(\alpha\beta)}\,\mathcal{F}_{(\beta\gamma)} \,=\, \mathcal{F}_{(\alpha\gamma)}e^{\Delta_{(\alpha\beta\gamma)}}
\end{equation}
i.e. they satisfy the desired transitive property up to a local $\Delta$-transformation.
In a more mathematical language we can say that \textit{doubled vectors would be sections of a stack} on the $2d$-dimensional manifold $\mathcal{M}$. This is not surprising since the algebra of $G_{\mathrm{DFT}}$ is of the form $\big(\mathfrak{X}(U)\oplus\Omega^1(U)\big)/\!/\Omega^1(U)$, which then must be glued on overlaps of patches by $B$-shifts $\di \tilde{\lambda}$. Thus we could replace the concept of non-associative transformations with a gerbe-like structure.

\subsection{Doubled-yet-gauged space proposal}

The idea of \textit{doubled-yet-gauged space} was proposed by \cite{Park13} as a solution for the discrepancy between finite gauge transformations and diffeomorphisms of the doubled space, in alternative to the non-associative proposal. Then it was further explored in \cite{Par13x}, where a covariant action was obtained, and in \cite{Par16x}, where it was generalized to the super-string case. Very intriguingly this formalism led to novel non-Riemannian backgrounds in \cite{Par17xx}, \cite{Par18xx} and \cite{Par19xx}. Recently a BRST formulation for the action of a particle on the doubled-yet-gauged space has been proposed by \cite{Par19x} and related to the NQP-geometry involved by other proposals.

\paragraph{The coordinate gauge symmetry.}
In doubled-yet-gauged space proposal the doubled space $\mathcal{M}$ is, at least locally, a smooth manifold. A local $2d$-dimensional coordinate patch $\mathcal{U}$ is characterized by \textit{coordinate symmetry}, i.e. there exists a canonical gauge action on its local coordinates expressed by
\begin{equation}
    x^M \;\sim\; x^M + \sum_i \rho_i \partial^M \chi_i(x)
\end{equation}
for any choice of functions $\rho_i,\chi_i\in\Coo(\mathcal{U})$.
This observation is motivated by the fact that any strong constrained tensor satisfies the identity $T_{A_1\dots A_n}\big(x+\lambda(x)\big) = T_{A_1\dots A_n}(x)$ where we called $\lambda^M := \sum_i \rho_i \partial^M \chi_i$ at any point $x\in\mathcal{U}$. \vspace{0.25cm}

\noindent Let us choose coordinates for our doubled patch $\mathcal{U}$ such that the strong constraint is solved by letting all the fields and parameters depend only on the $d$-dimensional subpatch $U\subset\mathcal{U}$. Then the coordinate symmetry on the doubled space reduces to
\begin{equation}
    \big(x^\mu,\,\tilde{x}_\mu\big) \;\sim\; \big(x^\mu,\,\tilde{x}_\mu+\tilde{\lambda}_\mu(x)\big)
\end{equation}
where $\tilde{\lambda}_\mu = \sum_i \rho_i \partial_\mu \chi_i$. This coordinate symmetry, similarly to the $\Delta$-transformations in the previous proposal, can be identified with the local gauge symmetry of the Kalb-Ramond field by $\delta_{\tilde{\lambda}}B=\di\tilde{\lambda}$, where the parameter is exactly $\tilde{\lambda}:=\tilde{\lambda}_\mu\di x^\mu$. We can thus identify the physical $d$-dimensional patches with the quotients $U_\alpha\,\cong\,\mathcal{U}_\alpha/\sim$. Thus, as argued by \cite{Park13}, physical spacetime points must be identified with gauge orbits of the doubled-yet-gauged space.\vspace{0.25cm}

\noindent The coordinate gauge symmetry is also the key to solve the discrepancy between DFT gauge transformations $e^{\mathfrak{L}_V}$ and diffeomorphisms $e^V$. Indeed, as argued by \cite{Park13}, the two exponentials induce two finite coordinate transformations $x^M\mapsto x^{\prime M}$ and $x^M\mapsto x^{\prime\prime M}$ whose ending points are coordinate gauge equivalent, i.e. $x^{\prime M}\sim x^{\prime\prime M}$. Therefore, upon section constraint, they differ just by a Kalb-Ramond field gauge transformation. 

\paragraph{How can we globalize the doubled-yet-gauged space?}
Now, the doubled-yet-gauged formalism encompasses the local geometry of the doubled space. However in this review we are interested in the global aspects of DFT, so we may try to understand how these doubled patches can be glued together. Let us first try a na\"{i}ve approach, for pedagogical reasons: we will try to glue our doubled patches by diffeomorphisms that respect the section condition, i.e. on two-fold overlaps of patches $\mathcal{U}_\alpha \cap \mathcal{U}_\beta$ we will have
\begin{equation}\label{eq:patchingpa}
    x_{(\beta)}\,=\,f_{(\alpha\beta)}\big(x_{(\alpha)}\big), \qquad \tilde{x}_{(\beta)} \,=\, \tilde{x}_{(\alpha)} + \Lambda_{(\alpha\beta)}\big(x_{(\alpha)}\big).
\end{equation}
This would imply the patching conditions $B_{(\beta)} = f^\ast_{(\alpha\beta)}B_{(\alpha)} + \di\Lambda_{(\alpha\beta)}$ where the local $1$-forms $\Lambda_{(\alpha\beta)}:=\Lambda_{(\alpha\beta)\mu} \di x_{(\beta)}^\mu$ are given by the gluing conditions \eqref{eq:patchingpa}. But then, with these assumptions, the doubled space $\mathcal{M}$ would become just the total space $(\mathbb{R}^{d})^\ast$-bundle on the physical $d$-dimensional spacetime $M$. If we compose the transformations \eqref{eq:patchingpa} on three-fold overlaps of patches $\mathcal{U}_\alpha \cap \mathcal{U}_\beta \cap \mathcal{U}_\gamma$ we immediately obtain the cocycle condition $\Lambda_{(\alpha\beta)} + \Lambda_{(\beta\gamma)} + \Lambda_{(\gamma\alpha)} = 0$, which is the cocycle describing a topologically trivial gerbe bundle with $[H]=0\in H^3(M,\mathbb{Z})$ and not a general string background. Therefore this na\"{i}ve attempt at gluing is not enough.

\paragraph{Further discussion.}
The doubled-yet-gauged formalism gives us an unprecedented interpretation of the coordinates of DFT. Upon choice of coordinates which are compatible with the section constraint, indeed, the coordinate gauge symmetry can be identified with the gauge transformations of the Kalb-Ramond field. This is a fundamental link between the geometry of the bundle gerbe formalizing the Kalb-Ramond field and the geometry of the doubled space. This also provides an interesting link with ordinary Kaluza-Klein geometry, where the points of the base manifold of a $G$-bundle are in bijection with the gauge $G$-orbits of the bundle. As we will see in section \ref{s4}, the local coordinate gauge symmetry which was discovered by \cite{Park13} will be also recovered as fundamental property of the double space which arises from the Higher Kaluza-Klein perspective. We will see that the Higher Kaluza-Klein formalism recovers a globalized version of the doubled-yet-gauged space with gluing conditions which are a gerby version of the na\"{i}ve patching conditions \eqref{eq:patchingpa}. Therefore the Higher Kaluza-Klein proposal can be seen also as a proposal of globalization of the doubled-yet-gauged space approach.

\subsection{Finite gauge transformations proposal}

In \cite[pag.$\,$23]{Hull14} it was proposed that, given a geometric background $M$, the group of gauge transformations of DFT should be just
\begin{equation}\label{eq:ghull}
    G_{\mathrm{DFT}}\;=\;\Diff(M)\ltimes \Omega^2_{\mathrm{cl}}(M),
\end{equation}
i.e. diffeomorphisms of the manifold $M$ and $B$-shifts. 
In particular it was argued that any try of realizing the group of gauge transformations of DFT as diffeomorphisms of a $2d$-dimensional space should fail, because it is not homomorphic to the group of diffeomorphisms.

\paragraph{Finite gauge transformations.}
In \cite[pag.$\,$20]{Hull14} it was then proposed that \textit{double vectors on a geometric background $M$ are just sections of a Courant algebroid $E\twoheadrightarrow M$ twisted by a bundle gerbe}. In other words, on any patch $U_\alpha$ of the manifold $M$, a doubled vector would be of the form
\begin{equation}\label{eq:vec}
    V_{(\alpha)} \;=\; \begin{pmatrix}1 & 0 \\ -B_{(\alpha)} & 1\end{pmatrix}  \begin{pmatrix} v_{(\alpha)} \\ \tilde{v}_{(\alpha)} \end{pmatrix} \;=\; \begin{pmatrix} v^\mu_{(\alpha)} \\[0.3em] \tilde{v}_{(\alpha)\mu} + B_{(\alpha)\mu\nu}v^\nu_{(\alpha)}\end{pmatrix}
\end{equation}
It was also shown by \cite[pag.$\,$23]{Hull14} that the $O(d,d)$-matrix \eqref{eq:fmatrix} transforming double vectors under a finite gauge transformation of DFT, i.e. a diffeomorphism $x_{(\beta)}=f_{(\alpha\beta)}\big(x_{(\alpha)}\big)$ and a $B$-shift $\di\lambda_{(\alpha\beta)}$, reduces to 
\begin{equation}
    \mathcal{F}_{(\alpha\beta)} \;=\; \begin{pmatrix}j_{(\alpha\beta)} & 0 \\ 0 & j^{-\mathrm{T}}_{(\alpha\beta)}\end{pmatrix} \begin{pmatrix}1 & 0 \\ -B_{(\alpha)} & 1\end{pmatrix}\begin{pmatrix}1 & 0 \\ \di\lambda_{(\alpha\beta)} & 1\end{pmatrix}
\end{equation}
where we called $j_{(\alpha\beta)}:= \partial x_{(\beta)}/\partial x_{(\alpha)}$ the Jacobian matrix of the diffeomorphism. This way it is natural to recover equation \eqref{eq:bbb}, i.e.
\begin{equation}
    \mathcal{F}_{(\alpha\beta)}\mathcal{F}_{(\beta\gamma)}\mathcal{F}_{(\gamma\alpha)} \;=\; e^{\Delta_{(\alpha\beta\gamma)}}
\end{equation}
where $e^{\Delta_{(\alpha\beta\gamma)}}$ will generally be a non-trivial local $B$-shift.

\paragraph{Further discussion.} 
This proposal clarifies the previous ones by illustrating that, whenever the strong constraint can be globally solved by letting the fields depend on a $d$-dimensional submanifold $M$, doubled vectors must be seen as sections of a Courant algebroid twisted by a gerbe on $M$. For non-geometric backgrounds, however, this picture holds only locally.

\subsection{C-space proposal}

The idea of C-spaces was born in \cite{Pap13}, matured in \cite{Pap14} and further explored in \cite{HowPap17} in relation to topological T-duality. This was the first proposal to suggest that a global double space should consist in the total space of a bundle gerbe, equipped with a particular notion of coordinates, which was renamed \textit{C-space}.

\paragraph{C-spaces.}
The notation $\mathscr{C}_M^{[H]}$ for a C-space makes explicit that it is topologically classified only by the base manifold $M$ and by the Dixmier-Douady class $[H]\in H^3(M,\mathbb{Z})$, i.e. the H-flux.
According to \cite{Pap14} we can introduce two sets of coordinates for a C-space $\mathscr{C}_M^{[H]}\longtwoheadrightarrow M$ on some base manifold $M$. We must consider new coordinates $y^1_{(\alpha)}$ on each patch $U_\alpha$ and $\theta_{(\alpha\beta)}$ on each two-fold overlap of patches $U_\alpha\cap U_\beta$ of $M$. Let us now recall that the differential data of a bundle gerbe on $M$ is specified by a \v{C}ech cocycle $\big(B_{(\alpha)},\Lambda_{(\alpha\beta)},G_{(\alpha\beta\gamma)}\big)$, with $B_{(\alpha)}\in\Omega^2(U_\alpha)$, $\Lambda_{(\alpha\beta)}\in\Omega^1(U_\alpha\cap U_\beta)$ and $G_{(\alpha\beta\gamma)}\in\Coo(U_\alpha\cap U_\beta\cap U_\gamma)$ which satisfy
\begin{equation}\label{eq:cgerbe}
    \begin{aligned}
        B_{(\beta)}-B_{(\alpha)} &= \mathrm{d}\Lambda_{(\alpha\beta)} \\
        \Lambda_{(\alpha\beta)}+\Lambda_{(\beta\gamma)}+\Lambda_{(\gamma\alpha)} &= \mathrm{d}G_{(\alpha\beta\gamma)} \\
        G_{(\alpha\beta\gamma)}-G_{(\beta\gamma\delta)}+G_{(\gamma\delta\alpha)}-G_{(\delta\alpha\beta)}&\in 2\pi\mathbb{Z}
    \end{aligned}
\end{equation}
Then the extra coordinates $\big(y^1_{(\alpha)},\, \theta_{(\alpha\beta)}\big)$, which have "the degree" of a $1$-form and of a scalar, must be then glued on two-fold and three-fold overlaps of patches of $M$ by using the transition functions of the gerbe, i.e. by
\begin{equation}\label{eq:papapatch}
    \begin{aligned}
        - y^1_{(\alpha)} + y^1_{(\beta)} +\di\theta_{(\alpha\beta)} \,&=\, \Lambda_{(\alpha\beta)}, \\
        \theta_{(\alpha\beta)} + \theta_{(\beta\gamma)} + \theta_{(\gamma\alpha)} \,&=\, G_{(\alpha\beta\gamma)}\;\mathrm{mod}\,2\pi\mathbb{Z}.
    \end{aligned}
\end{equation}
With this identification, a change of coordinates $\big(y^1_{(\alpha)},\, \theta_{(\alpha\beta)}\big) \mapsto\big(y^1_{(\alpha)}+\eta_{(\alpha)},\, \theta_{(\alpha\beta)}+\eta_{(\alpha\beta)}\big)$ induces a gauge transformation for the Kalb-Ramond field given by
\begin{equation}
    \begin{aligned}
        B_{(\alpha)} &\mapsto B_{(\alpha)} + \mathrm{d}\eta_{(\alpha)}, \\
        \Lambda_{(\alpha\beta)} &\mapsto \Lambda_{(\alpha\beta)}+\eta_{(\alpha)}-\eta_{(\beta)}+\mathrm{d}\eta_{(\alpha\beta)} \\
        G_{(\alpha\beta\gamma)} &\mapsto G_{(\alpha\beta\gamma)} + \eta_{(\alpha\beta)}+\eta_{(\beta\gamma)}+\eta_{(\gamma\alpha)}
    \end{aligned}
\end{equation}
in analogy with the extra coordinate of ordinary Kaluza-Klein Theory.\vspace{0.25cm}

\noindent Moreover, if we take the differential of the first patching condition in \eqref{eq:papapatch}, we obtain the condition $ - \di y^1_{(\alpha)} + \di y^1_{(\beta)} = \di\Lambda_{(\alpha\beta)}$ for the differentials. This means that if we rewrite in components $y^1_{(\alpha)}=y^1_{(\alpha)\mu}\di x^\mu $, we can also rewrite $ - \di y^1_{(\alpha)\mu} + \di y^1_{(\alpha)\mu} = \di\Lambda_{(\alpha\beta)\mu}$. If we define the dual vectors $\partial/\partial y^1_{(\alpha)\mu}$ to the $1$-forms $\di y^1_{(\alpha)\mu}$ as vectors satisfying $\big\langle \partial/\partial y^1_{(\alpha)\mu},\, \di y^1_{(\alpha)\nu} \big\rangle =\delta^\mu_{\;\nu}$, we obtain doubled vector of the following form:
\begin{equation}
    V_{(\alpha)} \,=\, v^\mu_{(\alpha)} \frac{\partial}{\partial x^\mu_{(\alpha)}} + \bigg( \tilde{v}_{(\alpha)\mu} + B_{(\alpha)\mu\nu}\,v^\nu_{(\alpha)} \bigg)\frac{\partial}{\partial y^1_{(\alpha)\mu}}
\end{equation}
which are exactly the same as the ones in \eqref{eq:vec}.
Therefore the analogue of \textit{the tangent bundle of the C-space can be identified with a Courant algebroid $E\twoheadrightarrow M$ twisted by the gerbe \eqref{eq:cgerbe}}.

\paragraph{Further discussion.} The proposal seems to capture something quite fundamental of the geometry of DFT, by suggesting that the doubled space should be the total space of the gerbe itself. This looks consistent with the existing idea that doubled vectors should belong to a Courant algebroid twisted by a gerbe, which is the analogous to the tangent bundle for a gerbe. However this intuition is still waiting for a proper formalization: for example it is not clear how to construct coordinates that are $1$-forms on $M$. Moreover it is still not clear what is the relation with the new extra coordinates and the T-dual spacetime.

\subsection{Pre-NQP manifold proposal}

The pre-NQP manifold proposal was developed by \cite{DesSae18}, generalized to Heterotic DFT by \cite{DesSae18x} and then applied to the particular example of nilmanifolds by \cite{DesSae19}. This approach to DFT is based on the fact that $n$-algebroids can be equivalently described by differential-graded manifolds, including the Courant algebroid, which describes the local symmetries of the bundle gerbe of the Kalb-Ramond field. The idea is thus that we can describe the geometry of DFT by considering the differential graded manifold which geometrizes the Courant algebroid and by relaxing some of the conditions.

\paragraph{$L_\infty$-algebroids as NQ-manifolds.}
Given a $L_\infty$-algebroid $\mathfrak{a}\twoheadrightarrow M$ on some base manifold $M$, we can always associate to $\mathfrak{a}$ its Chevalley-Eilenberg algebra $\mathrm{CE}(\mathfrak{a})$, which is essentially the differential graded algebra of its sections. This is defined by
\begin{equation}
    \mathrm{CE}(\mathfrak{a})\;:=\; \Big(\wedge^\bullet\Gamma(M,\mathfrak{a}_\bullet^\ast),\,\di_{\mathrm{CE}}\Big)
\end{equation}
where the underlying complex is defined by 
\begin{equation}
    \wedge^\bullet\Gamma(M,\mathfrak{a}_\bullet^\ast) \;\,:=\;\, \underbrace{\Coo(M)}_{\text{degree 0}} \,\oplus\, \underbrace{\Gamma(M,\mathfrak{a}_0^\ast)}_{\text{degree 1}} \,\oplus\, \underbrace{\Gamma\big(M,\mathfrak{a}_1^\ast \oplus(\mathfrak{a}_0^\ast\wedge\mathfrak{a}_0^\ast)\big)}_{\text{degree 2}} \,\oplus\; \dots
\end{equation}
where the $\mathfrak{a}_k$ for any $k\in\mathbb{N}$ are the ordinary vector bundles underlying the $L_\infty$-algebroid. In the definition $\di_{\mathrm{CE}}$ is a degree $1$ differential operator on the graded complex $\wedge^\bullet\Gamma(M,\mathfrak{a}_\bullet^\ast)$ which encodes the $L_\infty$-bracket structure of the original $L_\infty$-algebroid $\mathfrak{a}$.
\vspace{0.25cm}

\noindent Now a NQ-manifold is defined as a graded manifold $\mathcal{M}$ equipped with a degree $1$ vector field $Q$ satisfying $Q^2=0$. The fundamental feature of NQ-manifolds is that the algebra of functions of any NQ-manifold $\mathcal{M}$ is itself a differential graded algebra $\left(\Coo(\mathcal{M}),\,Q\right)$ where the role of the differential operator is played by the vector $Q$, which is thus called \textit{cohomological}. 
\vspace{0.25cm}

\noindent Crucially there exists an equivalence between $L_\infty$-algebroids and NQ-manifolds given by
\begin{equation}
    \mathrm{CE}(\mathfrak{m})\;=\; \Big(\Coo(\mathcal{M}),\,Q\Big)
\end{equation}
any $L_\infty$-algebroid $\mathfrak{m}$ can be equivalently seen as a NQ-manifold $\mathcal{M}$. In the particular case which is relevant for DFT we consider the $2$-algebroid $\mathfrak{at}_{\mathcal{G}}$ of infinitesimal gauge transformation of the bundle gerbe of the Kalb-Ramond field on  a manifold $M$. This is notoriously given by a NQ-manifold $T^\ast[2]T[1]M$ by the usual identification
\begin{equation}\boxed{\quad\label{eq:CEat}
    \mathrm{CE}(\mathfrak{at}_{\mathcal{G}})\;=\; \Big(\Coo\big(T^\ast[2]T[1]M\big),\,Q_H\Big) \quad}
\end{equation}
where $Q_H$ is the cohomological vector twisted by the curvature $H\in\Omega^3_{\mathrm{cl}}(M)$ of the gerbe. To show this, notice first that in this case the differential graded algebra of functions on our NQP-manifold will be truncated at degree $< 2$. The degree $1$ sections will be sums of a vector and a $1$-form $X+\xi\in\Gamma(M,TM\oplus T^\ast M)$ and the degree $0$ sections will be just functions $f\in\Coo(M)$ on the base manifold. Now we can explicitly rewrite the underlying chain complexes of the two differential graded algebras \eqref{eq:CEat} by
\begin{equation}
    \begin{aligned}
    \mathrm{CE}(\mathfrak{at}_{\mathcal{G}}) \;&=\;  \Big(\Coo(M) \xrightarrow{\;\,\di\,\;}\Gamma(M,\,TM\oplus T^\ast M)\Big)\\[0.25em]
    \Big(\Coo\big(T^\ast[2]T[1]M\big),\,Q_H\Big) \;&=\; \Big(\Coo(M) \xrightarrow{\;\,\di\,\;}\Gamma(M,\,TM\oplus T^\ast M)\Big),
    \end{aligned}
\end{equation}
moreover the derived bracket structure (see \cite{DesSae18} for details) defined by the cohomological vector $Q_H$ on $\Coo\big(T^\ast[2]T[1]M\big)$ is exactly the bracket structure of the Courant $2$-algebroid:
\begin{equation}
    \begin{aligned}
    \ell_1(f) \;&=\; \di f \\
    \ell_2(X+\xi,\,Y+\eta) \;&=\; [X,Y] + \mathcal{L}_X\eta-\mathcal{L}_Y\xi - \frac{1}{2}\di\langle X+\xi,\,Y+\eta \rangle+ \iota_X\iota_YH \\
    \;&=\; [X+\xi,\,Y+\eta]_{\mathrm{Cou}} \\
    \ell_2(X+\xi,\,f) \;&=\; \mathcal{L}_Xf \\
    \ell_2(X+\xi,\,Y+\eta,\,Z+\zeta) \;&=\; \frac{1}{3!}\big\langle X+\xi,\,[Y+\eta,\,Z+\zeta]_{\mathrm{Cou}} \big\rangle+\mathrm{cycl.}
    \end{aligned}
\end{equation}
where $[-,-]_{\mathrm{Cou}}$ is the Courant bracket and $\langle-,-\rangle$ is the bundle metric defined by the contraction $\langle X+\xi,\,Y+\eta\rangle= \iota_X\eta+\iota_Y\xi$ for every sections $X+\xi,Y+\eta\in\Gamma(M,TM\oplus T^\ast M)$. 

\paragraph{Symplectic $L_\infty$-algebroids as NQP-manifolds.}
The Courant $2$-algebroid is canonically a symplectic $2$-algebroid (see \cite{FRS18} for details), i.e. it can be equipped with a canonical symplectic form $\omega$, which can be easily expressed in local coordinates on the corresponding NQ-manifold. 
On each local patch of the NQ-manifold $T^\ast[2]T[1]M$ we can choose local coordinates $(x^\mu,\,  e^\mu,\, \bar{e}_\mu,\, p_\mu)$ where the $x^\mu$ are in degree $0$, while the $(e^\mu,\, \bar{e}_\mu)$ are both in degree $1$ and the $p_\mu$ are in degree $2$. On the local patches we can express the symplectic form $\omega\in\Omega^2(T^\ast[2]T[1]M)$ in local coordinates by
\begin{equation}
    \omega = \di x^\mu \wedge \di p_\mu + \di e^\mu \wedge \di \bar{e}_\mu.
\end{equation}
We can also use Hamilton's equations $\iota_{Q_H}\omega=\mathcal{Q}_H$ to express the vector $Q_H$ by an Hamiltonian function $\mathcal{Q}_H$. We find
\begin{equation}
    \mathcal{Q}_H = e^\mu p_\mu + H_{\mu\nu\lambda} e^\mu e^\nu e^\lambda
\end{equation}
where $H\in\Omega^3_{\mathrm{cl}}(M)$ is a representative of the Dixmier-Douady class $[H]\in H^3(M,\mathbb{Z})$ of the original bundle gerbe $\mathcal{G}\twoheadrightarrow M$. This class in the literature of differential graded manifolds changes name in \textit{\v{S}evera class}.

\paragraph{A pre-NQP-manifold for DFT.}
By following \cite{DesSae18}, we choose as $2d$-dimensional base manifold $M=T^\ast U$ the cotangent bundle of some $d$-dimensional local patch. This is because we are interested in the local geometry of the doubled space and we have still no information about how to patch together these local $2d$-dimensional $T^\ast U$ manifolds. Thus the Courant algebroid on $T^\ast U$ will be given by the NQP-manifold $T^\ast[2]T[1](T^\ast U)$, as we have seen.
This will have coordinates $(x^M,\,  e^M,\, \bar{e}_M,\, p_M)$ still respectively in degrees $0$, $1$, $1$ and $2$, but with $M=1,\dots,2d$. We must then think the local coordinates $x^M = (x^\mu, \tilde{x}_\mu)$ to be the doubled coordinates of DFT. 
\vspace{-0.2cm}

\noindent Since $T^\ast U$ is canonically equipped with the tensor $\eta_{MN}$, we can make a change of degree $1$ coordinates by
\begin{equation}
    E^M \,:=\, \frac{1}{\sqrt{2}}(e^M + \eta^{MN}\bar{e}_N), \qquad  \bar{E}_M \,:=\, \frac{1}{\sqrt{2}}(\bar{e}_M - \eta_{MN}e^N)
\end{equation}
Now we must restrict ourselves to the submanifold $\mathcal{M}:=\{\bar{E}_M=0\}$ of the original manifold $T^\ast[2]T[1](T^\ast U)$. It is not hard to check that this submanifold will be $\mathcal{M}=(T^\ast[2]\oplus T[1])(T^\ast U)$. The degree $1$ functions on $\mathcal{M}$ will then be doubled vectors of the form
\begin{equation}
    X^\mu(x,\tilde{x})\left(\frac{\partial}{\partial x^\mu}+\di \tilde{x}_\mu\right) + \xi_\mu(x,\tilde{x})\left(\frac{\partial}{\partial \tilde{x}_\mu}+\di x^\mu\right)
\end{equation}
and the degree $0$ functions will be just ordinary functions of the form $f\in\Coo(T^\ast U)$. The symplectic form restricted to the submanifold $\mathcal{M}$ will now be
\begin{equation}
    \omega|_{\mathcal{M}}\,=\, \di x^M \wedge \di p_M + \frac{1}{2}\eta_{MN}\di E^M \wedge \di E^N.
\end{equation}
The new Hamiltonian function will be $\mathcal{Q}|_{\mathcal{M}}=E^M p_M + H_{MNL}E^ME^NE^L$, where $H_{MNL}$ now is the curvature of a bundle gerbe on the $2d$-dimensional base $T^\ast U$, which we should think as the extended fluxes of DFT. Crucially our $\mathcal{M}$ will still be a symplectic graded manifold, however it will not be a NQP-manifold since the new restricted vector $Q$ is not nilpotent on $\mathcal{M}$, i.e. we have that $Q^2\neq 0$. This is exactly the reason why \cite{DesSae18} named $\mathcal{M}$ \textit{pre-NQP manifold} and therefore this cannot be seen an $L_\infty$-algebroid.
\vspace{0.2cm}

\noindent However this pre-NQP manifold satisfies a very interesting property: the pre-NQP-manifold has a number of sub-manifold which are proper NQP-manifolds and thus well-defined sub-$2$-algebroids. Schematically we have
\begin{equation}
    \mathrm{CE}(\mathfrak{a}) \;\subset\; \big(\Coo(\mathcal{M}),\,Q\big)
\end{equation}
where $\mathfrak{a}$ is one of these sub-$2$-algebroids.
On any of these, the bracket of doubled vectors in degree $1$ will be exactly the D-bracket of DFT, which will be given by $\llbracket X,Y \rrbracket_{\mathrm{D}} := \{QX,Y\}$. For instance we can choose the differential graded algebra of functions which are pullbacks from the submanifold $\mathcal{N}:=\{\tilde{x}_\mu=\tilde{p}^\mu=0\}\subset\mathcal{M}$, which is exactly the Courant $2$-algebroid $\mathcal{N}= T^\ast[2]T[1]U$. This corresponds to choosing a sub-$2$-algebroid which satisfies the strong constraint and therefore this restriction reduces the pre-NQP-geometry to bare Generalized Geometry on the manifold $U$. Any other solution of the strong constraint will correspond to a viable choice of sub-$2$-algebroid. \vspace{0.25cm}

\noindent We can also introduce tensors of the form $\mathcal{H}_{MN}E^M\otimes E^N$ on $\mathcal{M}$ and use the Poisson bracket to define a natural notion of D- and C-bracket on tensors. This allows to define a notion of generalized metric, curvature and torsion in analogy with Riemannian geometry.

\paragraph{An example of global pre-NQP manifold.}
It is well-known that higher geometry is the natural framework for geometric T-duality, see the formalization by \cite{BunNik13}, \cite{FSS16x}, \cite{FSS17x}, \cite{FSS18}, \cite{FSS18x} and \cite{NikWal18}. Assume that we have two $T^n$-bundle spacetimes $M\xrightarrow{\pi}M_0$ and $\widetilde{M}\xrightarrow{\tilde{\pi}}M_0$ over a common $(d-n)$-dimensional base manifold $M_0$. A couple of bundle gerbes $\mathscr{G}\xrightarrow{\Pi}M$ and $\widetilde{\mathscr{G}}\xrightarrow{\tilde{\Pi}}\widetilde{M}$, formalizing two Kalb-Ramond fields respectively on $M$ and $\widetilde{M}$, are geometric T-dual if the following isomorphism exists
\begin{equation}
    \begin{tikzcd}[row sep={12ex,between origins}, column sep={12ex,between origins}]
    & \mathscr{G}\times_{M_0} \widetilde{M}\arrow[rr, "\cong"', "\text{T-duality}"]\arrow[dr, "\Pi"']\arrow[dl, "\tilde{\pi}"] & & M\times_{M_0}\widetilde{\mathscr{G}}\arrow[dr, "\pi"']\arrow[dl, "\tilde{\Pi}"] \\
    \mathscr{G}\arrow[dr, "\Pi"'] & & M\times_{M_0}\widetilde{M}\arrow[dr, "\pi"']\arrow[dl, "\tilde{\pi}"] & & [-2.5em]\widetilde{P}\arrow[dl, "\tilde{\Pi}"] \\
    & M\arrow[dr, "\pi"'] & & \widetilde{M}\arrow[dl, "\tilde{\pi}"] & \\
    & & M_0 & &
    \end{tikzcd}
\end{equation}
This picture is nothing but the finite version of T-duality between Courant algebroids illustrated by \cite{CavGua11}. Now, in \cite{DesSae19} it is proposed that we should consider the fiber product of the pull-back of both the gerbes $\mathscr{G}$ and $\widetilde{\mathscr{G}}$ to the correspondence space $M\times_{M_0}\widetilde{M}$ of the T-duality, which will be itself a gerbe of the form
\begin{equation}\label{eq:dsgerbe}
    \Pi\otimes\tilde{\Pi}:\; \mathscr{G}\otimes\widetilde{\mathscr{G}}\;\longtwoheadrightarrow\; M\times_{M_0}\widetilde{M}.
\end{equation}
Now, as previously explained, we can take the algebroid of infinitesimal gauge transformations of this gerbe $\mathfrak{at}_{\mathscr{G}\otimes\widetilde{\mathscr{G}}}\longtwoheadrightarrow M\times_{M_0}\widetilde{M}$ and express it as a differential graded manifold $\big( T^\ast[2]T[1](M\times_{M_0}\widetilde{M}),\,Q\big)$ with local coordinates $(x^\mu,x^I,e^\mu, e^I, \bar{e}_\mu, \bar{e}_I, p_\mu, p_I)$ with indices $\mu=1,\dots,d-n$ and $I=1,\dots,n$. Now, as we explained for the local doubled space, we can change coordinates to $E^I:=(e^I+\eta^{IJ}\bar{e}_J)/\sqrt{2}$ and $\bar{E}_I:=(\bar{e}_I+\eta_{IJ}e^J)/\sqrt{2}$ and set $\bar{E}_I=0$ to zero so that we obtain a new differential graded manifold $\mathcal{M}$. This new manifold will be locally isomorphic to $(T^\ast[2]\oplus T[1])T^{2n} \oplus T^\ast[2]T[1]U$ on each patch $U\subset M_0$ of the base manifold, but which is globally well-defined.
In \cite{DesSae19} this machinery is applied for fiber dimension $n=1$ to the particular case where $M$ and $\widetilde{M}$ are nilmanifolds on a common base torus $M_0=T^2$.

\paragraph{Further discussion.}
This is the first proposal to interpret strong constrained doubled vectors as sections of the $2$-algebroid of the local symmetries of a gerbe: the Courant $2$-algebroid. This suggests that it could be a complementary approach to the ones attempting to realize the doubled space as a geometrization the bundle gerbe itself.
\vspace{0.25cm}

\noindent However there are still some open problems. 
The only non-trivial global case that was constructed in this framework was, as we saw, on the correspondence space $M\times_{M_0}\widetilde{M}$ equipped with the pullback of both the gerbe $\mathscr{G}$ and its dual $\widetilde{\mathscr{G}}$. But, for this construction, the correspondence space of the T-duality is not derived from the pre-NQP manifold theory, but it must be assumed and prepared by using the machinery of topological T-duality. Besides, the total gerbe \eqref{eq:dsgerbe} has "repeated" information: for example, if we start from a gerbe $\mathscr{G}_{i,j}$ with Dixmier-Douady number $i$ on a nilmanifold with $1$st Chern number $j$, its dual will be a gerbe $\widetilde{\mathscr{G}}_{j,i}$ on a nilmanifold with inverted Dixmier-Douady and $1$st Chern number. Now the total gerbe $\mathscr{G}_{i,j}\otimes\widetilde{\mathscr{G}}_{j,i}$ contain each number twice: as $1$st Chern number and as Dixmier-Douady number. Moreover, in literature, a globally defined pre-NQP manifold for a non-trivially fibrated spacetime $M$ was proposed only for the case of geometric T-duality. Recently \cite{Wat19} applied pre-NQP geometry to the case of DFT on group manifolds. However the extension of this formalism to general T-dualizable backgrounds is not immediate.

\subsection{Tensor hierarchies proposal}

The idea of tensor hierarchy was introduced in \cite{HohSam13KK} in the context of the dimensional reduction of DFT, then further formalized in \cite{Hohm19DFT}, \cite{Hohm19} and \cite{Hohm19x} as a higher gauge structure. See also work by \cite{Ced20a} and \cite{Ced20b}. 

\paragraph{Embedding tensor and Leibniz-Loday algebra.}

Let $\rho:\mathfrak{o}(d,d)\otimes R \rightarrow R$ be the fundamental representation of the Lie algebra $\mathfrak{o}(d,d)$ of the Lie group $O(d,d)$. The vector space underlying the fundamental representation of $O(d,d)$ is nothing but $R \cong \mathbb{R}^{2d}$. Let us use the notation $\mathbf{x}\otimes Y\mapsto \rho_\mathbf{x}Y \in R$. The \textit{embedding tensor} of DFT is defined as a linear map $\Theta: \mathfrak{X}(\mathbb{R}^{2d}) \,\hookrightarrow\, \Coo(\mathbb{R}^{2d},\,\mathfrak{o}(d,d))$ which satisfies the following compatibility condition, usually called \textit{quadratic constraint}:
\begin{equation}
    [\Theta(X),\Theta(Y)] \,=\, \Theta(\rho_{\Theta(X)}Y),
\end{equation}
where $[-,-]$ are the Lie bracket of the Lie algebra $\mathfrak{o}(d,d)$.
Concretely the embedding tensor maps a vector field by $X^M\mapsto (X^M\!,\,\partial_{[M}X_{N]})\in\Coo(\mathbb{R}^{2d},\,\mathfrak{o}(d,d))$. Now the embedding tensor defines a natural action of $\mathfrak{X}(\mathbb{R}^{2d})$ on itself by
\begin{align}
    \circ:\,\mathfrak{X}(\mathbb{R}^{2d}) \otimes \mathfrak{X}(\mathbb{R}^{2d}) \;&\longrightarrow\; \mathfrak{X}(\mathbb{R}^{2d})\\
    (X,Y) \;&\longmapsto\; X \circ Y \,:=\, \rho_{\Theta(X)}Y.
\end{align}
This is exactly the D-bracket of DFT. Thus the anti-symmetric part will be the C-bracket
\begin{equation}
\begin{aligned}
    \frac{1}{2}(X\circ Y - Y\circ X) \,=\, \llbracket X,Y\rrbracket_{\mathrm{C}}.
\end{aligned}
\end{equation}
On the other hand the symmetric part of the D-bracket is given by $X\circ Y + Y\circ X \,=\, \mathfrak{D}\langle X,Y \rangle$, where $\mathfrak{D}:\Coo(\mathbb{R}^{2d}) \longrightarrow \mathfrak{X}(\mathbb{R}^{2d})$ is defined by $f\mapsto \partial^M\!{f}$ and the metric is defined by the contraction $\langle X,Y\rangle:=\eta_{MN}X^MY^N$. Therefore the D-bracket can be expressed in terms of these operators by
\begin{equation}
    X\circ Y \,=\, \llbracket X,Y\rrbracket_{\mathrm{C}} + \frac{1}{2}\mathfrak{D}\langle X,Y \rangle
\end{equation}
An interesting consequence is that the couple $\big(\mathfrak{X}(\mathbb{R}^{2n}),\,\circ\;\big)$ is not a Lie algebra, since the D-bracket is not anti-symmetric, but it is a Leibniz-Loday algebra, since it satisfies the Leibniz property $X\circ (Y\circ Z) = (X \circ Y)\circ Z + Y \circ (X \circ Z)$ for any triple of vectors $X,Y,Z\in\mathfrak{X}(\mathbb{R}^{2n})$. \vspace{0.25cm}

\noindent Now something remarkable happens: the Leibniz-Loday algebra $\big(\mathfrak{X}(\mathbb{R}^{2n}),\,\circ\;\big)$ of infinitesimal DFT gauge transformations naturally defines a Lie $2$-algebra $\big(\mathscr{D}(\mathbb{R}^{2n}),\,\ell_i\big)$ of infinitesimal DFT gauge transformations. This is given by the underlying cochain complex
\begin{equation}\label{eq:dofd}
    \mathscr{D}(\mathbb{R}^{2n}) \;:=\; \Big( \Coo(\mathbb{R}^{2n}) \xrightarrow{\;\;\mathfrak{D}\;\;} \mathfrak{X}(\mathbb{R}^{2n})\Big)
\end{equation}
equipped with the following $L_\infty$-bracket structure:
\begin{align}
    \ell_1(f) \;&=\; \mathfrak{D}f \\
    \ell_2(X,Y) \;&=\; \llbracket X,Y\rrbracket_{\mathrm{C}} \\
    \ell_2(X,f) \;&=\; \langle X,\mathfrak{D}f\rangle \\
    \ell_3(X,Y,Z) \;&=\; -\frac{1}{2}\langle\llbracket X,Y\rrbracket_{\mathrm{C}},Z\rangle + \mathrm{cycl.}
\end{align}
for any $f\in\Coo(\mathbb{R}^{2d})$ and $X,Y,Z\in\mathfrak{X}(\mathbb{R}^{2n})$.
Now notice that the quadratic constraint, which is the condition controlling the closure of the Leibniz bracket $X\circ Y$, requires to impose an additional constraint: this condition is nothing but the strong constraint. This makes the underlying complex of sheaves reduce to the one of sections of the standard Courant $2$-algebroid
\begin{equation}
    \mathscr{D}_{\mathrm{sc}}(\mathbb{R}^{2n}) \;=\; \Big( \Coo(\mathbb{R}^{n}) \xrightarrow{\;\;\di\;\;} \mathfrak{X}(\mathbb{R}^{n}) \oplus \Omega^1(\mathbb{R}^{d})\Big)
\end{equation}
Hence if we want $(\mathfrak{X}(\mathbb{R}^{2n}),\,\circ\,)$ to be a well-defined Leibniz-Loday algebra we need to restrict to Generalized Geometry and the D-bracket $\circ$ must reduce to the Dorfman bracket of Generalized Geometry, not twisted by any flux. At the present time no ways to generalize this construction beyond the strong constraint have been found.

\paragraph{Tensor hierarchies.}
Now that we have our well-defined $L_\infty$-algebra $\big(\mathscr{D}(\mathbb{R}^{2n}),\,\ell_n\big)$, we can ask ourselves what happens if we use it to construct an \textit{higher gauge field theory} on a $(d-n)$-dimensional manifold $M$. The answer is that the theory resulting from this gauging process is exactly a tensor hierarchy, which is supposed to describe DFT truncated at codimension $n$.
\vspace{0.25cm}

\noindent Luckily for our gauging purposes, there exists a well-defined notion of the tensor product of a differential graded algebra with an $L_\infty$-algebra (see \cite{JRSW19} for the formal definition). Thus we can define the prestack of local tensor hierarchies $\Omega\!\left(U,\,\mathscr{D}(\mathbb{R}^{2n})\right)$ by the tensor product of the differential graded algebra of the de Rham complex $(\Omega^\bullet(U),\,\di)$ with the $L_\infty$-algebra $\big(\mathscr{D}(\mathbb{R}^{2n}),\,\ell_i\big)$. In other words we define $\Omega\!\left(U,\,\mathscr{D}(\mathbb{R}^{2n})\right) := \Omega^\bullet(U) \otimes \mathscr{D}(\mathbb{R}^{2n})$ for any contractible open set $U\subset M$. Its underlying complex of sheaves of this prestack will be
\begin{equation}
    \Omega\!\left(U,\,\mathscr{D}(\mathbb{R}^{2n})\right) \,=\, \underbrace{\Coo(U\times \mathbb{R}^{2n})}_{\text{degree }0} \,\oplus\, \bigoplus_{k>0} \bigg(\underbrace{ \Omega^k(U)\otimes\Coo(\mathbb{R}^{2n}) \,\oplus\, \Omega^{k-1}(U)\otimes\mathfrak{X}(\mathbb{R}^{2n})}_{\text{degree }k}\bigg)
\end{equation}
and the bracket structure is found by applying the definition by \cite{JRSW19}. Explicitly, for any elements $\mathcal{A}_p\in\Omega^\bullet(U)\otimes \mathfrak{X}(\mathbb{R}^{2n})$ and $\mathcal{B}_p\in\Omega^\bullet(U)\otimes \Coo(\mathbb{R}^{2n})$, we have the following bracket structure:
\begin{equation}\begin{aligned}
   \ell_1(\mathcal{A}+\mathcal{B}) \;&=\; (\di \mathcal{A} + \mathfrak{D}\mathcal{B}) + \di \mathcal{B}\\
   \ell_2(\mathcal{A}_1,\mathcal{A}_2) \;&=\; -\llbracket \mathcal{A}_1\,\overset{\wedge}{,}\, \mathcal{A}_2 \rrbracket_{\mathrm{C}}\\
   \ell_2(\mathcal{A}_1,\mathcal{B}_2) \;&=\; \langle \mathcal{A}\,\overset{\wedge}{,}\, \mathfrak{D}\mathcal{B} \rangle \\
   \ell_2(\mathcal{B}_1,\mathcal{B}_2) \;&=\; 0 \\
   \ell_3(\mathcal{A}_1,\mathcal{A}_2,\mathcal{A}_3) \;&=\; -\frac{1}{2}\langle\llbracket \mathcal{A}_1\,\overset{\wedge}{,}\, \mathcal{A}_2 \rrbracket_{\mathrm{C}}\,\overset{\wedge}{,}\, \mathcal{A}_3 \rangle + \mathrm{cycl.} \\
   \ell_3(\mathcal{A}_1,\mathcal{A}_2,\mathcal{B}_3) \;&=\;  \ell_3(\mathcal{A}_1,\mathcal{B}_2,\mathcal{B}_3)\;=\;  \ell_3(\mathcal{B}_1,\mathcal{B}_2,\mathcal{B}_3) \;=\; 0  
\end{aligned}\end{equation}
where we introduced the following compact notation for $\mathscr{D}(\mathbb{R}^{2n})$-valued differential forms:
\begin{itemize}
    \item $\llbracket - \,\overset{\wedge}{,}\, -\rrbracket_{\mathrm{C}}$ is a wedge product on $\Omega^\bullet(U)$ and a C-bracket on $\mathfrak{X}(\mathbb{R}^{2n})$,
    \item $\langle - \,\overset{\wedge}{,}\, - \rangle$ is a wedge product on $\Omega^\bullet(U)$ and a contraction $\langle X,Y\rangle=\eta_{IJ}X^IY^J$ on $\mathfrak{X}(\mathbb{R}^{2n})$.
\end{itemize}
The prestack $\Omega\!\left(U,\,\mathscr{D}(\mathbb{R}^{2n})\right)$ encodes the local fields of a tensor hierarchy on a local doubled space of the form $U\times \mathbb{R}^{2n}$ with base manifold $\mathrm{dim}(U)=d-n$.
In our degree convention the connection data of a tensor hierarchy is given by a degree $2$ multiplet
\begin{equation}
    \begin{aligned}
    \mathcal{A}^I_\mu \;&\in\; \Omega^1(U)\otimes \mathfrak{X}(\mathbb{R}^{2n}) \\
    \mathcal{B}_{\mu\nu} \;&\in\; \Omega^2(U)\otimes \Coo(\mathbb{R}^{2n}) 
    \end{aligned}
\end{equation}
while its curvature is given by the degree $3$ multiplet 
\begin{equation}
    \begin{aligned}
    \mathcal{F}^I_{\mu\nu} \;&\in\; \Omega^2(U)\otimes \mathfrak{X}(\mathbb{R}^{2n}) \\
    \mathcal{H}_{\mu\nu\lambda} \;&\in\; \Omega^3(U)\otimes \Coo(\mathbb{R}^{2n}) 
    \end{aligned}
\end{equation}
Notice that all the fields of the hierarchy depend not just on the coordinates $x$ of the base manifold $U$, but also on the coordinates $(y, \tilde{y})$ of the vector space $\mathbb{R}^{2n}$. The curvature of the tensor hierarchy can be expressed in terms of the connection, as it is found in \cite{HohSam13KK}, by
\begin{equation}\label{eq:tensorhi}\boxed{\quad
    \begin{aligned}
    \mathcal{F} \;&=\; \di \mathcal{A} - \llbracket \mathcal{A} \,\overset{\wedge}{,}\, \mathcal{A}\rrbracket_{\mathrm{C}} + \mathfrak{D}\mathcal{B} \\[0.2em]
    \mathcal{H} \;&=\;  \mathrm{D}\mathcal{B} + \frac{1}{2}\langle \mathcal{A} \,\overset{\wedge}{,}\, \di A\rangle - \frac{1}{3!} \langle \mathcal{A} \,\overset{\wedge}{,}\,   \llbracket \mathcal{A} \,\overset{\wedge}{,}\, \mathcal{A}\rrbracket_{\mathrm{C}} \rangle 
    \end{aligned}\quad}
\end{equation}
where we introduced  the covariant derivative $\mathrm{D} := \di -  \mathcal{A}\circ\wedge$ defined by the $1$-form connection $A$, which acts explicitly by $\mathrm{D}\mathcal{A} = \di \mathcal{A} + \llbracket \mathcal{A} \,\overset{\wedge}{,}\, \mathcal{A}\rrbracket_{\mathrm{C}}$ and $\mathrm{D}\mathcal{B} = \di \mathcal{B} + \langle A\,\overset{\wedge}{,}\, \mathfrak{D} \mathcal{B}\rangle$.
Notice the characteristic C-bracket Chern-Simons term in the expression of $3$-form curvature. We will call $\mathrm{CS}_3(\mathcal{A})$, so we will be able to write the curvature of the tensor hierarchy in a compact fashion:
\begin{equation}
    \begin{aligned}
    \mathcal{F} \;&=\; \mathrm{D} \mathcal{A} + \mathfrak{D}\mathcal{B} \\
    \mathcal{H} \;&=\;  \mathrm{D}\mathcal{B} + \frac{1}{2}\mathrm{CS}_3(\mathcal{A}) 
    \end{aligned}
\end{equation}
By calculating the differential of the field curvature multiplet, this immediately gives the Bianchi identities of the tensor hierarchy:
\begin{equation}\boxed{\quad
    \begin{aligned}
    \mathrm{D}\mathcal{F} +\mathfrak{D}\mathcal{H}\;&=\; 0 \\
    \mathrm{D}\mathcal{H} - \frac{1}{2}\langle \mathcal{F} \;\overset{\wedge}{,}\; \mathcal{F} \rangle\;&=\;  0 
    \end{aligned}\quad} 
    \begin{aligned}
    \quad&\in\; \Omega^3(U)\otimes \mathfrak{X}(\mathbb{R}^{2n})\\
    \quad&\in\; \Omega^4(U)\otimes \Coo(\mathbb{R}^{2n})
    \end{aligned}
\end{equation}
The infinitesimal gauge transformations of a tensor hierarchy are given by degree $1$ multiplets
\begin{equation}
    \begin{aligned}
    \lambda^I \;&\in\; \Coo(U)\otimes \mathfrak{X}(\mathbb{R}^{2n}) \\
    \Xi_{\mu} \;&\in\; \Omega^1(U)\otimes \Coo(\mathbb{R}^{2n}) 
    \end{aligned}
\end{equation}
so that
\begin{equation}\boxed{\quad
    \begin{aligned}
    \mathcal{A} \;&\longmapsto\; \mathcal{A} + \mathrm{D}\lambda + \mathfrak{D}\Xi \\
    \mathcal{B} \;&\longmapsto\; \mathcal{B} + \mathrm{D}\Xi - \langle\lambda,\mathcal{F}\rangle
    \end{aligned}\quad}
\end{equation}
where the covariant derivative explicitly acts as $\mathrm{D}\lambda \;=\; \di \lambda + \llbracket \mathcal{A} , \lambda \rrbracket_{\mathrm{C}}$ and $\mathrm{D}\Xi \;=\; \di \Xi + \langle \mathcal{A}\,\overset{\wedge}{,}\, \mathfrak{D} \Xi\rangle$.
Notice the extraordinary similarity of these equations to the ones defining a principal $\mathrm{String}$-bundle. (This similarity will be discussed in section \ref{s5}).

\paragraph{Recovering doubled torus bundles.}
Notice that, in the particular case of a tensor hierarchy where all the fields do not depend on the internal space $\mathbb{R}^{2n}$, the field equations reduce to the familiar equations of a doubled torus bundle, i.e.
\begin{equation}
    \begin{aligned}
    \mathcal{F} \;&=\; \di \mathcal{A} && \in\;\Omega^2_{\mathrm{cl}}(U, \mathbb{R}^{2n}) \\
    \mathcal{H} \;&=\;  \di \mathcal{B} + \frac{1}{2}\langle \mathcal{A} \,\overset{\wedge}{,}\, \di \mathcal{A}\rangle && \in\;\Omega^3_{\mathrm{cl}}(U)
    \end{aligned}
\end{equation}
which is exactly the curvature of the $\mathrm{String}(T^n\times T^n)$-bundle raising in the case of a globally geometric T-duality, as explained in \cite{Alf19}. Also the gauge transformations reduce to
\begin{equation}
    \begin{aligned}
    \mathcal{A}^{ I} \;&\mapsto\; \mathcal{A}^{I} + \di\lambda^I \\
    \mathcal{B} \;&\mapsto\; \mathcal{B} + \di\Xi - \langle\lambda,\mathcal{F}\rangle
    \end{aligned}
\end{equation}
The local field $\mathcal{F}\in\Omega^2_{\mathrm{cl}}(U, \mathbb{R}^{2n})$ can thus be globalized to the curvature of a doubled torus bundle with $1$st Chern class $[\mathcal{F}]\in H^2(M,\mathbb{Z}^{2n})$. At this point topological T-duality is immediately encompassed by the $O(n,n;\mathbb{Z})$-rotation $[\widetilde{\mathcal{F}}]_I := \eta_{IJ}[\mathcal{F}]^J$ of the $1$st Chern class of the doubled torus bundle. \vspace{0.25cm}

\noindent This particular example of tensor hierarchy allows a globalization to a principal $2$-bundle with gauge $2$-group $\mathrm{String}(T^n\times T^n)$. Moreover, if we forget the higher form field, we stay with a well-defined $T^{2n}$-bundle on the $(d-n)$-dimensional base manifold $M$. This leads to the question about how to geometrically globalize and interpret general tensor hierarchies. 

\paragraph{Further discussion: the doubled space as a higher object.}
This proposal is the first to understand that the doubled connections $\mathcal{A}^I_\mu$, which we have also for the doubled torus bundles, are just a part of the full connection of the prestack $\Omega(-,\mathscr{D}(\mathbb{R}^{2n}))$ including also $\mathcal{B}_{\mu\nu}$. Thus the doubled space is intrinsically a higher geometric object.

\paragraph{Further discussion: what global picture?}
In \cite{Hohm19} it was proposed that the global higher gauge theory of tensor hierarchies on a $(d-n)$-dimensional manifold $M$ should consist in the $L_\infty$-algebra of $\mathscr{D}(\mathbb{R}^{2n})$-valued differential forms on $M$, i.e. the $L_\infty$-algebra we called $\Omega\big(M,\,\mathscr{D}(\mathbb{R}^{2n})\big)$ in our notation. However this must be taken as a local statement, since we know that gauge and $p$-form fields are not generally global differential forms on $M$, unless their underlying principal bundles are topologically trivial. Exactly like gauge fields, the global stack of tensor hierarchies must be instead given by the \textit{stackification} of the prestack of local tensor hierarchies $\Omega\big(-,\,\mathscr{D}(\mathbb{R}^{2n})\big)$. For a formal definition of stackification see \cite{topos}. This is true, at least, if we want to formalize tensor hierarchies as higher gauge theories. (In section \ref{s5} we will discuss a different perspective).
\vspace{0.25cm}

\noindent Let us thus define the \textit{stack of DFT tensor hierarchies} $\tenhie(-)$ as stackification of the prestack $\Omega(-,\mathscr{D}(\mathbb{R}^{2n}))$ of local tensor hierarchies. By construction this means that on any set $U\subset M$ of a good cover of our $(d-n)$-dimensional manifold $M$ we will have the isomorphism
\begin{equation}
    \tenhie(U) \;\cong\; \Omega\big(U,\,\mathscr{D}(\mathbb{R}^{2n})\big),
\end{equation}
This conveys the intuition that $\tenhie(-)$ is a globalization of $\Omega(-,\mathscr{D}(\mathbb{R}^{2n}))$, but not necessarily a topologically-trivial one. By construction $\tenhie(-)$ maps any $(d-n)$-dimensional smooth manifold $M$ to the $2$-groupoid $\tenhie(M)$ whose objects are tensor hierarchies and whose morphisms are gauge transformations of tensor hierarchies on $M$. 
\vspace{0.25cm}

\noindent In more concrete terms a global tensor hierarchy, which is an object of the $2$-groupoid $\tenhie(M)$, can be expressed in a local trivialization by a \v{C}ech cocycle. Given any good cover $\{U_\alpha\}$ for the $(d-n)$-dimensional manifold $M$, such a cocycle will be of the form
\begin{equation}
    \big(\mathcal{F}^I_{(\alpha)},\,\mathcal{H},\,\mathcal{A}_{(\alpha)}^I,\,\mathcal{B}_{(\alpha)},\,\lambda^I_{(\alpha\beta)},\, \Xi_{(\alpha\beta)},\,g_{(\alpha\beta\gamma)}\big) \;\in\; \tenhie(M)
\end{equation}
where the fields are of the following differential forms:
\begin{equation}
    \begin{aligned}
    \mathcal{F}^I_{(\alpha)\mu\mu} \;&\in\; \Omega^2(U_{\alpha})\otimes \mathfrak{X}(\mathbb{R}^{2n}) \\
    \mathcal{H}_{\mu\nu\lambda} \;&\in\; \Omega^3(M)\otimes \Coo(\mathbb{R}^{2n}) \\[0.5em]
    A^I_{(\alpha)\mu} \;&\in\; \Omega^1(U_{\alpha})\otimes \mathfrak{X}(\mathbb{R}^{2n}) \\
    B_{(\alpha)\mu\nu} \;&\in\; \Omega^2(U_{\alpha})\otimes \Coo(\mathbb{R}^{2n}) \\[0.5em]
    \lambda^I_{(\alpha\beta)} \;&\in\; \Coo(U_\alpha\cap U_\beta)\otimes \mathfrak{X}(\mathbb{R}^{2n}) \\
    \Xi_{(\alpha\beta)\mu} \;&\in\; \Omega^1(U_\alpha\cap U_\beta)\otimes \Coo(\mathbb{R}^{2n}) \\[0.5em]
    g_{(\alpha\beta\gamma)}  \;&\in\; \Coo(U_\alpha\cap U_\beta\cap U_\gamma \times \mathbb{R}^{2n})
    \end{aligned}
\end{equation}
and they are glued on two-fold, three-fold and four-fold overlap of patches as it follows:  
\begin{equation}\label{eq:tensorhipatched}
    \begin{aligned}
        \mathcal{F}_{(\alpha)} \;&=\; \mathrm{D} \mathcal{A}_{(\alpha)} + \mathfrak{D}\mathcal{B}_{(\alpha)} \\
    \mathcal{H} \;&=\;  \mathrm{D}\mathcal{B}_{(\alpha)} + \frac{1}{2}\mathrm{CS}_3(\mathcal{A}_{(\alpha)}) \\[0.8em]
    \mathcal{A}_{(\alpha)} \;&=\;  e^{-\lambda_{(\alpha\beta)}} (\mathcal{A}_{(\beta)}+\di)e^{\lambda_{(\alpha\beta)}} + \mathfrak{D}\Xi_{(\alpha\beta)} \\
    \mathcal{B}_{(\alpha)} - \mathcal{B}_{(\beta)} \;&=\; \mathrm{D}\Xi_{(\alpha\beta)} - \langle\lambda_{(\alpha\beta)},\mathcal{F}_{(\alpha)}\rangle \\[0.8em]
    e^{\lambda_{(\alpha\beta)}} e^{\lambda_{(\beta\gamma)}} e^{\lambda_{(\gamma\alpha)}} \;&=\; e^{\mathfrak{D} g_{(\alpha\beta\gamma)}} \\
    \Xi_{(\alpha\beta)} + \Xi_{(\beta\gamma)} + \Xi_{(\gamma\alpha)} \;&=\; \di g_{(\alpha\beta\gamma)} \\[0.8em]
    g_{(\alpha\beta\gamma)} - g_{(\beta\gamma\delta)} +  g_{(\gamma\delta\alpha)} -  g_{(\delta\alpha\beta)}\;&\in\;2\pi\mathbb{Z}
    \end{aligned}
\end{equation}
where the covariant derivatives are $\mathrm{D}=\di- \mathcal{A}_{(\alpha)} 
\circ\wedge$. Notice the similarity, at least locally, of the potential $\mathcal{A}_{(\alpha)}\in\Omega^1\big(U_\alpha\times\mathbb{R}^{2n},\,U_\alpha\times T\mathbb{R}^{2n}\big)$ with the non-principal connection defined for instance in \cite[pag.$\,$77]{Nat93} for a general bundle. \vspace{0.25cm}

\noindent In section \ref{s5} we discuss the possibility of a more general definition of global tensor hierarchy, which can be obtained by directly dimensionally reducing the bundle gerbe and not as a higher gauge theory.

\paragraph{Further discussion.}
However, if we accept that the global picture of tensor hierarchies is a higher gauge algebra, we would still have some open questions. From \cite{HohSam13KK} we know that a tensor hierarchy is supposed to be a split version of DFT with a $(d-n)$-dimensional base manifold $M$ and $2n$-dimensional fibers for an arbitrary $n$. But since tensor hierarchies are higher gauge theories, this hints that the full $2d$-dimensional doubled space should carry a bundle gerbe structure. Such structure, as we have seen for previous proposals, still needs to be clarified.

\subsection{Born Geometry}

The first proposal of interpretation of DFT geometry as para-K\"{a}hler manifold is developed by \cite{Vais12} and then generalized to para-Hermitian manifolds by \cite{Vai13}. These ideas were further elaborated by \cite{Svo17}, \cite{Svo18}, \cite{MarSza18}, \cite{Svo19}, \cite{MarSza19} and \cite{Svo20}. This proposal sees the doubled space as a $2d$-dimensional smooth manifold, whose tangent bundle is canonically split in two rank $d$ lagrangian subbundles. The fluxes of DFT are then interpreted as the obstruction for the integrability of this structure.

\paragraph{Doubled space as an almost para-Hermitian manifold.}
An \textit{almost para-complex manifold} $(\mathcal{M},K)$ is a $2d$-dimensional manifold which is equipped with a $(1,1)$-tensor field $K\in\mathrm{End}(T\mathcal{M})$ such that $K^2=\mathrm{id}_{T\mathcal{M}}$, called \textit{almost para-complex structure}, and such that the $\pm 1$-eigenbundles $L_\pm\subset T\mathcal{M}$ of $K$ have both $\mathrm{rank}(L_\pm)=d$. Thus, since the para-complex structure defines a splitting $T\mathcal{M}=L_+\oplus L-$, the structure group of the tangent bundle $T\mathcal{M}$ of the almost para-complex manifold is $GL(d,\mathbb{R})\times GL(d,\mathbb{R})$. \vspace{0.25cm}

\noindent 
The almost para-complex structure $K$ is said to be $\pm$\textit{-integrable} if $L_\pm$ is closed under Lie bracket, i.e. if it satisfies the property
\begin{equation}
    \big[\Gamma(\mathcal{M},L_\pm),\,\Gamma(\mathcal{M},L_\pm)\big]_{\mathrm{Lie}} \,\subseteq\, \Gamma(\mathcal{M},L_\pm)
\end{equation}
The $\pm$-integrability of $K$ implies that there exists a foliation $\mathcal{F}_\pm$ on the manifold $\mathcal{M}$ such that $L_\pm = T \mathcal{F}_\pm$. An almost para-complex manifold $(\mathcal{M},K)$ is a proper \textit{para-complex manifold} if and only if $K$ is both $+$-integrable and $-$-integrable at the same time.
\vspace{0.25cm}

\noindent Our almost para-complex manifold $(\mathcal{M},K)$ becomes an \textit{almost para-Hermitian manifold} if we equip it with a metric $\eta \in \bigodot^2T^\ast\mathcal{M}$ of Lorentzian signature $(d,d)$ which is compatible with the almost para-complex structure by
\begin{equation}
    \eta(K-,K-) \,=\, - \eta(-,-).
\end{equation}
Now we have a natural $2$-form which is defined by $\omega(-,-) := \eta(K-,-)\in\Omega^2(\mathcal{M})$, called \textit{fundamental $2$-form}. Notice that the subbundles $L_\pm$ are both \textit{maximal isotropic} subbundles respect to $\eta$ and \textit{Lagrangian} subbundles respect to $\omega$. An almost para-Hermitian manifold $(\mathcal{M},K,\eta)$ becomes a \textit{para-K\"{a}hler manifold} if the fundamental $2$-form is closed, i.e. $\di\omega=0$. \vspace{0.25cm}

\noindent The closed $3$-form $\mathcal{K}\in\Omega^3_{\mathrm{cl}}(\mathcal{M})$ defined by $\mathcal{K}:=\di\omega$ is interpreted as encoding the \textit{fluxes} of DFT.

\paragraph{Born Geometry.}
A \textit{Born Geometry} is the datum of an almost para-Hermitian manifold $(\mathcal{M},K,\eta)$ equipped with a Riemannian metric $\mathcal{G}\in \bigodot^2T^\ast\mathcal{M}$ which is compatible with both the metric $\eta$ and the fundamental $2$-form $\omega$ by
\begin{equation}
    \eta^{-1}\mathcal{G} \,=\, \mathcal{G}^{-1}\eta \quad\;\text{and}\;\quad \omega^{-1}\mathcal{G} \,=\, -\mathcal{G}^{-1}\omega.
\end{equation}
This Riemannian metric must be identified with the generalized metric of DFT.

\paragraph{Generalized T-dualities.}
The generalized diffeomorphisms of DFT are now identified with diffeomorphisms of $\mathcal{M}$ preserving the metric $\eta$, i.e isometries $\mathrm{Iso}(\mathcal{M},\eta)$. Notice that the push-forward of a generalized diffeomorphism $f\in\mathrm{Iso}(\mathcal{M},\eta)$ is nothing but an $O(d,d)$-valued function $f_\ast\in\Coo(\mathcal{M},O(d,d))$, as expected. This group of symmetries can be further extended to the the group of \textit{generalized T-dualities}, which are general bundle automorphisms of $T\mathcal{M}$ preserving the metric $\eta$.
\vspace{0.25cm}

\noindent A generalized T-duality $f\in\mathrm{Aut}(T\mathcal{M})$ induces a morphism of Born geometries on $\mathcal{M}$ by
\begin{equation}
    (\mathcal{M},\,K,\,\eta,\,\mathcal{G})\;\longmapsto\;(\mathcal{M},\, f\circ K\circ f^{-1},\,\eta,\, f^\ast\mathcal{G}),
\end{equation}
which implies also the transformation $\omega \mapsto f^\ast\omega$ of the fundamental $2$-form. 
\vspace{0.2cm}

\noindent Particularly interesting is the case of $b$-transformations which we can see as a bundle morphisms $e^b:T\mathcal{M}\rightarrow T\mathcal{M}$ covering the identity $\mathrm{id}_{\mathcal{M}}$ of the base manifold. $K\mapsto K+b$. Therefore a $b$-transformation maps the splitting $T\mathcal{M} = L_+\oplus L_-$ to a new one $T\mathcal{M} =L_+^b\oplus L_-$ which preserves $L_-$, but does not preserve $L_+$ and $+$-integrability. This also implies $\omega\mapsto \omega+2b$.

\paragraph{Further discussion.}
We can notice that Born Geometry is not (at least immediately) related to bundle gerbes, even if theory of foliations is closely related to higher structures as seen by \cite{Vit14}. In the next subsection we will mostly discuss the relation between Born Geometry and the bundle gerbe of the Kalb-Ramond field, trying to clarify it.

\subsection{Can DFT actually recover bosonic supergravity?}

\paragraph{Recovering physical spacetime.}
We will now try to recover a general bosonic string background, consisting in a pseudo-Riemannian manifold $(M,g)$ equipped with a non-trivial H-flux $[H]\in H^3(M,\mathbb{Z})$, from Born Geometry as prescribed by \cite{MarSza19} and \cite{Svo20}.
\vspace{0.25cm}

\noindent Let us start from the almost para-Hermitian manifold $(\mathcal{M},K,\eta)$. The para-complex structure $K$ splits the tangent bundle $T\mathcal{M}= L_+ \oplus L_-$ where $L_\pm$ are its $\pm 1$-eigenbundles. Since we want to recover a conventional supergravity background let us firstly assume that $L_-$ is integrable (physically this corresponds to set the $R$-flux to zero, see \cite{MarSza19}). This implies that there exists a foliation $\mathcal{F}_-$ of $\mathcal{M}$ such that $L_- = T\mathcal{F}_-$. Secondly, since we want to recover a conventional supergravity background, let us require that the leaf space $M:=\mathcal{M}/\mathcal{F}_-$ of this foliation is a smooth manifold. Indeed, according to \cite{MarSza19} and \cite{Svo20}, physical spacetime must be identified with the leaf space $M$. Thus the foliation $\mathcal{F}_-$ is simple and the canonical quotient map $\pi:\mathcal{M}\twoheadrightarrow M= \mathcal{M}/\mathcal{F}_-$ is a surjective submersion, making $\mathcal{M}$ a fibered manifold.
\vspace{0.25cm}

\noindent Now we can use adapted (or fibered) coordinates $(x_{(\alpha)},\tilde{x}_{(\alpha)})$ on each patch $\mathcal{U}_{\alpha}$ of a good cover of the manifold $\mathcal{M}=\bigcup_\alpha \mathcal{U}_{\alpha}$. Thus there exist a frame $\{Z_\mu,\tilde{Z}^\mu\}$ and a dual coframe $\{e^\mu,\tilde{e}_\mu\}$, given on local patches $\mathcal{U}_{\alpha}$ as follows
\begin{equation}
    \begin{aligned}
    Z_{(\alpha)\mu} \,&=\, \frac{\partial}{\partial x^\mu_{(\alpha)}} + N_{(\alpha)\mu\nu}\frac{\partial}{\partial \tilde{x}_{(\alpha)\nu}}, \quad & \tilde{Z}^\mu_{(\alpha)} \,&=\, \frac{\partial}{\partial \tilde{x}_{(\alpha)\mu}} \\
    e^\mu_{(\alpha)} \,&=\, \di x^\mu_{(\alpha)}, \quad & \tilde{e}_{(\alpha)\mu} \,&=\, \di \tilde{x}_{(\alpha)\mu} + N_{(\alpha)\mu\nu}\di x^\nu_{(\alpha)}
    \end{aligned}
\end{equation}
such that they diagonalize the tensor $K$ and such that $\{Z_\mu\}$ is a local completion of the holonomic frame for $\Gamma(L_-)$. Notice that the $N_{(\alpha)\mu\nu}\in\Coo\!\left(\mathcal{U}_{\alpha}\right)$ are local functions. In this frame we can express the global $O(d,d)$-metric $\eta = \eta^\mu_\nu \tilde{e}_\mu \odot e_\nu$ and the fundamental $2$-form $\omega = \eta^\mu_\nu \tilde{e}_\mu \wedge e_\nu$. In local coordinates $(x_{(\alpha)},\tilde{x}_{(\alpha)})$ the latter can be written on each patch $\mathcal{U}_{\alpha}$ as
\begin{equation}
    \omega|_{\mathcal{U}_{\alpha}} \,=\, \eta^\mu_\nu \di \tilde{x}_{\mu(\alpha)} \wedge \di x^\nu_{(\alpha)} + \eta^\mu_\nu N_{(\alpha)\mu\lambda}\di x^\lambda_{(\alpha)}\wedge \di x^\nu_{(\alpha)}
\end{equation}
Now, by following \cite{MarSza19}, we can define a local $2$-form $B_{(\alpha)}\in\Omega^2\!\left(\mathcal{U}_{\alpha}\right)$ by the second term of the $2$-form $\omega|_{\mathcal{U}_{\alpha}}$, i.e.
\begin{equation}
    B_{(\alpha)} \,:=\, \eta^\mu_\nu N_{(\alpha)\mu\lambda}\di x^\lambda_{(\alpha)}\wedge \di x^\nu_{(\alpha)}.
\end{equation}
Now we must ask: what is the condition to make the local $2$-form $B_{(\alpha)}$ on $\mathcal{U}_\alpha$ descend to a proper local $2$-form on the leaf space $M=\mathcal{M}/\mathcal{F}_-$ (which is the physical spacetime)? By following \cite{MarSza19} we can impose the condition that $N_{(\alpha)\mu\lambda}$ are basic functions, i.e.
\begin{equation}
    \mathcal{L}_{X_-}N_{(\alpha)\mu\lambda} =0 \quad \forall X_-\in\Gamma(L_-),
\end{equation}
which assures exactly this. In local coordinates on $\mathcal{U}_{\alpha}$ this condition can be rewritten as
\begin{equation}
    \frac{\partial N_{(\alpha)\mu\nu}}{\partial \tilde{x}_{(\alpha)\lambda}} =0
\end{equation}
which is solved by $N_{(\alpha)\mu\nu}=N_{(\alpha)\mu\nu}\!\left(x_{(\alpha)}\right)$, i.e. by asking that the $N_{(\alpha)\mu\nu}$ are local functions only of the $x_{(\alpha)}$-coordinates on each patch $\mathcal{U}_{\alpha}$. Therefore, if our local $2$-form is of the form $B_{(\alpha)}= B_{(\alpha)\mu\nu}\!\left(x_{(\alpha)}\right)\di x^\mu_{(\alpha)}\wedge \di x^\nu_{(\alpha)}$ it will descend to a local $2$-form $\pi_{\ast} B_{(\alpha)}\in\Omega^2(U_{\alpha})$ on $M$.

\paragraph{Papadopoulos' puzzle revised.}
In the adapted (or fibered) coordinates the transition functions of $\mathcal{M}$ on two-fold overlaps of patches $\mathcal{U}_{\alpha}\cap\mathcal{U}_{\beta}$ will have the simple following form:
\begin{equation}\label{eq:introadapted}
    x_{(\alpha)} = f_{(\alpha\beta)}\!\left(x_{(\beta)}\right), \quad \tilde{x}_{(\alpha)} = \tilde{f}_{(\alpha\beta)}\!\left(x_{(\beta)},\tilde{x}_{(\beta)}\right).
\end{equation}
An adapted atlas will be also provided with the property that the sets $U_{\alpha}:=\pi\!\left(\mathcal{U}_{\alpha}\right)$, where $\pi:\mathcal{M}\twoheadrightarrow M= \mathcal{M}/\mathcal{F}_-$ is the quotient map, are patches of the leaf space $M$ with local coordinates $(q_{(\alpha)})$ defined by the equation $x_{(\alpha)}^\mu = q_{(\alpha)}^\mu\circ \pi$. These charts $(U_{\alpha},q_{(\alpha)})$ are uniquely defined. The local $2$-form $B_{(\alpha)}$ will then descend to the local $2$-form $\pi_{\ast} B_{(\alpha)} = B_{(\alpha)\mu\nu}\!\left(q_{(\alpha)}\right)\di q^\mu_{(\alpha)}\wedge \di q^\nu_{(\alpha)}$.
\vspace{0.25cm}

\noindent Since $\omega\in\Omega^2(\mathcal{M})$ is a global $2$-form, on each two-fold overlap of patches $\mathcal{U}_{\alpha}\cap\mathcal{U}_{\beta}$ we have
\begin{equation}\label{eq:introdiff}
    \begin{aligned}
         \di \tilde{x}_{\mu(\alpha)} \wedge \di x^\mu + B_{(\alpha)} \,=\, \di \tilde{x}_{\mu(\beta)} \wedge \di x^\mu + B_{(\beta)},
    \end{aligned}
\end{equation}
where we suppressed the patch indices on the $1$-forms $\{\di x^\mu\}$: this is because $\{e^\mu\}$ are global $1$-forms on $\mathcal{M}$ and thus we can slightly abuse the notation by calling $e^\mu\equiv \di x^\mu$. Thus we have
\begin{equation}\label{eq:introdx}
(\di \tilde{x}_{(\alpha)\mu} - \di \tilde{x}_{(\beta)\mu}) \wedge \di x^\mu  =  B_{(\beta)} - B_{(\alpha)}.
\end{equation}
Since the local $2$-forms $B_{(\alpha)}$ descend to local $2$-forms on patches $U_{\alpha}\subset M=\mathcal{M}/\mathcal{F}_-$ of the leaf space and these, according to \cite{MarSza19} and \cite{Svo20}, must be physically identified with the local data of the Kalb-Ramond Field, we must have bundle gerbe local data of the following form: 
\begin{equation}\label{eq:introgerbe}
    \begin{aligned}
    \pi_{\ast} B_{(\beta)} - \pi_{\ast} B_{(\alpha)} &= \di \Lambda_{(\alpha\beta)} &&\text{ on } U_{\alpha}\cap U_{\beta} \\
    \Lambda_{(\alpha\beta)} + \Lambda_{(\beta\gamma)} + \Lambda_{(\gamma\alpha)} &=\di G_{(\alpha\beta\gamma)}  &&\text{ on }U_{\alpha}\cap U_{\beta}\cap U_{\gamma}\\
    G_{(\alpha\beta\gamma)} + G_{(\beta\alpha\delta)} + G_{(\gamma\beta\delta)} + G_{(\delta\alpha\gamma)} &\in 2\pi\mathbb{Z} &&\text{ on }U_{\alpha}\cap U_{\beta}\cap U_{\gamma}\cap U_{\delta}.
    \end{aligned}
\end{equation}
Now, from the patching relations \eqref{eq:introadapted} of the adapted coordinates we obtain
\begin{equation}
    \di \tilde{x}_{(\alpha)\mu} \,=\, \frac{\partial \tilde{f}_{(\alpha\beta)\mu}}{\partial x_{(\beta)}^\nu} \,\di x_{(\beta)}^\nu \,+\, \frac{\partial \tilde{f}_{(\alpha\beta)\mu}}{\partial \tilde{x}_{(\beta)\nu}} \,\di \tilde{x}_{(\beta)\nu}
\end{equation}
This, combined with \eqref{eq:introdx} and \eqref{eq:introgerbe}, implies the following equations
\begin{equation}
    \frac{\partial \tilde{f}_{(\alpha\beta)\mu}}{\partial \tilde{x}_{(\beta)\nu}} = \delta_{\mu}^{\;\,\nu}, \qquad
    \frac{\partial \tilde{f}_{(\alpha\beta)\mu}}{\partial x_{(\beta)}^\nu} = \left(\di\Lambda_{(\alpha\beta)}\right)\!_{\mu\nu}
\end{equation}
We can immediately solve the first equation by decomposing $\tilde{f}_{(\alpha\beta)}\!\left(x_{(\beta)},\tilde{x}_{(\beta)}\right) = \tilde{x}_{(\beta)} + \tilde{f}'_{(\alpha\beta)}(x_{(\beta)})$, where $\tilde{f}'_{(\alpha\beta)}(x_{(\beta)})$ is a new basic function of the $x_{(\beta)}$-coordinates only. Now the second equation is equivalent to the new equation $\di (\tilde{f}'_{(\alpha\beta)\mu}\di x^\mu)=-\di\Lambda_{(\alpha\beta)}$ on $U_\alpha\cap U_\beta$, which is solved by
\begin{equation}
    \tilde{f}_{(\alpha\beta)\mu}'\di x^\mu \,=\, -\Lambda_{(\alpha\beta)\mu}\di x^\mu + \di\eta_{(\alpha\beta)}
\end{equation}
where $\di\eta_{(\alpha\beta)}\in\pi^\ast\Omega^1_{\mathrm{ex}}(U_{\alpha}\cap U_{\beta})$ are local exact basic $1$-forms on overlaps of patches. The cocycle condition for transition functions of a manifold on three-fold overlaps of patches implies then
\begin{equation}\label{eq:introlambda}
    \di(\eta_{(\alpha\beta)} + \eta_{(\beta\gamma)} + \eta_{(\gamma\alpha)}) \,=\, \Lambda_{(\alpha\beta)} + \Lambda_{(\beta\gamma)} + \Lambda_{(\gamma\alpha)}.
\end{equation}
Since $\Lambda_{(\alpha\beta)} + \Lambda_{(\beta\gamma)} + \Lambda_{(\gamma\alpha)} = \di G_{\alpha\beta\gamma}$ from \eqref{eq:introgerbe}, then we must have the trivialization
\begin{equation}\label{eq:introtriv}
    G_{(\alpha\beta\gamma)} \,=\, \left(\eta_{(\alpha\beta)} + \eta_{(\beta\gamma)} + \eta_{(\gamma\alpha)}\right) + c_{(\alpha\beta\gamma)}
\end{equation}
where $c_{(\alpha\beta\gamma)}\in\mathbb{R}$ are local constants which must satisfy the following cocycle condition
\begin{equation}
    c_{(\alpha\beta\gamma)} + c_{(\beta\alpha\delta)} + c_{(\gamma\beta\delta)} + c_{(\delta\alpha\gamma)} \in 2\pi\mathbb{Z}
\end{equation}
on each four-fold overlaps of patches of the leaf space. 
\vspace{0.25cm}

\noindent This implies that the gerbe $(\pi_{\ast} B_{(\alpha)},\Lambda_{(\alpha\beta)},G_{(\alpha\beta\gamma)})$ on our physical spacetime $M=\mathcal{M}/\mathcal{F}_-$ is \textit{flat}, i.e. that the curvature of the Kalb-Ramond field $H\in\Omega^3_{\mathrm{ex}}(M)$ is exact and that the topological Dixmier-Douady class $[H]\in H^3(M,\mathbb{Z})$ of the gerbe is torsion. To see this, it is enough to check that the equation \eqref{eq:introlambda} implies that $\Lambda_{(\alpha\beta)}=\di\eta_{(\alpha\beta)}+\tau_{(\alpha)}-\tau_{(\beta)}$ for some local $1$-forms $\tau_{(\alpha)}\in \Omega^1\!\left(U_{\alpha}\right)$. But this implies that $\pi_{\ast} B_{(\alpha)}-\pi_{\ast} B_{(\beta)}=\tau_{(\beta)}-\tau_{(\alpha)}$ so that, in other words, there exist a global $2$-form on the leaf space $M=\mathcal{M}/\mathcal{F}_-$ given by gauge transformations of the Kalb-Ramond field and expressed on overlaps of patches by
\begin{equation}
    \pi_{\ast} B_{(\alpha)} + \di\tau_{(\alpha)} \,=\, \pi_{\ast} B_{(\beta)} + \di\tau_{(\beta)}.
\end{equation}
If we call $B'|_{U_{\alpha}} := \pi_{\ast} B_{(\alpha)} + \di\tau_{(\alpha)}$ the gauge transformed Kalb-Ramond field, we immediately see that it satisfies $H=\di B'$ globally on $M=\mathcal{M}/\mathcal{F}_-$. Therefore in the de Rham cohomology we have the class $[H]=0\in H^3_{\mathrm{dR}}(M)$, which is mapped to a torsion element of the integral cohomology $H^3(M,\mathbb{Z})$. In this context the constants $c_{(\alpha\beta\gamma)}$ are interpreted as a representative of the flat holonomy class $[c_{(\alpha\beta\gamma)}]\in H^2(M,\mathbb{R}/2\pi\mathbb{Z})$ of the flat bundle gerbe.

\paragraph{Open problem.}
Therefore it does not seem possible to recover a general geometric string background made of a smooth manifold $M$ equipped with a non-trivial Kalb-Ramond field $[H]\in H^3(M,\mathbb{Z})$. And, since DFT was introduced to extend supergravity, the impossibility of recovering supergravity poses a problem. This means that the original argument by \cite{Pap13} is still relevant whenever we try to construct the doubled space as a manifold.\vspace{0.25cm}

\noindent However, as we will see, Born geometry is still extremely efficient in dealing with doubled group manifold and, in particular, Drinfel'd doubles. Remarkable results from the application of para-Hermitian geometry to group manifolds can be found in \cite{MarSza18}, \cite{DFTWZW19} and \cite{MarSza19}. These groups, where fluxes are constant, allow a simple geometrization of the gerbe with a group manifold which is not possible in the general case. Notice that a link between Drinfel'd doubles and bundle gerbes was firstly found by \cite{Wil08}.

\paragraph{Higher Kaluza-Klein perspective on the problem.}
The Higher Kaluza-Klein proposal is an attempt to attack this problem and allow the geometrization of general bundle gerbes. In the Higher Kaluza-Klein perspective the doubled space would be identified with the total space of a gerbe $M\twoheadrightarrow M$. Thus the quotient $\pi:\mathcal{M}\twoheadrightarrow M=\mathcal{M}/\mathcal{F}_-$ is reinterpreted as a local version of the projector of the gerbe as a principal $\infty$-bundle, i.e.
\begin{equation}
    \bbpi:\,\mathcal{M}\; \longtwoheadrightarrow \; M \simeq \mathcal{M}/\!/\mathbf{B}U(1)
\end{equation}
which is the higher geometric version of the statement $\pi:P\twoheadrightarrow M\cong P/G$ for any $G$-bundle $P$. See the next section for an introduction.

\section{Introduction to Higher Kaluza-Klein Theory}\label{s3}

In this section we will give a very brief introduction to the Higher Kaluza-Klein perspective on the geometry of DFT we started to develop in \cite{Alf19}. 

\subsection{The doubled space as a bundle gerbe}

In the Higher Kaluza-Klein proposal (see \cite{Alf19} for details) the doubled space of DFT is identified with the total space of a bundle gerbe with connection. In this section we will mostly describe the geometry of the bundle gerbe.
Let us now give a concrete geometric characterization to the concept of bundle gerbe.

\paragraph{The bundle gerbe.}
Let $\{U_\alpha\}$ be a cover for a smooth manifold $M$.
We define (see \cite{Hit99} for details) a \textit{bundle gerbe} $\bbpi:\mathcal{M}\longtwoheadrightarrow M$ on the base manifold $M$ by a collection of circle bundles $\{P_{\alpha\beta}\twoheadrightarrow U_\alpha\cap U_\beta\}$ on each overlap of patches $U_\alpha\cap U_\beta\subset M$ such that:
\begin{itemize}
    \item there exists an isomorphism $P_{\alpha\beta}\cong P^{-1}_{\beta\alpha}$ on any two-fold overlap of patches $U_\alpha\cap U_\beta$,
    \item there exists an isomorphism $P_{\alpha\beta}\otimes P_{\beta\gamma}\cong P_{\alpha\gamma}$ on any three-fold overlap of patches $U_\alpha\cap U_\beta\cap U_\gamma$ given by the gauge transformation $G_{(\alpha\beta\gamma)}\in\Coo(U_\alpha\cap U_\beta\cap U_\gamma)$,
    \item the trivialization satisfies $G_{(\alpha\beta\gamma)}-G_{(\beta\gamma\delta)}-G_{(\gamma\delta\alpha)}+G_{(\delta\alpha\beta)}\in 2\pi\mathbb{Z}$ on any four-fold overlaps of patches $U_\alpha\cap U_\beta\cap U_\gamma \cap U_\delta$.
\end{itemize}
where for a given circle bundle $P$ we denote with $P^{-1}$ the circle bundle with opposite $1$st Chern class, i.e. with $\mathrm{c}_1(P^{-1})=-\mathrm{c}_1(P)$. \vspace{0.25cm}

\noindent Notice that the trivialization we introduced defines a \v{C}ech cocycle corresponding to an element $\big[G_{(\alpha\beta\gamma)}\big]\in H^{3}(M,\mathbb{Z})$ of the $3$rd cohomology group of the base manifold $M$. Thus bundle gerbes are topologically classified by classes $[H]\in H^{3}(M,\mathbb{Z})$, which physically correspond to the H-flux.\vspace{0.25cm}

\noindent More recently, in \cite{Principal1}, the bundle gerbe has been reformulated as a principal $\infty$-bundle, where the gauge $2$-group is $G=\mathbf{B}U(1)$, i.e. the \textit{group-stack of circle bundles}. To see that the set of circle bundles on any manifold $M$ carries a group-stack structure, notice that we have the isomorphisms
\begin{equation}
    \begin{aligned}
    &P^{-1}\otimes P\cong M\times U(1), \quad P\otimes P^{-1}\cong M\times U(1), \\
    &P_1\otimes(P_2\otimes P_3)\cong (P_1\otimes P_2)\otimes P_3
    \end{aligned}
\end{equation}
where thus the tensor product $\otimes$ plays the role of the group multiplication, while the trivial bundle $M\times U(1)$ plays the role of the \textit{identity element} and $P^{-1}$ plays the role of the \textit{inverse element} of $P$.

\paragraph{Automorphisms of the bundle gerbe.}
As seen also in \cite{BMS20}, the $2$-group of symmetries of a bundle gerbe $\mathcal{M}\twoheadrightarrow M$ is
\begin{equation}
    \mathrm{Aut}(\mathcal{M}) \;=\; \Diff(M)\,\ltimes\, U(1)\text{-Bundles}(M)
\end{equation}
where $U(1)\text{-Bundles}(M)$ is the $2$-group of $U(1)$-bundles with connection on $M$. Notice that this is nothing but the higher geometric version of the gauge group $G_{\mathrm{DFT}}\;=\;\Diff(M)\ltimes \Omega^2_{\mathrm{cl}}(M)$ of DFT proposed by \cite{Hull14} in \eqref{eq:ghull}. In fact, the natural map $U(1)\text{-Bundles}(M) \rightarrow \Omega^2_{\mathrm{cl}}(M)$ is just the curvature map sending a $U(1)$-bundle to its curvature $b\in\Omega^2_{\mathrm{cl}}(M)$.

\paragraph{Sections of the bundle gerbe.}
Let us fix a good cover $\{U_\alpha\}$ of $M$. A section of the bundle gerbe $\mathcal{M}\twoheadrightarrow M$ is given by a collection of sections $(x,\exp i\phi_{(\alpha\beta)}):U_\alpha\cap U_\beta\rightarrow P_{\alpha\beta}$ of the local circle bundles on two-fold overlaps of patches, which must satisfy at any $x\in U_\alpha\cap U_\beta\cap U_\gamma$
\begin{equation}\label{eq:coords}
    \begin{aligned}
       \phi_{(\alpha\gamma)} \,&=\,  \phi_{(\alpha\beta)}+\phi_{(\beta\gamma)}+G_{(\alpha\beta\gamma)}(x) \quad\! \mathrm{mod}\,2\pi\mathbb{Z}.
    \end{aligned}
\end{equation}
In \cite{Principal1} it was understood that the sections of a bundle gerbe are equivalently twisted $U(1)$-bundles.
Then the local connection data of the twisted bundle can be expressed by local $1$-forms patched by  $\tilde{x}_{(\alpha)} - \tilde{x}_{(\beta)}  \,=\, -\Lambda_{(\alpha\beta)} + \mathrm{d}\phi_{(\alpha\beta)}$. Then locally on each patch, a section of the connective bundle gerbe looks like a section of $T^\ast U_\alpha$, which is a property expected by the doubled space.

\paragraph{Geometric strong constraint.}
The bundle gerbe $\mathcal{M}\twoheadrightarrow M$, being a particular example of principal $\infty$-bundle, will come equipped with a natural principal action. Crucially, the principal action $\rho:\mathbf{B}U(1)\times \mathcal{M}\rightarrow \mathcal{M}$ reproduces exactly the $2$-group of gauge transformations and gauge-of-gauge transformations of the Kalb-Ramond field. Moreover the homotopy quotient of the bundle gerbe by the principal action
\begin{equation}
    M \;\simeq\; \mathcal{M}/\!/_{\!\rho\,}\mathbf{B}U(1)
\end{equation}
is just the base manifold. This is totally analogous to an ordinary $U(1)$-bundle $P\twoheadrightarrow M$ satisfying $P/U(1)\cong M$.
In \cite{Alf19} we show that we can define a generalized metric as a principal action-invariant structure on our bundle gerbe (see also section \ref{s4} for more details).

\paragraph{NS5-brane as Higher Kaluza-Klein monopole.}
The Kalb-Ramond field of an NS5-brane is modelled by a bundle gerbe on the smooth manifold $\mathbb{R}^{1,5} \times \left(\mathbb{R}^4-\{0\}\right) \,\simeq\, \mathbb{R}^{1,5} \times \mathbb{R}^+ \times S^3$ which is non-trivial only on the transverse space $\mathbb{R}^+ \times S^3$. The transverse space is topologically equivalent to the $3$-sphere $S^3$ and the third cohomology group of the $3$-sphere is $H^3(S^3,\mathbb{Z})\cong\mathbb{Z}$. Thus bundle gerbes on the $3$-spheres will be classified by elements $[m\mathrm{Vol}(S^3)/2]\in H^3(S^3,\mathbb{Z})$ with H-charge $m\in\mathbb{Z}$ and where $\mathrm{Vol}(S^3)\in\Omega^3(S^3)$ is the volume form on $S^3$.\vspace{0.25cm}

\noindent We can explicitly construct a bundle gerbe $\mathcal{M}\twoheadrightarrow S^3$ on the $3$-sphere as it follows. Let us consider a cover $\{U_\mathrm{N}, U_{\mathrm{S}}\}$ for $S^3$ where the two open sets are open neighbourhoods of the north and south hemispheres. Their intersection can be retracted to a $2$-sphere $S^2$ and then we can choose a circle bundle $P_{\mathrm{NS}}\twoheadrightarrow S^2$ with $1$st Chern class $m\in\mathbb{Z}$, making it a Lens space.\vspace{0.25cm}

\noindent In \cite{Alf19} we showed that, on a gerbe $\mathcal{M}\twoheadrightarrow\mathbb{R}^{1,5} \times \mathbb{R}^+ \times S^3$ classified by $[m\mathrm{Vol}(S^3)/2]\in H^3(S^3,\mathbb{Z})$, there is a natural generalized metric which encompasses the metric and Kalb-Ramond field of an NS5-brane with H-charge $m\in\mathbb{Z}$. This is a direct generalization of the Gross-Perry monopole in ordinary Kaluza-Klein Theory.

\subsection{Doubled geometry from reduction of bundle gerbes}

We will use this subsection for a a deeper discussion of generalized correspondence spaces and of how they emerge from the bundle gerbe picture.

\paragraph{Correspondence space from reduction of the gerbe.}
In \cite{Alf19} we applied to DFT the \textit{dimensional reduction} of bundle gerbes, which was defined in terms of $L_\infty$-algebras by \cite{FSS17x}. The statement is that, if $M$ is the total space of a $G$-bundle on some base manifold $M_0$, the dimensional reduction of a bundle gerbe $\mathcal{M}$ on $M$ specified by the cocycle $M\rightarrow\mathbf{B}^2U(1)$ will be a certain higher geometric structure on $M_0$. More in detail, the dimensional reduction will be a map
\begin{equation}\label{eq:red}
    \left(\begin{tikzcd}[row sep=7ex, column sep=5ex]
    M \arrow[r] & \mathbf{B}^2U(1)
    \end{tikzcd}\right) \;\;\overset{\cong}{\longmapsto}\;\; \left(\begin{tikzcd}[row sep=7ex, column sep=5ex]
    & \left[G,\mathbf{B}^2U(1)\right]\!/G \arrow[d]\\
    M_0 \arrow[ru]\arrow[r, ""] & \mathbf{B}G
    \end{tikzcd}\right)
\end{equation}
where $M_0\rightarrow\mathbf{B}G$ is the cocycle which specifies the $G$-bundle $M\twoheadrightarrow M_0$. The stack $\left[G,-\right]\!/G$ is a generalization for stacks of the \textit{cyclic loop space}: in fact in the particular case where $G=U(1)$ and $N$ is a smooth manifold, this reduces to the space $\left[U(1),N\right]\!/U(1) \,=\, \Coo(N,\,S^1)/S^1$. \vspace{0.25cm}

\noindent Crucially, a particular example of this reduction is well-known in DFT since the work by \cite{Hull06} and by \cite{BelHulMin07}. In fact, if we consider the particular example of a $T^n$-equivariant gerbe on a $T^n$-bundle, we automatically have that it dimensionally reduce as follows:
\begin{equation}\label{eq:string}
    \Big(\; T^n\text{-equivariant gerbe on }M \;\Big) \;\;\overset{\cong}{\longmapsto}\;\; \Big(\; \String(T^n\times\tilde{T}^n)\text{-bundle on }M_0 \;\Big).
\end{equation}
In fact a $\String(T^n\times\tilde{T}^n)$-bundle on $M_0$ is a particular principal $\infty$-bundle with the following curvature forms (see \cite{Alf19} for details):
\begin{equation}
    \begin{aligned}
    \mathcal{F} \;&=\; \di \mathcal{A}_{(\alpha)} &&\in\, \Omega^2(M_0,\,\mathbb{R}^{2n}),\\[0.2em]
    \mathcal{H} \;&=\;  \di \mathcal{B}_{(\alpha)} + \frac{1}{2}\langle \mathcal{A}_{(\alpha)} \,\overset{\wedge}{,}\, \di \mathcal{A}_{(\alpha)}\rangle &&\in\, \Omega^3(M_0),
    \end{aligned}
\end{equation}
which are the usual equations of doubled torus bundles we find in \cite{Hull06}.
Notice that, indeed, if we forget the higher form field, we stay with a $T^{2n}$-bundle on $M_0$, which is exactly \textit{the correspondence space of a T-duality}. In other words $K=M\times_{M_0}\widetilde{M}$ and we have a diagram of the form
\begin{equation}
\begin{tikzcd}[row sep=4.5ex, column sep=4ex]
 & & K \arrow[dl, two heads]\arrow[dr, two heads] & &\\
 T^n \arrow[r, hook] & M \arrow[dr, "\pi"', two heads] & & \widetilde{M}\arrow[dl, two heads] &T^n \arrow[l, hook'] \\
 & & M_0 & &
\end{tikzcd}
\end{equation}
where the $1$st Chern class of the new $T^n$-bundle $\widetilde{M}\twoheadrightarrow M_0$ is the pullback of the curvature of the gerbe along $\pi$, i.e. it is $\mathrm{c}_1(\widetilde{M})=\big[\pi^\ast H\big]\in H^2(M,\mathbb{Z}^n)$. Now a topological T-duality in the $i$-th direction of the fiber exchanges the $1$st Chern classes of the two bundles:
\begin{equation}
    \begin{tikzcd}[row sep=13ex, column sep=10ex]
    \mathrm{c}_1(\widetilde{M})_i \arrow[r, leftrightarrow, "\mathcal{T}_i"] & \mathrm{c}_1(M)^i 
\end{tikzcd}
\end{equation}
Now we can ask ourselves: what is the most general doubled geometry that we can obtain by reduction of bundle gerbes? We will deal with this question in the rest of this subsection.

\paragraph{Topology and non-geometry: the generalized correspondence space.}
As firstly noticed by \cite{BelHulMin07}, a bundle gerbe $\mathcal{M}\twoheadrightarrow M$ which satisfies the T-dualizability condition $\mathcal{L}_{\partial_i}H=0$ gives rise to a well-defined $T^n$-bundle $K\twoheadrightarrow M$. Thus we obtain the following diagram
\begin{equation}
    \begin{tikzcd}[row sep=4.5ex, column sep=5ex]
    \widetilde{T}^n\arrow[r, hook] & K \arrow[d, two heads] \\
    T^n\arrow[r, hook] & M \arrow[d, two heads, "\pi"] \\
    & M_0 &
    \end{tikzcd}
\end{equation}
Now $K$ is called \textit{generalized correspondence space} and it can be interpreted as the correspondence space for a non-geometric T-duality, i.e. the case where the T-dual background is non-geometric and hence a T-fold. We can then write
\begin{equation}\label{eq:tfolddiagram}
\begin{tikzcd}[row sep=4.5ex, column sep=2.8ex]
 & & K \arrow[dl, two heads]\arrow[dr, dotted] & \\
 T^n \arrow[r, hook] & M \arrow[dr, "\pi"', two heads] & & [-1ex]\text{T-fold}\arrow[dl, dotted] \\
 & & M_0 &
\end{tikzcd}
\end{equation}
where the dotted arrows only indicative ones, in this context. For a proper formalization of the concept of T-fold as a \textit{non-commutative} $T^n$\textit{-bundle} on $M_0$ see \cite{Bou08} and more recently \cite{AscSza20}. \vspace{-0.2cm}

\noindent Notice that, on any patch $U_\alpha$ of the base manifold $M_0$, the generalized correspondence space is locally $T^{2n}$-bundle $K|_{U_\alpha}\cong U_\alpha \times T^{2n}$. The difference of the non-geometric case with the geometric case is in how these patches are globalized. Therefore non-geometry is a global property of the topology of $K$ or, equivalently, of the topology of $\mathcal{M}$.

\paragraph{The generalized correspondence space of Poisson-Lie T-duality.}
In \cite{Alf19} we derived that something very similar happens for non-abelian T-duality and more generally for Poisson-Lie T-duality. A bundle gerbe $\mathcal{M}\twoheadrightarrow M$ on a $G$-bundle which satisfies the Poisson-Lie T-dualizability condition we get a generalized correspondence space of the form 
\begin{equation}
    \begin{tikzcd}[row sep=4.5ex, column sep=5ex]
    \widetilde{G}\arrow[r, hook] & K \arrow[d, two heads] \\
    G\arrow[r, hook] & M \arrow[d, two heads, "\pi"] \\
    & M_0 &
    \end{tikzcd}
\end{equation}
where $\widetilde{G}$ is the Poisson-Lie T-dual fiber of the starting group $G$. See \cite{Alf19} for more details.

\section{The atlas of Higher Kaluza-Klein Theory}\label{s4}

The aim of this section is finding an answer to the following question: if the doubled space $\mathcal{M}$ is not a smooth manifold, but a bundle gerbe, then how can we define local coordinates?
\vspace{0.2cm}

\noindent We will also show that \textit{the natural notion of local coordinates for the bundle gerbe coincides with the notion of local coordinates for DFT}. 
\vspace{0.2cm}

\noindent We will need first to investigate what gluing charts means in theoretical physics. The notions we are going to use were introduced in String Theory by \cite{FSS18}, \cite{FSS19x} and \cite{FSS20}\footnote{The author thanks Urs Schreiber for explaining the notion of atlas of a stacks and, in particular, how it was firstly applied in the context of Super-Exceptional Geometry by \cite{FSS18} and \cite{FSS19x}. We will try here to apply the definition to a simpler bundle gerbe and we will focus more on its global aspects.}.

\subsection{Review: atlases in higher geometry}

For a geometric stack $\mathscr{G}$ an \textit{atlas} is a smooth manifold $\mathcal{U}$ equipped with a morphism of stacks
\begin{equation}
    \phi:\,\mathcal{U}\; \longtwoheadrightarrow\; \mathscr{G}
\end{equation}
which is, in particular, an \textit{effective epimorphism} (see \cite{Hei05} and \cite{topos} for details).
This formalizes the idea that any geometric stack $\mathscr{G}$ looks locally like an ordinary manifold $\mathcal{U}$.

\paragraph{Example: atlas for a smooth manifold.}
If our geometric stack is an ordinary smooth manifold $\mathscr{G}:=M$, we can choose an atlas given by $\mathcal{U}:=\bigsqcup_{\alpha\in I}\mathbb{R}^{d}$ and by a surjective map $\phi:\bigsqcup_{\alpha\in I}\mathbb{R}^{d}\xtwoheadrightarrow{\;\;\{\phi_\alpha\}_{\alpha\in I}\;\;} M$ given by the local charts $\{\phi_\alpha: \mathbb{R}^{d}\twoheadrightarrow U_\alpha\}_{\alpha\in I}$ of any cover $\{U\}_{\alpha\in I}$ of the manifold $M$. This formalizes the intuitive idea that any smooth manifold looks locally like a Cartesian space $\mathbb{R}^d$. 
\vspace{0.2cm}

\noindent We physicists, in fact, work not directly on a manifold $M$, but on local charts of the form $\mathbb{R}^d$.

\paragraph{Example: atlas for ordinary Kaluza-Klein.}
In ordinary Kaluza-Klein Theory the space to consider is a circle bundle $P$ on the base manifold $M$. On an open cover $\{U_\alpha\}_{\alpha\in I}$ of the base manifold $M$ this is locally trivialized by a collection of local trivial bundles $\{U_\alpha\times U(1)\}_{\alpha\in I}$. From ordinary differential geometry we know that the total space $P$ of the bundle can be covered by local charts $\{\phi_\alpha:\,\mathbb{R}^{d+1}\twoheadrightarrow U_\alpha\times U(1)\subset P\}_{\alpha\in I}$. These charts define an atlas for the total space of the bundle $P$ of the form $\bigsqcup_{\alpha\in I}\mathbb{R}^{d+1}\xtwoheadrightarrow{\;\;\{\phi_\alpha\}_{\alpha\in I}\;\;} P$.
This corresponds to the well-known idea in differential geometry that the total space $P$ locally looks like a Cartesian space $\mathbb{R}^{d+1}$.
\vspace{0.2cm}

\noindent Any such map $\phi_\alpha$ uniquely factorizes as $\phi_\alpha:\,\mathbb{R}^{1,d+1}\xtwoheadrightarrow{\;F\;} \mathbb{R}^{1,d}\times U(1) \xtwoheadrightarrow{\;\varphi_\alpha\;} U_\alpha\times U(1)$, where the first map is just the surjection
\begin{equation}
    F:\,\mathbb{R}^{d+1}\; \longtwoheadrightarrow\; \mathbb{R}^{d}\times U(1)
\end{equation}
which is the identity on $\mathbb{R}^{1,d}$ and the quotient map $\mathbb{R}\twoheadrightarrow U(1)=\mathbb{R}/2\pi\mathbb{Z}$. Crucially, the map $F$ is an atlas of ordinary Lie groups.
\vspace{0.2cm}

\noindent The surjective map $\bigsqcup_{\alpha\in I}\mathbb{R}^{d}\times U(1)\xtwoheadrightarrow{\;\;\{\varphi_\alpha\}_{\alpha\in I}\;\;} P$ is an atlas for the total space $P$, in the stacky sense of the word. This corresponds to the intuitive idea that the total space of a circle bundle locally looks like the Lie group $\mathbb{R}^{d}\times U(1)$.

\paragraph{The \v{C}ech nerve of an atlas.}
When we defined the atlas $\phi:\,\mathcal{U}\; \longtwoheadrightarrow\; \mathscr{G}$ for the stack $\mathscr{G}$, we said that it must be an \textit{effective epimorphism}. An effective epimorphism is defined as the colimit of a certain \textit{simplicial object} which is called \textit{\v{C}ech nerve}. In other words we have
\begin{equation}
\Big(\begin{tikzcd} \mathcal{U}  \arrow[r, two heads, "\phi"] & \mathscr{G} \end{tikzcd}\Big) \,\;=\;\, \varinjlim\bigg( \begin{tikzcd} \cdots \;\arrow[r, yshift=1.7ex, two heads] \arrow[r, yshift=-0.6ex, two heads] \arrow[r, yshift=0.6ex, two heads] \arrow[r, yshift=-1.7ex, two heads] & \mathcal{U}\times_{\mathscr{G}}\mathcal{U}\times_{\mathscr{G}}\mathcal{U} \arrow[r, two heads] \arrow[r, yshift=1.2ex, two heads] \arrow[r, yshift=-1.2ex, two heads] & \mathcal{U}\times_{\mathscr{G}}\mathcal{U} \arrow[r, yshift=0.7ex, two heads] \arrow[r, yshift=-0.7ex, two heads] & \mathcal{U}
\end{tikzcd} \bigg)
\end{equation}
The \v{C}ech nerve of an atlas can be interpreted as a $\infty$-groupoid, which we will call \textit{\v{C}ech groupoid}. This groupoid encodes the global geometry of the stack in terms of the smooth manifold $\mathcal{U}$, which makes it easier to deal with. Besides the original stack can always be recovered by the colimit of the nerve. \vspace{0.25cm}

\noindent How do we construct such a simplicial object? Let us firstly consider the \textit{kernel pair} of the map $\phi$, which is defined as the pullback (in the category theory meaning) of two copies of the map $\phi$. The \textit{coequalizer} diagram of this kernel pair will thus be of the following form:
\begin{equation}\label{eq:kernelpairdef}
\begin{tikzcd}
\mathcal{U}\times_{\mathscr{G}}\mathcal{U} \arrow[r, yshift=0.7ex, two heads] \arrow[r, yshift=-0.7ex, two heads] & \mathcal{U}  \arrow[r, two heads, "\phi"] & \mathscr{G}
\end{tikzcd}
\end{equation}
This was the first step. By iterating this construction we obtain all the rest of the \v{C}ech nerve.

\paragraph{Example: \v{C}ech nerve of the atlas of a smooth manifold.}
Let us consider, like in the first example, the case where our stack $\mathscr{G}=M$ is just a smooth manifold with an atlas $\bigsqcup_{\alpha\in I}\mathbb{R}^{d}\xtwoheadrightarrow{\;\;\{\phi_\alpha\}_{\alpha\in I}\;\;} M$. In this case the kernel pair, defined in \eqref{eq:kernelpairdef}, is the following:
\begin{equation}
\begin{tikzcd}[column sep=9ex]
\displaystyle\bigsqcup_{\alpha,\beta\in I}\mathbb{R}^d \cap_{\phi} \mathbb{R}^d \arrow[r, yshift=0.7ex, two heads] \arrow[r, yshift=-0.7ex, two heads] & \displaystyle\bigsqcup_{\alpha\in I}\mathbb{R}^d  \arrow[r, two heads, "\{\phi_\alpha\}_{\alpha\in I}"] & M
\end{tikzcd}
\end{equation}
where we called $\mathbb{R}^d \cap_{\phi} \mathbb{R}^d := \big\{(x,y)\in\mathbb{R}^d\times\mathbb{R}^d|\,  \phi_{\alpha}(x)=\phi_\beta(y)\big\}$. We notice that the kernel pair of the atlas encodes nothing but the information about how the charts are glued together over the manifold $M$. We intuitively have that the global geometry of a smooth manifold $M$ is entirely encoded in its \v{C}ech groupoid $\big(\!\begin{tikzcd}
\bigsqcup_{\alpha,\beta\in I}\mathbb{R}^d \cap_{\phi} \mathbb{R}^d \arrow[r, yshift=0.7ex, two heads] \arrow[r, yshift=-0.7ex, two heads] & \bigsqcup_{\alpha\in I}\mathbb{R}^d
\end{tikzcd}\!\big)$. 
\vspace{0.25cm}

\noindent We physicists are actually very familiar with this perspective: in fact we usually describe our fields as functions on the local charts $\mathbb{R}^d$ of a manifold $M$ and, if we want to describe how they behave globally, we simply write how they transform on the overlaps $\mathbb{R}^d \cap_{\phi} \mathbb{R}^d$ of these charts.
In the next paragraph we will formalize exactly this perspective on fields.

\paragraph{Gluing a field on a stack.}
Let $\mathcal{U}\xtwoheadrightarrow{\;\phi\;}\mathscr{G}$ be an atlas for the stack $\mathscr{G}$ and let $\mathscr{F}$ be another stack, which we will interpret as the moduli-stack of some physical field. Now let $A:\mathscr{G}\rightarrow\mathscr{F}$ be a morphism of stacks (i.e. a physical field on $\mathscr{G}$). We obtain an induced morphism $A\circ\phi:\mathcal{U}\rightarrow \mathscr{F}$ together with an isomorphism between the two induced morphisms $\begin{tikzcd}\mathcal{U}\times_{\mathscr{G}}\mathcal{U} \arrow[r, yshift=0.7ex] \arrow[r, yshift=-0.7ex] &\mathscr{F}\end{tikzcd}$ which satisfies the cocycle condition on $\mathcal{U}\times_{\mathscr{G}}\mathcal{U}\times_{\mathscr{G}}\mathcal{U}$.

\paragraph{Example: gluing a gauge field on a smooth manifold.}
Let $\bigsqcup_{\alpha\in I}\mathbb{R}^{d}\xtwoheadrightarrow{\,\{\phi_\alpha\}_{\alpha\in I}\,}M$ be an atlas for the smooth manifold $M$ and let $\mathscr{F}:=\mathbf{B}G_{\mathrm{conn}}$ be the moduli-stack of Yang-Mills fields with gauge group $G$. Then a gauge field $A:M\rightarrow\mathbf{B}G_{\mathrm{conn}}$ on the smooth manifold induces a local $1$-form $A_{(\alpha)}:=A\circ\phi_{\alpha}\in\Omega^1(\mathbb{R}^d,\mathfrak{g})$ on each chart of the atlas. Notice that these $1$-forms $A_{(\alpha)}(x)$ depends on local coordinates $x\in\mathbb{R}^d$, like we physicists are used. On overlaps of charts we must also have an isomorphism between $A_{(\alpha)}$ and $A_{(\beta)}$ given by a gauge transformation $A_{(\alpha)}= h_{(\alpha\beta)}^{-1}(A_{(\beta)}+\di)h_{(\alpha\beta)}$ with $h_{(\alpha\beta)}\in\Coo(\mathbb{R}^d \cap_{\phi} \mathbb{R}^d,\,G)$. Again, these $h_{(\alpha\beta)}(x)$ are not $G$-valued functions directly on the manifold, but on the atlas. Finally, these isomorphisms must satisfy the cocycle condition $h_{(\alpha\beta)}h_{(\beta\gamma)}h_{(\gamma\alpha)}=1$.
\vspace{0.25cm}

\noindent In this subsection we explained in an almost pedantic way how geometric structures on smooth manifolds become the familiar and more treatable objects on local $\mathbb{R}^d$ coordinates we physicists use. We will see in the next subsection that these intermediate steps become much less trivial if we want to glue local charts for DFT.

\subsection{The atlas of a bundle gerbe}

The aim of this section will be finding an atlas for a bundle gerbe $\mathcal{M}$, seen as a stack. 

\paragraph{An atlas for the $2$-algebra.}
Let us call $\mathbb{R}^{d}\oplus\mathbf{b}\mathfrak{u}(1)$ the $2$-algebra of the abelian $2$-group $\mathbb{R}^{d}\times\mathbf{B}U(1)$. It is well-known that an $L_\infty$-algebra is equivalently described in terms of its Chevalley-Eilenberg differential graded algebra. In our particular case this will be
\begin{equation}
    \mathrm{CE}\!\left(\mathbb{R}^{d}\oplus\mathbf{b}\mathfrak{u}(1)\right) \;=\; \mathbb{R}[e^a,B]/\langle\di e^a=0,\;\di B = 0\rangle,
\end{equation}
where the $\{e^a\}$ with $a=0,\dots,d-1$ are elements in degree $1$ and $B$ is an element in degree $2$. Since the $2$-algebra is abelian, the differentials of the generators of its Chevalley-Eilenberg algebra are trivial. Now recall that an atlas is an effective epimorphism from our manifold to our stack. In this case it will be an effective epimorphism from an ordinary Lie algebra $\mathfrak{a}$ to our $2$-algebra of the form
\begin{equation}
    f:\,\mathfrak{a}\; \longtwoheadrightarrow\; \mathbb{R}^{d}\oplus\mathbf{b}\mathfrak{u}(1).
\end{equation}
Dually this can be given as an effective monomorphism between the respective Chevalley-Eilenberg differential graded algebras
\begin{equation}
    f^\ast:\,\mathrm{CE}\!\left(\mathbb{R}^{d}\oplus\mathbf{b}\mathfrak{u}(1)\right) \;\longhookrightarrow\; \mathrm{CE}(\mathfrak{a})
\end{equation}
In other words we want  to identify an ordinary Lie algebra $\mathfrak{a}$ such that its Chevalley-Eilenberg algebra contains an element $\omega:=f^\ast (B)\in\mathrm{CE}(\mathfrak{a})$ in degree $2$ which is the image of the degree $2$ generator of $\mathrm{CE}\!\left(\mathbb{R}^{d}\oplus\mathbf{b}\mathfrak{u}(1)\right)$ and which must satisfy the same equation
\begin{equation}
    \di \omega \,=\, 0,
\end{equation}
since a homomorphism of differential graded algebras maps $f^\ast(0)=0$. Since $\mathfrak{a}$ must be an ordinary Lie algebra, its Chevalley-Eilenberg algebra $\mathrm{CE}(\mathfrak{a})$ will have only degree $1$ generators. Thus its generators must consist not only in the $e^a:=f^\ast(e^a)$, but also in an extra set $\tilde{e}_a$ for $a=0,\dots,d-1$ which satisfies
\begin{equation}
    \omega \,=\, \tilde{e}_a \wedge e^a
\end{equation}
Now the equation $\di \omega =0$, combined with the equation $\di e^a = 0$, implies that the differential of the new generator is zero, i.e. $\di \tilde{e}_a=0$. Thus we found the differential graded algebra
\begin{equation}
    \mathrm{CE}(\mathfrak{a}) \;=\; \mathbb{R}[e^a,\tilde{e}_a]/\langle\di e^a=0,\;\di \tilde{e}_a = 0\rangle
\end{equation}
which must come from the ordinary Lie algebra
\begin{equation}
    \mathfrak{a} \;=\; \mathbb{R}^{d}\oplus (\mathbb{R}^{d})^\ast.
\end{equation}
Let us now call $\mathbb{R}^{d,d}:=\mathbb{R}^{d}\oplus (\mathbb{R}^{d})^\ast$ and notice that the underlying vector space is $2d$-dimensional.
The differential graded algebra can be thought as
\begin{equation}
    \mathrm{CE}(\mathbb{R}^{d,d}) \;\cong\; \big(\Omega_{\mathrm{li}}^\bullet (\mathbb{R}^{d,d}),\,\di\big)
\end{equation}
where the notation $\Omega_{\mathrm{li}}^\bullet(G)$ means the complex of the left invariant differential forms on a Lie group $G$. But $\mathbb{R}^{d,d}$ is also a smooth manifold, with functions $\Coo(\mathbb{R}^{d,d})$ generated by coordinate functions $x^a,$ and $\tilde{x}_a$. Thus the left invariant $1$-forms on $\mathbb{R}^{d,d}$ are just
\begin{equation}
    \begin{aligned}
    e^a \,=\, \di x^a \\
    \tilde{e}_a \,=\, \di \tilde{x}_a
    \end{aligned}
\end{equation}
In conclusion we constructed a homomorphism of $2$-algebras
\begin{equation}\label{eq:algebraatlas}
    f:\,\mathbb{R}^{d}\oplus (\mathbb{R}^{d})^\ast\; \longtwoheadrightarrow\; \mathbb{R}^{d}\oplus\mathbf{b}\mathfrak{u}(1)
\end{equation}
which is a well-defined atlas for our $2$-algebra.
\vspace{0.25cm}

\noindent Now let us discuss the \textit{kernel pair} of the atlas \eqref{eq:algebraatlas}. As we have seen, this is defined as the \textit{pullback} (in the category theory sense) of two copies of the map $f$ of the atlas  \eqref{eq:algebraatlas}. The coequalizer diagram of these maps is
\begin{equation}\label{eq:kernelpair}
\begin{tikzcd}
\mathbb{R}^{d,d}\times_{\mathbb{R}^{d}\oplus\mathbf{b}\mathfrak{u}(1)}\mathbb{R}^{d,d} \arrow[r, yshift=0.7ex, two heads] \arrow[r, yshift=-0.7ex, two heads] & \mathbb{R}^{d,d}  \arrow[r, two heads, "f"] & \mathbb{R}^{d}\oplus\mathbf{b}\mathfrak{u}(1).
\end{tikzcd}
\end{equation}
To deal with it, we can consider the Chevalley-Eilenberg algebras of all the involved $L_\infty$-algebras and look at the equalizer diagram of the \textit{cokernel pair} which is dual to the starting kernel pair \eqref{eq:kernelpair}. This will be given by the following maps of differential graded algebras:
\begin{equation}
\begin{tikzcd}
\mathrm{CE}(\mathbb{R}^{d,d})\sqcup_{\mathrm{CE}\left(\mathbb{R}^{d}\oplus\mathbf{b}\mathfrak{u}(1)\right)}\mathrm{CE}(\mathbb{R}^{d,d})  & \mathrm{CE}(\mathbb{R}^{d,d}) \arrow[l, yshift=0.7ex, hook'] \arrow[l, yshift=-0.7ex, hook']  & \mathrm{CE}(\mathbb{R}^{d}\oplus\mathbf{b}\mathfrak{u}(1)) \arrow[l, hook', "f^\ast"'].
\end{tikzcd}
\end{equation}
Let us describe this in more detail. If composed with $f^\ast$, the two maps send the generators $e^a$ to $e^a$ and the generator $B$ to a couple $\omega=\tilde{e}_a\wedge e^a$ and $\omega'=\tilde{e}'_a\wedge e^a$, where $\tilde{e}_a$ and $\tilde{e}'_a$ are such that they both satisfy the same equation $\di \tilde{e}_a' = \di \tilde{e}_a$. This implies that they are related by a gauge transformation $\tilde{e}'_a = \tilde{e}_a + \di\lambda_a$. This fact can be seen as a consequence of the gauge transformations $B'=B+\di\lambda$ with parameter $\lambda:=\lambda_ae^a$.

\paragraph{An atlas for the $2$-group.}
Now $\mathbb{R}^{d}\times \mathbf{B}U(1)$ is the $2$-group which integrates the $2$-algebra $\mathbb{R}^{d}\oplus\mathbf{b}\mathfrak{u}(1)$. Let us now call $\mathbb{R}^{d,d}$ the manifold underlying the ordinary Lie group which integrates the ordinary abelian Lie algebra $\mathbb{R}^{d}\oplus (\mathbb{R}^{d})^\ast$. Then the map
\begin{equation}\label{eq:groupatlas}\boxed{\quad
    F:\,\mathbb{R}^{d,d}\; \longtwoheadrightarrow\; \mathbb{R}^{d}\times\mathbf{B}U(1) \quad}
\end{equation}
which exponentiates the homomorphism of algebras \eqref{eq:algebraatlas} is a well defined atlas for the Lie $2$-group $\mathbb{R}^{d}\times\mathbf{B}U(1)$. 

\paragraph{An atlas for the bundle gerbe.}
Recall that we proposed in \cite{Alf19} that the doubled space $\mathcal{M}$ is a bundle gerbe on a base manifold spacetime $M$. This implies that $\mathcal{M}$ can be locally trivialized to a collection of local trivial gerbes $\{U_\alpha \times \mathbf{B}U(1)\}_{\alpha\in I}$ on any given cover $\{U_\alpha\}_{\alpha\in I}$ of the base manifold $M$. 
Thus, similarly to the example of the ordinary circle bundle, we have a collection of effective epimorphism $\varphi_\alpha: \mathbb{R}^{d}\times \mathbf{B}U(1)\twoheadrightarrow U_\alpha \times \mathbf{B}U(1)$ for any chart. These induce a single effective epimorphism $\bigsqcup_{\alpha\in I}\mathbb{R}^{d}\times \mathbf{B}U(1)\xtwoheadrightarrow{\;\;\{\varphi_\alpha\}_{\alpha\in I}\;\;} \mathcal{M}$ which, however, \textit{is not an atlas}, since its source is itself a stack and not an ordinary manifold. This map is still interesting, because it formalizes the idea that the total space of the gerbe $\mathcal{M}$ locally looks like the Lie $2$-group $\mathbb{R}^{d}\times\mathbf{B}U(1)$. \vspace{0.25cm}

\noindent Now we know that each $\mathbb{R}^d\times \mathbf{B}U(1)$ has a natural atlas \eqref{eq:groupatlas}. Thus by composition we can construct maps
$\phi_\alpha:\mathbb{R}^{d,d} \xtwoheadrightarrow{\;\;F\;\;} \mathbb{R}^d\times\mathbf{B}U(1) \xtwoheadrightarrow{\;\;\varphi_\alpha\;\;} U_\alpha \times\mathbf{B}U(1)$. By combining them we can construct an atlas for the bundle gerbe:
\begin{equation}\label{eq:atlas}\boxed{\quad
    \bigsqcup_{\alpha\in I}\mathbb{R}^{d,d}\; \xtwoheadrightarrow{\;\; \{\phi_\alpha\}_{\alpha\in I} \;\;}\; \mathcal{M} \quad}
\end{equation}

\paragraph{\v{C}ech nerve of the atlas of the bundle gerbe.}
Let us assume that our bundle gerbe is specified by the \v{C}ech cocycle $\big(B_{(\alpha)},\,\Lambda_{(\alpha\beta)},\,G_{(\alpha\beta\gamma)}\big)$. We can now use the map \eqref{eq:atlas} to explicitly construct the \v{C}ech nerve of the atlas. We obtain the following simplicial object:
\begin{equation*}
\begin{tikzcd}[column sep=9ex]
\displaystyle\bigsqcup_{\alpha,\beta,\gamma\in I}\!\!\mathbb{R}^{d,d} \times_\mathcal{M} \mathbb{R}^{d,d} \times_\mathcal{M} \mathbb{R}^{d,d} \arrow[r, yshift=1.4ex, two heads]\arrow[r, two heads] \arrow[r, yshift=-1.4ex, two heads] & \displaystyle\bigsqcup_{\alpha,\beta\in I}\!\mathbb{R}^{d,d} \times_\mathcal{M} \mathbb{R}^{d,d} \arrow[r, yshift=0.7ex, two heads] \arrow[r, yshift=-0.7ex, two heads] & \displaystyle\bigsqcup_{\alpha\in I}\mathbb{R}^{d,d}  \arrow[r, two heads, "\{\phi_\alpha\}_{\alpha\in I} "] & \mathcal{M} 
\end{tikzcd}
\end{equation*}
Let us describe this diagram in more detail in terms of its dual diagram of Chevalley-Eilenberg algebras. The two maps of the kernel pair send the local degree $1$ generator to $\di x^\mu$ and the local degree $2$ generator to a couple of local $2$-forms $\omega_{(\alpha)}^{\mathrm{triv}}=\di\tilde{x}_{(\alpha)\mu}\wedge \di x^\mu$ and $\omega_{(\beta)}^{\mathrm{triv}}=\di\tilde{x}_{(\beta)\mu}\wedge \di x^\mu$ on the fiber product of the $\alpha$-th and $\beta$-th charts. Now the local $1$-forms $\di\tilde{x}_{(\alpha)\mu}$ and $\di\tilde{x}_{(\beta)\mu}$ are required to be related by a gauge transformation $\di\tilde{x}_{(\alpha)\mu} = \di\tilde{x}_{(\beta)\mu} + \di\Lambda_{(\alpha\beta)\mu}$ where the gauge parameters $\Lambda_{(\alpha\beta)\mu}$ are given by the cocycle of the bundle gerbe. Equivalently the two $2$-forms must be related by a gauge transformation $\omega_{(\alpha)}^{\mathrm{triv}}=\omega_{(\beta)}^{\mathrm{triv}}+\di\Lambda_{(\alpha\beta)}$  with gauge parameter $\Lambda_{(\alpha\beta)}=\Lambda_{(\alpha\beta)\mu}\di x^\mu$. The gauge parameters, as expected, are required to satisfy the cocycle condition $\Lambda_{(\alpha\beta)}+\Lambda_{(\beta\gamma)}+\Lambda_{(\gamma\alpha)}=\di G_{(\alpha\beta\gamma)}$ on three-fold fiber products of charts.

\subsection{The atlas of DFT}

The main problem of the traditional approaches to geometry of DFT is trying to glue the left-hand-side $\bigsqcup_{\alpha\in I}\mathbb{R}^{d,d}$ of the atlas \eqref{eq:atlas} to form a global $2d$-dimensional smooth manifold, not recognizing that it is actually the atlas of a bundle gerbe.

\paragraph{Natural interpretation for the extra dimensions.}
\textit{The $2d$-dimensional atlas of the bundle gerbe is the natural candidate for being an atlas for the doubled space of DFT}.
This means that we can avoid the conceptual issue of requiring a $2d$-dimensional spacetime (or even a much higher-dimensional one for Exceptional Field Theory), because the extra $d$ coordinates of the charts locally describe the remaining degree of freedom of the bundle gerbe. In this sense, DFT on a chart $\mathbb{R}^{d,d}$ is a local description for a \textit{field theory on the bundle gerbe}. 

\paragraph{Principal connection of the gerbe.}
On the atlas of a bundle gerbe we can define its \textit{principal connection} $\omega\in\Omega^2(\bigsqcup_{\alpha\in I}\mathbb{R}^{d,d})$ by the difference $\omega_{(\alpha)} := \omega^{\mathrm{triv}}_{(\alpha)} - B_{(\alpha)}$ of the local $2$-form $\omega^{\mathrm{triv}}_{(\alpha)}$ we obtained in the previous subsection and the pullback of the local connection $2$-form $B_{(\alpha)}$ of the bundle gerbe living on the base manifold. This definition assures that $\omega_{(\alpha)} =  \omega_{(\beta)}$ on overlaps of charts $\mathbb{R}^{d,d}\times_\mathcal{M}\mathbb{R}^{d,d}$. Thus in local coordinates we can write
\begin{equation}\boxed{\quad
    \omega \;=\; \big(\di\tilde{x}_{(\alpha)\mu}+B_{(\alpha)\mu\nu}\di x^\nu \big)\wedge \di x^\mu \quad}
\end{equation}
Notice that the form $\omega$ is invariant under gauge transformations of the bundle gerbe, i.e. of the Kalb-Ramond field.
In general it is also possible to express the principal connection $\omega=\tilde{e}_{a}\wedge e^a$ in terms of the globally defined $1$-forms $\tilde{e}_{a}=\di\tilde{x}_{(\alpha)a} + B_{(\alpha)a\nu}\di x^\nu$ and $e^a=\di x^a$ on the atlas.
\vspace{0.25cm}

\noindent We can also pack both the left invariant differential forms in a single $1$-form $E^A$ with index $A=1,\dots,2d$ which is defined by $E^a:=e^a$ and $E_{a}:=\tilde{e}_{a}$. In this notation we have that the connection can be expressed by
\begin{equation}
    \omega \;=\; \omega_{AB}\,E^A\wedge E^B
\end{equation}
where $\omega_{AB}$ is the $2d$-dimensional standard symplectic matrix. Notice that we recover the curvature of the bundle gerbe by
\begin{equation}
    H\;=\;-\di\omega \;\in\,\Omega^3_{\mathrm{cl}}(M).
\end{equation}
This is completely analogous to the curvature of a circle bundle $P\twoheadrightarrow M$ being the differential of its connection $\xi\in\Omega^1(P)$, i.e. it is $F=\di\xi\in\Omega^2_{\mathrm{cl}}(M)$.

\paragraph{Global generalized metric on the gerbe.}
A \textit{global generalized metric} can now be defined just as an orthogonal structure
\begin{equation}
    \mathcal{M} \; \longrightarrow \; GL(2d)/\!/O(2d),
\end{equation}
on the bundle gerbe itself, just like a Riemannian metric on a manifold.
As explained in \cite{Alf19}, if we require the generalized metric structure to be invariant under the principal action of the bundle gerbe, this will have to be of the form
\begin{equation}
\begin{aligned}
    \mathcal{G}_{(\alpha)} \;&=\; g_{ab}\,e^a\otimes e^b \,+\, g^{ab}\,\tilde{e}_{a} \otimes \tilde{e}_{b} \\[0.2em]
    &=\; \mathrm{diag}(g,g^{-1})_{AB}\,E^A \otimes E^B
    \end{aligned}
\end{equation}
and in in terms of local coordinates we find the usual expression
\begin{equation}
    \mathcal{G}_{(\alpha)MN} \;=\; \begin{pmatrix}g_{\mu\nu}- B_{(\alpha)\mu\lambda}g^{\lambda\rho}B_{(\alpha)\rho\beta} & B_{(\alpha)\mu\lambda}g^{\lambda\nu} \\-g^{\mu\lambda}B_{(\alpha)\lambda\nu} & g^{\mu\nu} \end{pmatrix}.
\end{equation}

\paragraph{Local para-complex geometry.}
Our local chart is canonically split by $\mathbb{R}^{d,d} = \mathbb{R}^d \oplus (\mathbb{R}^d)^\ast$, where the restriction $\mathbb{R}^d$ can be seen as a chart for the $d$-dimensional base manifold $M$ of the gerbe. This immediately implies that the tangent bundle of the local chart splits by
\begin{equation}
    T\mathbb{R}^{d,d} \; \cong\; T\mathbb{R}^d \,\oplus\,  T(\mathbb{R}^d)^\ast
\end{equation}
Then on each chart the gerbe connection becomes a projector to the vertical bundle
\begin{equation}
    \omega:\, T\mathbb{R}^{d,d} \,\longtwoheadrightarrow\, T(\mathbb{R}^d)^\ast
\end{equation}
Recall that an Ehresmann connection for an ordinary principal bundle defines a projection $\xi: TP\twoheadrightarrow VP$ onto the vertical subbundle: for the gerbe it is not so different. If we consider a vector $X=X^\mu \partial_\mu + \tilde{X}_\mu\tilde{\partial}^\mu$ on $\mathbb{R}^{d,d}$, this will be mapped by the connection to $X_V:=(\tilde{X}_\mu+B_{(\alpha)\mu\nu}X^\nu)\tilde{\partial}^\mu$. Thus, if we call $\{E_A\}$ a basis of left-invariant vectors on $\mathbb{R}^{d,d}$ dual to the $1$-forms $\{E^A\}$, we obtain vectors of the form
\begin{equation}
    X^AE_A\;=\; X^\mu \partial_\mu + \big(\tilde{X}_\mu+B_{(\alpha)\mu\nu}X^\nu\big)\tilde{\partial}^\mu
\end{equation}
Notice that, if we restrict ourselves to strong constrained vectors, these are immediately globalized to sections of a Courant algebroid twisted by the gerbe with connection $B_{(\alpha)}$. See \cite{Alf19} for more details about the tangent stack of the gerbe.
\vspace{0.25cm}

\noindent The para-complex structure can be defined by using the gerbe connection by $J := \mathrm{id}_{T\mathbb{R}^{d,d}}-2\omega$, in analogy with a principal connection. If we split a vector in horizontal and vertical projection $X=X_H+X_V$, this will be mapped to $J(X)= X_H-X_V$.\vspace{0.25cm}

\noindent Thus every chart $(\mathbb{R}^{d,d},J,\omega)$ is a para-Hermitian vector space. In this specific sense \textit{a bundle gerbe is naturally equipped with an atlas of para-Hermitian charts}, even if its total space is not a smooth manifold.

\paragraph{Local doubled-yet-gauged geometry.}
The principal action of the bundle gerbe will be given on a local chart of the atlas by a shift $(x^\mu,\tilde{x}_\mu)\mapsto(x^\mu,\tilde{x}_\mu+\lambda_\mu)$ in the unphysical coordinates, identified with a gauge transformation $B\mapsto B + \di(\lambda_\mu\di x^\mu)$ of the Kalb-Ramond field.
This matches with the coordinate gauge symmetry discovered by \cite{Park13}, upon application of the section condition.
Moreover the global bundle gerbe property $\mathcal{M}/\!/\mathbf{B}U(1)\cong M$, when written on a local chart of the atlas, can be identified with the property that physical points corresponds to gauge orbits of the doubled space: this gives a global geometric interpretation of the strong constraint. Thus the local charts of the bundle gerbe match with the doubled-yet-gauged patches by \cite{Park13}.
In this sense the Higher Kaluza-Klein formalism can also be seen as a globalization of the local geometry underlying the doubled-yet-gauged space proposal which we briefly reviewed in section \ref{s2}. 

\subsection{The T-dual spacetime is a submanifold of the gerbe}

If we accept the identification of the global doubled space with the total space of the bundle gerbe, then how can we obtain the T-dual spacetime to the starting one? Let us briefly explain it with a concrete example.

\paragraph{Abelian T-duality.}
Let our gerbe $\mathcal{M}\twoheadrightarrow M$ have a base manifold which is itself the total space of a $T^n$-bundle $M\twoheadrightarrow M_0$. Moreover we will assume that the gerbe bundle satisfies the \textit{T-duality condition} $\mathcal{L}_{\partial_i}H=0$. Now let $\big(x_{(\alpha)}^\mu,\theta^i_{(\alpha)}\big)$ be the local coordinates of a chart $\mathbb{R}^d=\mathbb{R}^{d-n}\times \mathbb{R}^n$ of $M$ adapted to the torus fibration and let $\xi^i=\di\theta_{(\alpha)}^i + A^i_{(\alpha)}\in\Omega^1(M)$ be the global connection $1$-form of the torus bundle.
This means that the gerbe connection will be 
\begin{equation}
    \omega \;=\; \big(\di\tilde{\theta}_{(\alpha)i}-\iota_{\partial_i}B_{(\alpha)} \big)\wedge \big(\di \theta_{(\alpha)}^i + A^i_{(\alpha)}\big) \,+\, \Big(\di\tilde{x}_{(\alpha)\mu}+B_{(\alpha)\mu\nu}^{(2)}\di x^\nu \Big)\wedge \di x^\mu
\end{equation}
where we called $B_{(\alpha)}^{(2)}$ the horizontal part of the $2$-form $B_{(\alpha)}$ respect to the torus fibration and where $\big(\tilde{x}_{(\alpha)i},\tilde{\theta}_{(\alpha)i}\big)$ are the local coordinates of $(\mathbb{R}^{d})^\ast$.
Thus we obtain the following forms
\begin{equation}
    \begin{aligned}
        e^\mu \;&=\; \di x^\mu &\quad e^i \;&=\; \di\theta_{(\alpha)}^i + A^i_{(\alpha)} \\[0.4em]
        \tilde{e}_{\mu} \;&=\; \di \tilde{x}_{(\alpha)\mu}+ B^{(2)}_{(\alpha)\mu\nu}\di x^\nu \quad & \tilde{e}_{i} \;&=\; \di\tilde{\theta}_{(\alpha)i} - \iota_{\partial_i}B_{(\alpha)} 
    \end{aligned}
\end{equation}
Thanks to the T-dualizability condition satisfied by the bundle gerbe, something special happens: the $1$-form $\tilde{e}_i=\di\tilde{\theta}_{(\alpha)i}-\iota_{\partial_i}B_{(\alpha)}$ becomes the global connection of a well-defined $T^n$-bundle $K\twoheadrightarrow M$. See \cite{Alf19} for more details about abelian T-duality in the bundle gerbe picture. 
\vspace{0.25cm}

\noindent Therefore, \textit{in the special case of a T-dualizable bundle gerbe, the $(\mathbb{R}^{n})^\ast\subset(\mathbb{R}^{d})^\ast$ part of the charts are glued together to form an extra manifold: an extra $\tilde{T}^n$-bundle. This manifold can be seen as the T-dual spacetime and this gives rise to T-duality.} On the other hand the remaining $(\mathbb{R}^{d-n})^\ast\subset(\mathbb{R}^{d})^\ast$ part of the charts still cannot be glued to form a manifold. Let us remark that in the general case the local charts $(\mathbb{R}^d)^\ast$ of the bundle gerbe cannot be glued to form a manifold at all. Whenever the gerbe contains such a fiber bundle $K\twoheadrightarrow M$, which we will call \textit{generalized correspondence space}, there is T-duality.
\vspace{0.25cm}

\noindent See section \ref{s5} for a deeper discussion of more general cases of T-duality and generalized correspondence space, including the ones whose fibers are Drinfel'd doubles.

\paragraph{Born Geometry of the internal space.}
This also explains the effectiveness of Born Geometry when dealing with internal spaces, such as tori or Drinfel'd doubles. In fact, the restriction of the gerbe connection $\omega$ to the fiber of the generalized correspondence space $K$ is the fundamental $2$-form of an almost symplectic structure. For example, for the previous example abelian T-duality, the connection restricts to $\omega|_{T^{2n}}= \big(\di\tilde{\theta}_{(\alpha)i} + B^{(0)}_{(\alpha)ij}\di\theta_{(\alpha)}^j \big)\wedge \di\theta_{(\alpha)}^i$, where we called $B^{(0)}_{(\alpha)ij} := \iota_{\partial_i}\iota_{\partial_j}B_{(\alpha)}$ the moduli field of the Kalb-Ramond field. It is not hard to see that the fibers of $K$ are almost para-Hermitian manifolds equipped with Born Geometry. 

\paragraph{Relaxation of the strong constraint.}
In principle, if we want to relax the strong constraint, we can consider fields on our bundle gerbe which are not necessarily invariant under its principal action. We will have to glue these new fields according to the rules that we exposed in the last subsection.

\section{Global tensor hierarchy in Higher Kaluza-Klein Theory}\label{s5}

In this section we will argue that strong constrained tensor hierarchies should be globalized and geometrized by reduction of the differential data of a bundle gerbe. As particular examples of this perspective we will deal with two example of T-folds.

\subsection{Global tensor hierarchies, topology and non-geometry}

We will try to examine the idea of tensor hierarchies from section \ref{s2} by using the globally-defined bundle gerbe machinery from section \ref{s3}.

\paragraph{Motivation.}
The global T-fold geometries we introduced in \cite{Alf19} (see section \ref{s3}), since they are obtained by dimensionally reducing the global doubled space (i.e. the bundle gerbe), should be considered global tensor hierarchies, in the spirit of \cite{HohSam13KK}. However it is not at all clear that T-folds can be obtained by gauging the local tensor hierarchy algebra of section \ref{s2}, i.e. the local prestack $\Omega\big(U,\mathscr{D}(\mathbb{R}^{2n})\big)$. In this section we will propose a definition of global strong constrained tensor hierarchies which is more general than the higher gauge theory definition of section \ref{s2}. Then we will show that abelian and Lie-Poisson T-folds match this definition.

\paragraph{Strong constrained tensor hierarchies.}
Recall that tensor hierarchies require the strong constraint to be well-defined. Let us thus replace the C-bracket with the anti-symmetrized Roytenberg bracket of Generalized Geometry. We can then solve the strong constraint and obtain locally the curvature as $\mathscr{D}_{\mathrm{sc}}(\mathbb{R}^{2n})$-valued differential forms
\begin{equation}
    \begin{aligned}
    \mathcal{F} \;&=\; \di \mathcal{A} - [ \mathcal{A} \,\overset{\wedge}{,}\, \mathcal{A}]_{\mathrm{Roy}} + \mathfrak{D}\mathcal{B} \\[0.2em]
    \mathcal{H} \;&=\;  \mathrm{D}\mathcal{B} + \frac{1}{2}\langle \mathcal{A} \,\overset{\wedge}{,}\, \di \mathcal{A}\rangle - \frac{1}{3!} \langle \mathcal{A} \,\overset{\wedge}{,}\,   [ \mathcal{A} \,\overset{\wedge}{,}\, \mathcal{A}]_{\mathrm{Roy}} \rangle 
    \end{aligned}
\end{equation}
where now the bracket $[ -,-]_{\mathrm{Roy}}$ is the anti-symmetrized Roytenberg bracket. In coordinates this corresponds to setting $\tilde{\partial}^i=0$ on any field so we will also have $(\mathfrak{D}\mathcal{B})_i = \partial_i\mathcal{B}$, which implies $\mathrm{D}\mathcal{B} \;=\; \di \mathcal{B} + \mathcal{A}^i\wedge \partial_i \mathcal{B}$. Analogously for all the others $\mathscr{D}_{\mathrm{sc}}(\mathbb{R}^{2n})$-valued differential forms.

\paragraph{Global tensor hierarchies.}
Here is our proposal of definition of global tensor hierarchy:

\noindent\textit{A global DFT tensor hierarchy on $M_0$ is the result of the dimensional reduction of the connection of a bundle gerbe on a principal $G$-bundle $M\twoheadrightarrow M_0$.} Therefore the $2$-groupoid $\tenhie^{\,G}_{\!\mathrm{sc}}(M_0)$ of global DFT tensor hierarchies on $M_0$ is given as follows 
\begin{equation}\label{eq:globtenhie}
    \tenhie^{\,G}_{\!\mathrm{sc}}(M_0) \;:=\; \left\{\begin{tikzcd}[row sep=7ex, column sep=5ex]
    & \left[G,\mathbf{B}^2U(1)\right]\!/G \arrow[d]\\
    M_0 \arrow[ru]\arrow[r, ""] & \mathbf{B}G
    \end{tikzcd}\right\}
\end{equation}
\noindent Let us call $\mathrm{Gerbes}(M) \,=\,\big\{\!\!\begin{tikzcd}[row sep=7ex, column sep=3ex]M \arrow[r] & \mathbf{B}^2U(1)\end{tikzcd}\!\!\big\}$ the $2$-groupoid of bundle gerbes on a manifold $M$. From our definition of global tensor hierarchy \eqref{eq:globtenhie} and from the definition of dimensional reduction \eqref{eq:red} we immediately have the natural isomorphism of $2$-groupoids
\begin{equation}\boxed{\quad
    \bigsqcup_{\begin{subarray}{c}M\text{ s.t. }
    M\twoheadrightarrow M_0\\\text{is a }G\text{-bundle}\end{subarray}} \!\!\!\!\mathrm{Gerbes}(M) \;\;\cong\;\; \tenhie^{\,G}_{\!\mathrm{sc}}(M_0)\quad}
\end{equation}

\paragraph{Example: topological T-duality.}
Let us now re-examine better the case \eqref{eq:string}.
For any Lie group $G$ let us call $G\text{-}\mathrm{equivGerbes}(M)$ the groupoid of $G$-equivariant gerbes on $M$ and for any $L_\infty$-group $H$ let us call $H\text{-}\mathrm{Bundles}(M_0)$ the groupoid of principal $H$-bundles on $M_0$. Thus we have the equivalence
\begin{equation}
    \bigsqcup_{\begin{subarray}{c}M\text{ s.t. }
    M\twoheadrightarrow M_0\\\text{is a }T^n\text{-bundle}\end{subarray}} \!\!\!\!T^n\text{-}\mathrm{equivGerbes}(M) \;\;\cong\;\; \String(T^n\!\times \tilde{T}^n)\text{-}\mathrm{Bundles}(M_0)
\end{equation}
This reads as follows: any gerbe on the total space of a $T^n$-bundle $M\twoheadrightarrow M_0$ which is equivariant under the principal $T^n$-action is equivalently a $\String(T^n\!\times \tilde{T}^n)$-bundle on the base manifold $M_0$. If we forget the higher form fields, we remain with a $T^n\times \tilde{T}^n$-bundle, which is nothing but the correspondence space $K=M\times_{M_0}\widetilde{M}$ of topological T-duality. \vspace{0.25cm}

\noindent Thus, unsurprisingly, our definition of global tensor hierarchy includes doubled torus bundles.

\subsection{T-folds as global tensor hierarchies}

In this subsection we will briefly explain the global geometry of a T-fold, which is obtained by dimensionally reducing a bundle gerbe on a torus bundle spacetime. Then we will explain how the geometric structure we obtain can be naturally interpreted as a particular case of the global tensor hierarchies we defined. It will be important to notice that these T-folds cannot be obtained by gauging the algebra of local tensor hierarchies from section \ref{s2}. This will give further motivation to the definition of the previous subsection.

\paragraph{The generalized correspondence space of T-duality.}
Let us start from the $T^n$-bundle $M\twoheadrightarrow M_0$, whose total space $M$ is equipped with a Riemannian metric $\bigsp{g}$ and gerbe structure with curvature $\bigsp{H}\in\Omega^3_{\mathrm{cl}}(M)$. In the following we will use the underlined notation for the fields living on the total space $M$. We can now use the principal connection $\xi\in\Omega^1(M,\mathbb{R}^n)$ of the torus bundle to expand metric and curvature in horizontal and vertical components respect to the fibration. We will obtain
\begin{align}
    \bigsp{g} &= g^{(2)} + g^{(0)}_{ij}\xi^i\odot\xi^j\\
    \bigsp{H} &= H^{(3)} + H^{(2)}_{i}\wedge\xi^i  + \frac{1}{2} H^{(1)}_{ij}\wedge\xi^i\wedge\xi^j +\frac{1}{3!} H^{(0)}_{ijk}\xi^i\wedge\xi^j\wedge\xi^k
\end{align}
where we can choose $H^{(3)},H^{(2)}_{i},H^{(1)}_{ij}, H^{(0)}_{ijk}$ as globally defined differential forms which are pullbacks from base manifold $M_0$, so that they do not depend on the torus coordinates. Now on patches and two-fold overlaps of patches of a good cover of $M$ we can use the connection of the torus bundle to split the differential local data of the connection of the gerbe in horizontal and vertical part too. We obtain
\begin{equation}\label{splithv}
\begin{aligned}
    \bigsp{B}_{(\alpha)} &= B_\alpha^{(2)} + B_{(\alpha)i}^{(1)}\wedge\xi^i + \frac{1}{2} B_{(\alpha)ij}^{(0)} \xi^i\wedge\xi^j \\
    \bigsp{\Lambda}_{(\alpha\beta)} &= \Lambda_{(\alpha\beta)}^{(1)} + \Lambda_{(\alpha\beta)i}^{(0)} \xi^i
\end{aligned}
\end{equation}
The Bianchi identity of the gerbe on the total space $M$ reduces to the base $M_0$ as it follows:
\begin{equation}\label{eq:bianchit}
    \bigsp{\di} \bigsp{H} \,=\, 0 \hspace{0.4cm} \implies \hspace{0.4cm} \left\{ \;
    \begin{aligned}
    \di H^{(0)}_{ijk} \,&=\, 0 \\[0.5em]
    \di H^{(1)}_{ij} + H^{(0)}_{ijk}\wedge F^k \,&=\, 0 \\[0.5em]
    \di H^{(2)}_{i} + H^{(1)}_{ij}\wedge F^j \,&=\, 0 \\[0.5em]
    \di H^{(3)} + H^{(2)}_i\wedge F^i \,&=\, 0
    \end{aligned} 
    \right.
\end{equation}
where $\bigsp{\di}$ and $\di$ are respectively the exterior derivative on the total space $M$ and on the base manifold $M_0$. Analogously, the expression of the curvature of bundle gerbe on local patches becomes 
\begin{equation}
    \bigsp{H} \,=\, \bigsp{\di} \bigsp{B}_{(\alpha)} \hspace{0.4cm} \implies \hspace{0.4cm} \left\{ \;
    \begin{aligned}
    H^{(0)}_{ijk} \,&=\, \partial_{[i}B^{(0)}_{(\alpha)jk]} \\[0.5em]
    H^{(1)}_{ij} \,&=\, \di B^{(0)}_{(\alpha)ij} - \partial_{[i}B^{(1)}_{(\alpha)j]} \\[0.5em]
    H^{(2)}_i \,&=\, \di B^{(1)}_{(\alpha)i} + \partial_iB^{(2)}_{(\alpha)} - B^{(0)}_{(\alpha)ij} F^j \\[0.5em]
    H^{(3)} \,&=\, \di B_{(\alpha)}^{(2)} - B^{(1)}_{(\alpha)i}\wedge F^i
    \end{aligned} 
    \right.
\end{equation}
where $\partial_i = \partial/\partial\theta^i$ is the derivative respect to the $i$-th coordinate of the torus fiber.
The patching conditions of the connection $2$-form on two-fold overlaps of patches are as following:
\begin{equation}\label{eq:bshifts}
    \bigsp{B}_{(\beta)} - \bigsp{B}_{(\alpha)} \,=\, \bigsp{\di} \bigsp{\Lambda}_{(\alpha\beta)} \hspace{0.4cm} \implies \hspace{0.4cm} \left\{ \;
    \begin{aligned}
    B_{(\beta)ij}^{(0)} - B_{(\alpha)ij}^{(0)} \,&=\, \partial_{[i}\Lambda^{(0)}_{(\alpha\beta)j]} \\[0.5em]
    B_{(\beta)i}^{(1)} - B_{(\alpha)i}^{(1)} \,&=\, \di \Lambda^{(0)}_{(\alpha\beta)i} - \partial_i\Lambda^{(1)}_{(\alpha\beta)} \\[0.5em]
    B_{(\beta)}^{(2)} - B_{(\alpha)}^{(2)} \,&=\, \di \Lambda^{(1)}_{(\alpha\beta)} +  \Lambda^{(0)}_{(\alpha\beta)i}F^i
    \end{aligned} 
    \right.
\end{equation}
And the patching conditions of the $1$-forms on three-fold overlaps of patches become
\begin{equation}
    \bigsp{\Lambda}_{(\alpha\beta)} + \bigsp{\Lambda}_{(\beta\gamma)} + \bigsp{\Lambda}_{(\gamma\alpha)} \,=\, \bigsp{\di} g_{(\alpha\beta\gamma)} \hspace{0.4cm} \implies \hspace{0.4cm} \left\{ \;
    \begin{aligned}
       \Lambda_{(\alpha\beta)i}^{(0)} + \Lambda_{(\beta\gamma)i}^{(0)} + \Lambda_{(\gamma\alpha)i}^{(0)} &= \partial_ig_{(\alpha\beta\gamma)} \\
    \Lambda_{(\alpha\beta)}^{(1)} + \Lambda_{(\beta\gamma)}^{(1)} + \Lambda_{(\gamma\alpha)}^{(1)} &= \di g_{(\alpha\beta\gamma)}
    \end{aligned}\right.
\end{equation}

\paragraph{The tensor hierarchy of a T-fold.}
To show that this geometric structure we obtained from the reduction of the bundle gerbe is a particular case of global tensor hierarchy, let us make the following redefinitions to match with the notation we used in section \ref{s2}:
\begin{equation}\label{eq:maptolanguage}
    \begin{aligned}
    \mathcal{F}^I_{(\alpha)} \,&:=\, \begin{pmatrix}\delta^i_j & 0 \\[0.6em] B^{(0)}_{(\alpha)ij} & \delta^j_i \end{pmatrix} \begin{pmatrix} F^j \\[0.6em] H^{(2)}_j \end{pmatrix}, \; &\mathcal{H}\,&:=\, H^{(3)} \\[0.4em]
    \mathcal{A}^I_{(\alpha)} \,&:=\, \begin{pmatrix} A_{(\alpha)} \\[0.6em] B^{(1)}_{(\alpha)} \end{pmatrix} , \; & \mathcal{B}_{(\alpha)}\,&:=\, B^{(2)}_{(\alpha)} \\[0.3em]
    \lambda_{(\alpha\beta)}^I \,&:=\, \begin{pmatrix} \lambda_{(\alpha\beta)}^i \\[0.7em] \Lambda^{(0)}_{(\alpha\beta)i} \end{pmatrix} , \; &\Xi_{(\alpha\beta)} \,&:=\, \Lambda^{(1)}_{(\alpha\beta)}.
    \end{aligned}
\end{equation}
where the local $1$-forms $A_{(\alpha)}^i$ and scalars $\lambda_{(\alpha\beta)}^i$ are respectively the local potential and the transition functions of the original torus bundle $M\twoheadrightarrow M_0$.
Since our fields are assumed to be strong constrained, we well have simply $\mathfrak{D}^I=(0,\,\partial_i)$ with $\partial_i=\partial/\partial\theta^i$ on horizontal forms. The Bianchi equation \eqref{eq:bianchit} of the gerbe curvature, together with the Bianchi equation $\di F = 0$ of the curvature of the torus bundle can now be equivalently rewritten as
\begin{equation}\boxed{\quad
    \begin{aligned}
    \di \mathcal{F}_{(\alpha)} \;&=\; 0 \\
    \di \mathcal{H} - \frac{1}{2}\langle \mathcal{F}_{(\alpha)} \;\overset{\wedge}{,}\; \mathcal{F}_{(\alpha)} \rangle\;&=\;  0
    \end{aligned}\quad}
\end{equation}
which are a particular case of the Bianchi equations of a tensor hierarchy. We can now rewrite all the patching conditions in the following equivalent form:
\begin{equation}\label{eq:tfoldpat}\boxed{\quad
    \begin{aligned}
    \mathcal{F}_{(\alpha)} \;&=\; \di \mathcal{A}_{(\alpha)}  + \mathfrak{D}\mathcal{B}_{(\alpha)} \\
    \mathcal{H} \;&=\;  \di \mathcal{B}_{(\alpha)} + \frac{1}{2}\langle \mathcal{A}_{(\alpha)} \,\overset{\wedge}{,}\, \mathcal{F}_{(\alpha)}\rangle \\[0.8em]
    \mathcal{A}_{(\alpha)} - \mathcal{A}_{(\beta)} \;&=\; \di\lambda_{(\alpha\beta)} + \mathfrak{D}\Xi_{(\alpha\beta)} \\
    \mathcal{B}_{(\alpha)} - \mathcal{B}_{(\beta)} \;&=\; \di \Xi_{(\alpha\beta)} - \langle\lambda_{(\alpha\beta)},\mathcal{F}_{(\alpha)}\rangle \\[0.8em]
    \lambda_{(\alpha\beta)}+\lambda_{(\beta\gamma)}+\lambda_{(\gamma\alpha)} \;&=\; \mathfrak{D} g_{(\alpha\beta\gamma)} \\
    \Xi_{(\alpha\beta)} + \Xi_{(\beta\gamma)} + \Xi_{(\gamma\alpha)} \;&=\; \di g_{(\alpha\beta\gamma)} \\[0.8em]
    g_{(\alpha\beta\gamma)} - g_{(\beta\gamma\delta)} +  g_{(\gamma\delta\alpha)} -  g_{(\delta\alpha\beta)}\;&\in\;2\pi\mathbb{Z}
    \end{aligned}\quad}
\end{equation}
which, at first look, appears a particular and strong constrained case of the global tensor hierarchy in \eqref{eq:tensorhipatched}. However we will see in the following that it is not completely the case. This will motivate more the identification of a T-fold with an element of $\tenhie^{\,T^n}_{\!\mathrm{sc}}\!(M_0)$. 
\vspace{0.25cm}

\noindent It is well-known that, to be T-dualizable, the string background we started with must satisfy the T-duality condition $\mathcal{L}_{\partial_i}\bigsp{H}=0$ on the curvature of the bundle gerbe. From now on we will assume a simple solution for this equation: the invariance of Kalb-Ramond field under the torus action. In other words we will require $\mathcal{L}_{\partial_i}\bigsp{B}_{(\alpha)}=0$, but the other differential data $\bigsp{\Lambda}_{(\alpha\beta)}$, $g_{(\alpha\beta\gamma)}$ of the gerbe are still allowed to depend on the torus coordinates. See \cite{Alf19} for the general solution. Notice that this immediately implies that $\mathcal{F}_{(\alpha)} = \di \mathcal{A}_{(\alpha)}$ in equation \eqref{eq:tfoldpat}.

\paragraph{Topology of the tensor hierarchy of a T-fold.}
Now it is important to show that the curvature $\mathcal{F}_{(\alpha)}^I$\textit{ is not in general the curvature of a }$T^{2n}$\textit{-bundle on the }$(d-n)$\textit{-dimensional base manifold }$M_0$. To see this let us split $\mathcal{F}_{(\alpha)}^I=(\mathcal{F}_{(\alpha)}^i,\,\widetilde{\mathcal{F}}_{(\alpha)i})$ and consider $[\pi^\ast H] \in H^2(M,\mathbb{Z}^n)$. We can see that
\begin{equation}\label{eq:bigneq}
    [(\pi^\ast H)_i] \,=\, \left[H^{(2)}_i-H^{(1)}_{ij}\xi^j\right] \;\neq\; \left[H^{(2)}_i+B^{(0)}_{(\alpha)ij}F^j\right] \,=\, [\widetilde{\mathcal{F}}_{(\alpha)i}]
\end{equation}
Notice that the inequality \eqref{eq:bigneq} becomes an equality if and only if $H^{(1)}$ is an exact form on the base manifold $M_0$, as it is showed in \cite{BelHulMin07} and \cite{Alf19}. In this case we would have $H^{(0)}=0$ and $H^{(1)}=\di B^{(0)}$, where $B^{(0)}$ would be a global $\wedge^2\mathbb{R}^n$-valued scalar on $M_0$, which indeed implies $[\di B^{(0)}_{ij}\xi^j]=-[B^{(0)}_{ij}F^j]$. As explained in \cite{Alf19}, this particular case corresponds to \textit{geometric T-duality}, which is the case where the T-dual spacetime is a well-defined manifold and not a non-geometric T-fold. Thus geometric T-duality is exactly the special case where $\mathcal{F}_{(\alpha)}^I$ is the curvature of a $T^{2n}$-bundle on $M_0$. But what is the geometric picture for a T-fold? 
\vspace{0.25cm}

\noindent In the T-fold case, as seen in \cite{Alf19}, we can think about $[\pi^\ast H] \in H^2(M_0,\mathbb{Z}^n)$ as the curvature of a $T^n$-bundle $K\twoheadrightarrow M$ over the total spacetime $M$. The total space $K$ is called \textit{generalized correspondence space}.
\begin{equation}
    \begin{tikzcd}[row sep=4ex, column sep=5ex]
    \tilde{T}^n\arrow[r, hook] & K \arrow[d, two heads] & \;\;\;\,\mathrm{c}_1(K)=[\pi^\ast H] \\
    T^n\arrow[r, hook] & M \arrow[d, two heads, "\pi"] & \mathrm{c}_1(M)=[F] \\
    & M_0 &
    \end{tikzcd}
\end{equation}

\noindent Now the picture of the doubled torus bundle holds only locally on $U_{\alpha}\times T^{2n}$ for each patch $U_{\alpha}\subset M_0$. Now, if we call collectively $\Theta_{(\alpha)}^I:=\big(\theta^i_{(\alpha)},\,\tilde{\theta}_{(\alpha)i}\big)$ the $2n$ coordinates of the fiber $T^{2n}$, we can construct the Ehresmann connection of any local doubled torus bundle $U_{\alpha}\times T^{2n}$ by 
\begin{equation}\label{eq:localconn}
    \di\Theta_{(\alpha)}^I +\mathcal{A}^I_{(\alpha)} \,\;\in \;\Omega^{1}\!\left(U_{\alpha}\times T^{2n}\right).
\end{equation}
The geometrical meaning of the curvature $\mathcal{F}_{(\alpha)}\in\Omega^{2}_{\mathrm{cl}}\!\left(U_{\alpha}\times T^{2n}\right)$ is being at every patch the curvature of the local torus bundle $U_{\alpha}\times T^{2n}$, even if these ones are not globally glued to be a $T^n$-bundle on $M_0$. This corresponds indeed to the well-known fact that a \textit{non-geometry is a global property}. In fact we can always perform geometric T-duality if we restrict ourselves on any local patch: the problem is that all these T-dualized patches will in general not glue together. 
\vspace{0.25cm}

\noindent As derived by \cite{BelHulMin07} and more recently by \cite{NikWal18}, T-folds are characterized by a \textit{monodromy matrix} cocycle $\big[n_{(\alpha\beta)}\big]$, which is a collection of an anti-symmetric integer-valued matrix $n_{(\alpha\beta)}$ at each two-fold overlap of patches, satisfying the cocycle condition $n_{(\alpha\beta)}+n_{(\beta\gamma)}+n_{(\gamma\alpha)}=0$ on each three-fold overlap of patches. The monodromy matrix cocycle is nothing but the gluing data for the local $B^{(0)}_{(\alpha)}$ moduli fields, i.e. it encodes integer $B$-shifts $B^{(0)}_{(\alpha)}-B^{(0)}_{(\beta)}=n_{(\alpha\beta)}$ on each two-fold overlap of patches. This arises from equation \eqref{eq:bshifts} combined with the T-dualizability condition $\mathcal{L}_{\partial_i}\bigsp{B}_{(\alpha)}$. The consequence of the presence of the monodromy matrix cocycle is that the local connections \eqref{eq:localconn} are glued on two-fold overlaps of patches $(U_{\alpha}\cap U_{\beta})\times T^{2n}$ by
\begin{equation}
    \big(\di\Theta_{(\alpha)} +\mathcal{A}_{(\alpha)}\big)^I \; = \; \big(e^{n_{(\alpha\beta)}}\big)^I_{\;J} \, \big(\di\Theta_{(\beta)} +\mathcal{A}_{(\beta)} \big)^J
\end{equation}
This immediately comes from the definition of these connections in equation \eqref{eq:maptolanguage}. Moreover this immediately implies that the curvature is glued by the monodromy matrix cocycle too as 
\begin{equation}
\mathcal{F}_{(\alpha)}^I \,=\, \big(e^{n_{(\alpha\beta)}}\big)^I_{\;J} \,\mathcal{F}_{(\beta)}^J
\end{equation}
In this sense, a T-fold is patched by a cocycle $e^{n_{(\alpha\beta)}}\in O(n,n;\mathbb{Z})$ valued in the T-duality group.
\vspace{-0.1cm}

\noindent If, instead, we want to look at the T-fold as a globally defined $T^n$-bundle $K\twoheadrightarrow M$ with first Chern class $[\pi^\ast H]\in H^2(M,\mathbb{Z}^n)$, we can easily construct its connection by noticing that the following $1$-form is global on the total space $K$ of the bundle:
\begin{equation}
    \Big(e^{-B_{(\alpha)}^{(0)}}\Big)^{\! I}_{\;J} \big(\di\Theta_{(\alpha)} +\mathcal{A}_{(\alpha)}\big)^J \;\, = \,\; \Big(e^{-B_{(\beta)}^{(0)}}\Big)^{\! I}_{\;J} \big(\di\Theta_{(\beta)} +\mathcal{A}_{(\beta)} \big)^J.
\end{equation}
We can thus define the global $1$-form:
\begin{equation}\label{eq:defrealconnection}
    \Xi^I \;:=\; \Big(e^{-B_{(\alpha)}^{(0)}}\Big)^{\! I}_{\;J} \big(\di\Theta_{(\alpha)} +\mathcal{A}_{(\alpha)}\big)^J \;\;\in\,\Omega^1(K,\mathbb{R}^{2n})
\end{equation}
whose first $n$ components are just the pullback of connection $\Xi^i=\xi^i$ of spacetime $M\twoheadrightarrow M_0$ and whose last $n$ components $\Xi_i$ are the wanted connection of the generalized correspondence space $K\twoheadrightarrow M$. As desired, the differential $\di\Xi_i$ on $K$ gives the pullback on $K$ of the globally-defined curvature $\pi^\ast H\in\Omega^2_{\mathrm{cl}}(M,\mathbb{R}^n)$.
\vspace{0.25cm}

\noindent Similarly to $\mathcal{F}_{(\alpha)}$, the moduli field $\mathcal{G}_{(\alpha)IJ}$ of the generalized metric is not a global $O(n,n)$-valued scalar on the base manifold $M_0$, but it is glued on two-fold overlaps of patches $U_\alpha\cap U_\beta \subset M_0$ by the integer $B$-shifts encoded by the monodromy matrix $n_{(\alpha\beta)}$ of the T-fold as
\begin{equation}
    \mathcal{G}_{(\alpha)IJ} \;\,=\,\; \big(e^{n_{(\alpha\beta)}}\big)^K_{\;I}\; \mathcal{G}_{(\beta)KL}\; \big(e^{n_{(\alpha\beta)}}\big)^L_{\;J}
\end{equation}
Only the $3$-form field $\mathcal{H}$ of the tensor hierarchy, as seen in \cite{Alf19}, is a globally defined (but not closed) differential form on the $(d-n)$-dimensional base manifold $M_0$.

\begin{figure}[!ht]\centering
\vspace{0.2cm}
\tikzset{every picture/.style={line width=0.75pt}} 
\begin{tikzpicture}[x=0.75pt,y=0.75pt,yscale=-1,xscale=1]
\draw  [fill={rgb, 255:red, 0; green, 0; blue, 0 }  ,fill opacity=0.05 ] (15.22,311.35) .. controls (6.75,279.73) and (46.17,241.7) .. (103.27,226.4) .. controls (160.36,211.1) and (213.52,224.33) .. (221.99,255.95) .. controls (230.46,287.56) and (191.04,325.59) .. (133.94,340.89) .. controls (76.85,356.19) and (23.69,342.96) .. (15.22,311.35) -- cycle ;
\draw  [fill={rgb, 255:red, 0; green, 0; blue, 0 }  ,fill opacity=0.05 ] (147.59,255.24) .. controls (156.06,223.63) and (209.21,210.4) .. (266.31,225.7) .. controls (323.41,241) and (362.83,279.03) .. (354.35,310.65) .. controls (345.88,342.26) and (292.73,355.49) .. (235.63,340.19) .. controls (178.54,324.89) and (139.12,286.86) .. (147.59,255.24) -- cycle ;
\draw  [dash pattern={on 0.84pt off 2.51pt}]  (8.65,153.88) -- (80.09,286.94) ;
\draw  [dash pattern={on 0.84pt off 2.51pt}]  (80.09,286.94) -- (143.12,150.37) ;
\draw  [dash pattern={on 0.84pt off 2.51pt}]  (254.33,169.67) -- (291.29,288.34) ;
\draw  [dash pattern={on 0.84pt off 2.51pt}]  (291.29,288.34) -- (324.33,162.33) ;
\draw   (8.65,146.1) .. controls (8.65,128.01) and (38.58,113.33) .. (75.49,113.33) .. controls (112.41,113.33) and (142.33,128.01) .. (142.33,146.1) .. controls (142.33,164.2) and (112.41,178.88) .. (75.49,178.88) .. controls (38.58,178.88) and (8.65,164.2) .. (8.65,146.1) -- cycle ;
\draw   (250.67,142.5) .. controls (250.67,105.96) and (267.31,76.33) .. (287.83,76.33) .. controls (308.36,76.33) and (325,105.96) .. (325,142.5) .. controls (325,179.04) and (308.36,208.67) .. (287.83,208.67) .. controls (267.31,208.67) and (250.67,179.04) .. (250.67,142.5) -- cycle ;
\draw  [draw opacity=0] (103.66,139.28) .. controls (100.05,147.07) and (88.51,152.77) .. (74.83,152.77) .. controls (61.17,152.77) and (49.65,147.1) .. (46.02,139.33) -- (74.83,134.1) -- cycle ; \draw   (103.66,139.28) .. controls (100.05,147.07) and (88.51,152.77) .. (74.83,152.77) .. controls (61.17,152.77) and (49.65,147.1) .. (46.02,139.33) ;
\draw  [draw opacity=0] (53.07,145.7) .. controls (57.98,140.67) and (65.6,137.44) .. (74.17,137.44) .. controls (83.55,137.44) and (91.81,141.32) .. (96.6,147.19) -- (74.17,159.05) -- cycle ; \draw   (53.07,145.7) .. controls (57.98,140.67) and (65.6,137.44) .. (74.17,137.44) .. controls (83.55,137.44) and (91.81,141.32) .. (96.6,147.19) ;
\draw  [draw opacity=0] (295.98,113.17) .. controls (295.66,122.96) and (292.17,130.68) .. (287.91,130.68) .. controls (283.66,130.68) and (280.17,122.97) .. (279.85,113.19) -- (287.91,111.72) -- cycle ; \draw   (295.98,113.17) .. controls (295.66,122.96) and (292.17,130.68) .. (287.91,130.68) .. controls (283.66,130.68) and (280.17,122.97) .. (279.85,113.19) ;
\draw  [draw opacity=0] (281.67,125.07) .. controls (282.97,119.07) and (285.2,115.11) .. (287.74,115.11) .. controls (290.48,115.11) and (292.87,119.75) .. (294.09,126.58) -- (287.74,137.05) -- cycle ; \draw   (281.67,125.07) .. controls (282.97,119.07) and (285.2,115.11) .. (287.74,115.11) .. controls (290.48,115.11) and (292.87,119.75) .. (294.09,126.58) ;
\draw  [color={rgb, 255:red, 208; green, 2; blue, 27 }  ,draw opacity=1 ][line width=0.75]  (67.33,165.27) .. controls (67.33,158.31) and (70.98,152.67) .. (75.49,152.67) .. controls (80,152.67) and (83.66,158.31) .. (83.66,165.27) .. controls (83.66,172.23) and (80,177.88) .. (75.49,177.88) .. controls (70.98,177.88) and (67.33,172.23) .. (67.33,165.27) -- cycle ;
\draw  [color={rgb, 255:red, 208; green, 2; blue, 27 }  ,draw opacity=1 ][line width=0.75]  (279.67,169.5) .. controls (279.67,147.87) and (283.32,130.33) .. (287.83,130.33) .. controls (292.34,130.33) and (296,147.87) .. (296,169.5) .. controls (296,191.13) and (292.34,208.67) .. (287.83,208.67) .. controls (283.32,208.67) and (279.67,191.13) .. (279.67,169.5) -- cycle ;
\draw  [color={rgb, 255:red, 74; green, 144; blue, 226 }  ,draw opacity=1 ] (26.6,144.76) .. controls (26.6,132.38) and (49.08,122.33) .. (76.8,122.33) .. controls (104.53,122.33) and (127,132.38) .. (127,144.76) .. controls (127,157.15) and (104.53,167.19) .. (76.8,167.19) .. controls (49.08,167.19) and (26.6,157.15) .. (26.6,144.76) -- cycle ;
\draw  [color={rgb, 255:red, 74; green, 144; blue, 226 }  ,draw opacity=1 ] (270,130.17) .. controls (270,111.67) and (278.36,96.67) .. (288.67,96.67) .. controls (298.98,96.67) and (307.33,111.67) .. (307.33,130.17) .. controls (307.33,148.67) and (298.98,163.67) .. (288.67,163.67) .. controls (278.36,163.67) and (270,148.67) .. (270,130.17) -- cycle ;
\draw [line width=1.5]    (158.67,143.67) -- (229,143.67) ;
\draw [shift={(233,143.67)}, rotate = 180] [fill={rgb, 255:red, 0; green, 0; blue, 0 }  ][line width=0.08]  [draw opacity=0] (6.97,-3.35) -- (0,0) -- (6.97,3.35) -- cycle    ;
\draw [shift={(154.67,143.67)}, rotate = 0] [fill={rgb, 255:red, 0; green, 0; blue, 0 }  ][line width=0.08]  [draw opacity=0] (6.97,-3.35) -- (0,0) -- (6.97,3.35) -- cycle    ;
\draw (83.05,314.99) node  [font=\small]  {$U_{\alpha }$};
\draw (292.26,312.72) node  [font=\small]  {$U_{\beta }$};
\draw (184.59,263.37) node  [font=\small]  {$U_{\alpha } \cap U_{\beta }$};
\draw (145.67,123) node [anchor=north west][inner sep=0.75pt]  [font=\footnotesize]  {$e^{n_{( \alpha \beta )}} \!\!\in\! O(n,n;\mathbb{Z})$};
\draw (175,150) node [anchor=north west][inner sep=0.75pt]  [font=\footnotesize]  {gluing};
\draw (-23,135) node [anchor=north west][inner sep=0.75pt]  [font=\small]  {$T^{2n}$};
\draw (334,135) node [anchor=north west][inner sep=0.75pt]  [font=\small]  {$T^{2n}$};
\draw (46,84) node [anchor=north west][inner sep=0.75pt]  [font=\small]  {$\mathcal{F}_{( \alpha )} ,\,\mathcal{G}_{( \alpha )}$};
\draw (258.67,48) node [anchor=north west][inner sep=0.75pt]  [font=\small]  {$\mathcal{F}_{( \beta )},\,\mathcal{G}_{( \beta )}$};
\end{tikzpicture}\caption{The gluing conditions on the $T^{2n}$ fibers and the fields $\mathcal{F}_{( \alpha )} $, $\mathcal{G}_{( \alpha )}$ for a simple T-fold.}\end{figure}
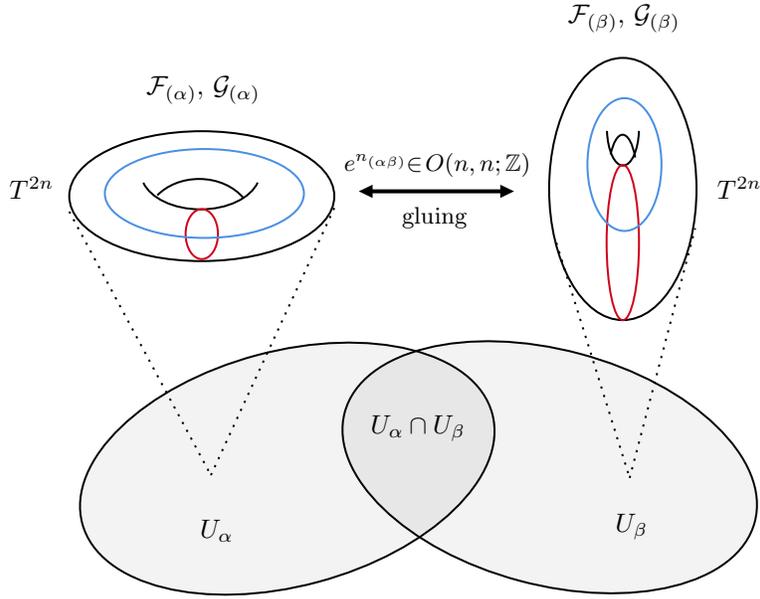

\paragraph{T-duality on the tensor hierarchy of a T-fold.}
Given any element $\mathcal{O}\in O(n,n;\mathbb{Z})$ of the T-duality group, we can see that there is a natural action on the local $2n$ coordinates of the torus by $\Theta_{(\alpha)}^I\mapsto \mathcal{O}^{I}_{\;J}\Theta_{(\alpha)}^J$ and on the fields of the tensor hierarchy by
\begin{equation}
     \mathcal{F}_{(\alpha)}^I \,\mapsto\, \mathcal{O}^{I}_{\;J}\mathcal{F}_{(\alpha)}^J, \qquad \mathcal{H} \,\mapsto\, \mathcal{H}, \qquad \mathcal{G}_{(\alpha)IJ} \,\mapsto\, \mathcal{O}^{K}_{\;\,I}\,\mathcal{G}_{(\alpha)KL}\,\mathcal{O}^{L}_{\;J}.
\end{equation}
If we include also the dimensionally reduced global pseudo-Riemannian metric $g$ on $M_0$, we can sum how the fields transform under T-duality in the table \ref{tab:t}.\vspace{0.2cm}

\begin{table}[h!]\begin{center}
\begin{center}
 \begin{tabular}{|| c |  c | c||} 
 \hline
 \multicolumn{3}{||c||}{Fields under $O(n,n;\mathbb{Z})$-action} \\
 \hline
 \hspace{0.8cm}Singlet rep.\hspace{0.8cm} & Fundamental rep. & \hspace{0.5cm}Adjoint rep.$\quad$ \\ [0.5ex] 
 \hline
 $g, \mathcal{B}_{(\alpha)}$ & $\mathcal{A}^I_{(\alpha)}$ & $\mathcal{G}_{(\alpha)IJ}$   \\[0.8ex]  
 \hline
\end{tabular}
\end{center}
\caption{\label{tab:t}A summary of the fields transforming under T-duality.}\vspace{0.0cm}
\end{center}\end{table}

\noindent Clearly, together with the torus coordinates, this will transform our monodromy matrix cocycle $(e^{n_{(\alpha\beta)}})^I_{\;J}$ to a new $O(d,d;\mathbb{Z})$-cocycle $\mathcal{O}^{I}_{\;K}(e^{n_{(\alpha\beta)}})^K_{\;J}$, but, since the fields depend only on the physical coordinates, the global geometric interpretation in the generalized correspondence space does not change.\vspace{0.25cm}

\noindent On the other hand, if we want an intrinsic and "non-doubled" description of the T-dual string backgrounds, as explained by \cite{Bou08}, we obtain a \textit{non-commutative} $T^n$\textit{-bundle} on $M_0$ which is classified by the cocycle $\left[n_{(\alpha\beta)}\right]\in H^1(M_0,\,\wedge^2\mathbb{Z}^2)$ on its base manifold. \vspace{0.25cm}

\noindent We can finally summarize these two equivalent descriptions within the table \ref{tab:t-fold}. \newpage

\begin{table}[h!]\begin{center}
\vspace{0.25cm}
  \setlength\extrarowheight{1.5ex}
 \begin{tabular}{|c||c|c||} 
 \hline
 & \multicolumn{2}{c||}{\makecell{\textbf{T-fold}}}\\[0.5ex] 
 \hline \hline
 & $\qquad$\makecell{Tensor\\hierarchy\\description}$\qquad$ & \makecell{Generalized\\correspondence space\\description} \\[0.5ex] 
 \hline\hline
 \makecell{Space of\\the T-fold} & \makecell{Local $U_\alpha\times T^{2n}$\\glued by $e^{n_{(\alpha\beta)}}$ } & \makecell{Global $T^n$-bundle\\$K\twoheadrightarrow M$\\ on spacetime $M$} \\[0.5ex]
 \hline
 \makecell{Connection\\data} & \makecell{$\di\Theta_{(\alpha)}^I+\mathcal{A}^I_{(\alpha)}$}  & \makecell{$\Xi^I$} \\[0.5ex]
 \hline
 \makecell{Curvature\\data} & \makecell{$\mathcal{F}^I_{(\alpha)}$}  & \makecell{$F^i,\,(\pi^\ast \bigsp{H})_i$} \\[0.5ex]
 \hline
 \makecell{Main\\ feature} & \makecell{Fields of the \\tensor hierarchy\\are manifest}  & \makecell{Topology (including\\non-geometry)\\is manifest} \\[0.5ex]
 \hline
 \makecell{Relation} & \multicolumn{2}{c||}{Related by a $e^{B_{(\alpha)}^{(0)}}$-twist on each patch}\\[0.5ex]
 \hline
\end{tabular}
\caption{\label{tab:t-fold}A brief summary of the two descriptions for the T-fold.}
\end{center}\end{table}

\paragraph{The tensor hierarchy from gerbe dimensional reduction.}
There is an important subtlety in this discussion: \textit{the most general tensor hierarchy of the T-fold does not arise by gauging the tensor hierarchy $2$-algebra}. This means that \textit{the tensor hierarchy of the T-fold is not a higher gauge theory on $M_0$}. This is due to the presence of the monodromy matrix cocycle $n_{(\alpha\beta)}$, which glues the curvatures by $\mathcal{F}_{(\alpha)}^I = \big(e^{n_{(\alpha\beta)}}\big)^{\!I}_{\,J} \mathcal{F}_{(\beta)}^J$ and which is not a gauge transformation of the connection.
\vspace{0.25cm}

\noindent Thus the most general global formalization of the tensor hierarchy of a T-fold must then be given by the dimensional reduction of a bundle gerbe, i.e. by a diagram of the form
\begin{equation}
    \tenhie^{\,T^n}_{\!\mathrm{sc}}\!(M_0) \;=\; \left\{\begin{tikzcd}[row sep=7ex, column sep=5ex]
    & \left[T^n,\mathbf{B}^2U(1)\right]\!/T^n \arrow[d]\\
    M_0 \arrow[ru]\arrow[r, ""] & \mathbf{B}T^n
    \end{tikzcd}\right\}
\end{equation}

\subsection{Poisson-Lie T-folds as non-abelian global tensor hierarchies}

Non-abelian T-duality is a generalization of abelian T-duality for string backgrounds whose group of isometries is non-abelian. \textit{Poisson-Lie T-duality} is a further generalization of this concept where the string background is not even required to have isometries, but which relies on the existence of a more subtle rigid group structure. See \cite{BPV20} for discussion of Poisson-Lie T-duality of a $\sigma$-model in a group manifold and \cite{Has17} for discussion of Poisson-Lie T-duality in DFT. For recent applications concerning the Drinfel'd double $SL(2,\mathbb{C})=SU(2)\bowtie SB(2,\mathbb{C})$ see \cite{Ba19}, \cite{Vit19} and \cite{Ba20}.

\paragraph{The generalized correspondence space of Poisson-Lie T-duality.}
As we have just seen, a bundle gerbe on a $T^n$-bundle spacetime $\pi:M\twoheadrightarrow M_0$ is abelian T-dualizable if $[\pi^\ast \bigsp{H}]\in H^2(M,\mathbb{Z}^n)$, so that $\pi^\ast \bigsp{H}$ becomes the curvature of a $T^n$-bundle $K\twoheadrightarrow M$, the generalized correspondence space. In the simplest case we examined, we had $(\pi^\ast \bigsp{H})_i = \di(-\iota_{e_i}\bigsp{B}_{(\alpha)})$, which makes $\tilde{\Xi}_i=\di\tilde{\theta}_{(\alpha)}-\iota_{e_i}\bigsp{B}_{(\alpha)}$ the global connection $1$-form of the bundle $K$. Now we want to study the generalization of this from abelian T-duality to Poisson-Lie T-duality.
\vspace{0.25cm}

\noindent We say that \textit{a bundle gerbe $\mathcal{M}\twoheadrightarrow M$ on a $G$-bundle spacetime  $\pi\!:\!M\twoheadrightarrow M_0$ is Poisson-Lie T-dualizable if $\pi^\ast \bigsp{H}$ is the curvature of a $\widetilde{G}$-bundle $K\twoheadrightarrow M$}, where $\widetilde{G}$ is some Lie group with the same dimension $\mathrm{dim}\, G = \mathrm{dim}\, \widetilde{G}$. In the simple case we examined in \cite{Alf19} this implies that
\begin{equation}
    (\pi^\ast \bigsp{H})_i \;=\; \di(-\iota_{e_i}\bigsp{B}_{(\alpha)}) + \frac{1}{2}\widetilde{C}_{i}^{\;jk}\, (-\iota_{e_j}\bigsp{B}_{(\alpha)}) \wedge (-\iota_{e_k}\bigsp{B}_{(\alpha)})
\end{equation}
where $\{e_i\}_{i=1,\dots,\mathrm{dim}\,G}$ is a basis of vertical $G$-left-invariant vectors on $M$ and where $\widetilde{C}_{i}^{\;jk}$ are the structure constants of the Lie algebra $\tilde{\mathfrak{g}}:=\mathrm{Lie}(\widetilde{G})$. Notice that the local $1$-form $-\iota_{e_i}\bigsp{B}_{(\alpha)}$ is now the local potential of a non-abelian principal bundle. In analogy with abelian T-duality we call the total space $K$ the \textit{generalized correspondence space} of the Poisson-Lie T-duality. Therefore we have a diagram of the following form:
\begin{equation}
    \begin{tikzcd}[row sep=4ex, column sep=5ex]
    \widetilde{G}\arrow[r, hook] & K \arrow[d, two heads] \\
    G\arrow[r, hook] & M \arrow[d, two heads] \\
    & M_0
    \end{tikzcd}
\end{equation}
Crucially the composition $K\twoheadrightarrow M_0$ is a fiber bundle on $M_0$ with fiber $G\times \widetilde{G}$, but it is not a principal bundle. However, for any good cover $\{U_\alpha\}$ of the base manifold $M_0$, the total space $K$ will be still locally of the form $K|_{U_\alpha}\cong U_\alpha \times G \times \widetilde{G}$.

\paragraph{The hidden Drinfel'd double fiber.}
The generalized correspondence space $K$, on any patch $U_\alpha$ of the base manifold $M_0$, can be restricted to a local trivial bundle $K|_{U_\alpha}\cong U_\alpha\times G \times \widetilde{G}$. For the fiber $G\times \widetilde{G}$ we can introduced the parametrization defined by $\gamma_{(\alpha)}=\exp(\theta_{(\alpha)}^ie_i)$ and $\tilde{\gamma}_{(\alpha)}=\exp(\tilde{\theta}_{(\alpha)i}\tilde{e}^i)$, where $\theta_{(\alpha)}$ and $\tilde{\theta}_{(\alpha)}$ are local coordinates on the group $G\times \widetilde{G}$ near the identity element.
Now, on each trivial local $G\times\widetilde{G}$-bundle $K|_{U_\alpha} \cong U_\alpha\times G \times \widetilde{G}$ we can construct the following local $\mathfrak{g}\oplus\tilde{\mathfrak{g}}$-valued differential $1$-form:
\begin{equation}\label{eq:locon}
    \begin{pmatrix}\xi_{(\alpha)}\\[0.4em] \tilde{\xi}_{(\alpha)} \end{pmatrix}\;:=\; \begin{pmatrix} \gamma_{(\alpha)}^{-1}\di\gamma_{(\alpha)} + \mathrm{Ad}_{\gamma_{(\alpha)}^{-1}}A_{(\alpha)}  \\[1em] \tilde{\gamma}_{(\alpha)}^{-1}\di\tilde{\gamma}_{(\alpha)} + \mathrm{Ad}_{\tilde{\gamma}_{(\alpha)}^{-1}}B^{(1)}_{(\alpha)}  \end{pmatrix} \;\;\in\, \Omega^1\big(U_\alpha\times G \times \widetilde{G},\,\mathfrak{g}\oplus\tilde{\mathfrak{g}}\big),
\end{equation}
where we called $A_{(\alpha)}=A_{(\alpha)}^i\otimes e_i$ and $B_{(\alpha)}^{(1)}=B_{(\alpha)i}^{(1)}\otimes \tilde{e}^i$. Here we used the vector notation for elements of $\mathfrak{g}\oplus\tilde{\mathfrak{g}}$.
Notice that this is a local $G\times\widetilde{G}$-connection $1$-form for our local bundle $U_\alpha\times G\times\widetilde{G}$.
\vspace{0.25cm}

\noindent Now, the global connection data of the generalized correspondence space $K$ is given by the connection of the $G$-bundle $M\twoheadrightarrow M_0$ and the one of the $\widetilde{G}$-bundle $K\twoheadrightarrow M$. We can combine them in a global $\mathfrak{g}\oplus\tilde{\mathfrak{g}}$-valued $1$-form on the total space $K$ as it follows
\begin{equation}
    \Xi\; := \; \begin{pmatrix}\gamma_{(\alpha)}^{-1}\di\gamma_{(\alpha)} + \mathrm{Ad}_{\gamma_{(\alpha)}^{-1}}A_{(\alpha)} \\[1em] \tilde{\gamma}_{(\alpha)}^{-1}\di\tilde{\gamma}_{(\alpha)} + \mathrm{Ad}_{\tilde{\gamma}_{(\alpha)}^{-1}}\big(B^{(1)}_{(\alpha)}-B^{(0)}_{(\alpha)k}\xi^k\big) \end{pmatrix} \;\;\in\, \Omega^1\big(K,\,\mathfrak{g}\oplus\tilde{\mathfrak{g}}\big)
\end{equation}
The relation between the global $1$-form $\Xi$ encoding the global connection data of the generalized correspondence space and the local $G\times\widetilde{G}$-connections $\xi_{(\alpha)},\tilde{\xi}_{(\alpha)}$ defined in \eqref{eq:locon} is given by 
\begin{equation}
\begin{aligned}
    \Xi\; = \; \begin{pmatrix}\xi \\[0.3em] \tilde{\xi}_{(\alpha)} - \mathrm{Ad}_{\tilde{\gamma}_{(\alpha)}^{-1}}\big(B^{(0)}_{(\alpha)k}\xi^k\big) \end{pmatrix}
    \end{aligned}
\end{equation}
We can rewrite the relation by making the generators $\{e_i,\,\tilde{e}^i\}_{i=1,\dots,n}$ of the algebra $\mathfrak{g}\oplus\tilde{\mathfrak{g}}$ explicit
\begin{equation}\label{eq:d1}
\begin{aligned}
    \Xi\; = \; \begin{pmatrix}\xi^i, & \tilde{\xi}_{(\alpha)i}\end{pmatrix} \begin{pmatrix}\delta_i^{\;j} & -B^{(0)}_{(\alpha)\ell i}\, \tilde{u}^{\;\;\;\ell}_{(\alpha)j} \\[0.3em] 0 & \delta^{i}_{\;j} \end{pmatrix}
    \begin{pmatrix}e_j \\[0.3em] \tilde{e}^{j}\end{pmatrix} 
    \end{aligned}
\end{equation}
where the matrix $\tilde{u}^{\;\;\;i}_{(\alpha)j}$ is defined by the adjoint action  $\mathrm{Ad}_{\tilde{\gamma}_{(\alpha)}^{-1}}(\tilde{e}^i)= \tilde{u}^{\;\;\;i}_{(\alpha)j}\,\tilde{e}^j$ of the group $\widetilde{G}$ on its algebra $\tilde{\mathfrak{g}}$ and depends only on the local coordinates $\tilde{\theta}_{\alpha}$.
\vspace{0.25cm}

\noindent Let us not introduce the concept of \textit{Drinfel'd double} $D\, :=\,  G \bowtie\widetilde{G}$. A Drinfel'd double $D$ is defined as an even-dimensional Lie group whose Lie algebra $\mathfrak{d}=\mathrm{Lie}(D)$ has underlying vector space $\mathfrak{g}\oplus\tilde{\mathfrak{g}}$ generated by the generators $\{e_i,\,\tilde{e}^i\}_{n=1,\dots,i}$ and bracket structure given by
\begin{equation}
    \begin{aligned}
    [e_i,\, e_j]_{\mathfrak{d}} \;&=\; C_{ij}^{\;\;\,k}e_k\\
    [e_i,\, \tilde{e}^j]_{\mathfrak{d}} \;&=\; C_{ki}^{\;\;\,j}\tilde{e}^k - \widetilde{C}^{kj}_{\;\;\;i}e^k\\
    [\tilde{e}^i,\, \tilde{e}^j]_{\mathfrak{d}} \;&=\; \widetilde{C}^{ij}_{\;\;\,k}\tilde{e}^k
    \end{aligned}
\end{equation}
Thus we can write its Lie algebra in the split form $\mathfrak{d}=\mathfrak{g}\bowtie\tilde{\mathfrak{g}}$, where $\mathfrak{g}$ is the subalgebra generated by the generators $\{e_i\}_{n=1,\dots,i}$ and $\tilde{\mathfrak{g}}$ is the subalgebra generated by the generators $\{\tilde{e}^i\}_{i=1,\dots,n}$. 
We can also write $D\, =\,  G \bowtie\widetilde{G}$, where $G$ is the Lie group integrating $\mathfrak{g}$ and $\widetilde{G}$ is the one integrating $\tilde{\mathfrak{g}}$. 
Notice that the manifold underlying the Drinfel'd double group $D\, =\,  G \bowtie\widetilde{G}$ is the same manifold underlying the direct product $G\times \widetilde{G}$, but it crucially comes equipped with a different Lie group structure. 
Now let us choose the parametrization $\Gamma_{(\alpha)}\,:=\,\gamma_{(\alpha)}\tilde{\gamma}_{(\alpha)}\in D$ for the Drinfel'd double $D=G\bowtie\widetilde{G}$, where we are still calling $\gamma_{(\alpha)}=\exp(\theta_{(\alpha)}^ie_i)\in G$ and $\tilde{\gamma}_{(\alpha)}=\exp(\tilde{\theta}_{(\alpha)i}\tilde{e}^i)\in \widetilde{G}$ with $\big(\theta_{(\alpha)},\,\tilde{\theta}_{(\alpha)}\big)$ local coordinates on the product manifold $G\times \widetilde{G}$.
As explained by \cite{Hull07}, the adjoint action of the subgroup $\widetilde{G}$ on the Lie algebra $\mathfrak{d}$ of the full Drinfel'd double $D=G\bowtie\widetilde{G}$ is specified on the generators by the following matrix 
\begin{equation}
    \tilde{\gamma}_{(\alpha)}^{-1} \begin{pmatrix}e_i \\[0.5em] \tilde{e}^i \end{pmatrix} \tilde{\gamma}_{(\alpha)} \;=\; \begin{pmatrix} (\tilde{u}_{(\alpha)}^{-\mathrm{T}})_i^{\;j} & \tilde{b}_{(\alpha)ij} \\[0.5em] 0 & \tilde{u}^{\;\;\;i}_{(\alpha)j} \end{pmatrix} \begin{pmatrix}e_j \\[0.5em] \tilde{e}^j \end{pmatrix}
\end{equation}
where the submatrices $\tilde{u}_{(\alpha)}$ and $\tilde{b}_{(\alpha)}$ depend only on the local coordinates of $\widetilde{G}$ and $\tilde{b}_{(\alpha)}$ is skew-symmetric.
Similarly, the adjoint action of the subgroup $G$ on the Lie algebra $\mathfrak{d}$ is given on generators by
\begin{equation}
    \gamma_{(\alpha)}^{-1} \begin{pmatrix}e_i \\[0.5em] \tilde{e}^i \end{pmatrix} \gamma_{(\alpha)} \;=\; \begin{pmatrix} (u_{(\alpha)})_i^{\;j} & 0 \\[0.5em] \beta_{(\alpha)}^{ij} & (u_{(\alpha)}^{-\mathrm{T}})_{\;j}^{i} \end{pmatrix} \begin{pmatrix}e_j \\[0.5em] \tilde{e}^j \end{pmatrix}
\end{equation}
where this time the matrices ${u}_{(\alpha)}$ and $\beta_{(\alpha)}$ depend only on the local coordinates of $G$ and $\beta_{(\alpha)}$ is skew-symmetric. \vspace{0.25cm}

\noindent Recall that we are parametrizing the points of our local bundle by $\big(x_{(\alpha)},\,\Gamma_{(\alpha)}\big)\in U_\alpha\times D$. It was shown by \cite{Hul09} that on each $D$ fiber the Maurer-Cartan $1$-form is given by
\begin{equation}
\begin{aligned}
    \Gamma_{(\alpha)}^{-1}\di\Gamma_{(\alpha)} \;&=\; \begin{pmatrix}\big(\gamma_{(\alpha)}^{-1}\di\gamma_{(\alpha)}\big)^i, & \big(\tilde{\gamma}_{(\alpha)}^{-1}\di\tilde{\gamma}_{(\alpha)}\big)_i\end{pmatrix} \begin{pmatrix}\big(\tilde{u}_{(\alpha)}^{-\mathrm{T}}\big)^{\;j}_{i} & \tilde{b}_{(\alpha)ij} \\[0.5em] 0 & \delta^{i}_{\;j} \end{pmatrix}
    \begin{pmatrix}e_j \\[0.3em] \tilde{e}^j\end{pmatrix}
    \end{aligned}
\end{equation}
where $\gamma_{(\alpha)}^{-1}\di\gamma_{(\alpha)}$ is the Maurer-Cartan $1$-form on the subgroup $G$ and $\tilde{\gamma}_{(\alpha)}^{-1}\di\tilde{\gamma}_{(\alpha)}$ is the one on $\widetilde{G}$. If we write the Maurer-Cartan $1$-form in terms of the generators of the Drinfel'd double we get
\begin{equation}
\begin{aligned}
    \left(\Gamma_{(\alpha)}^{-1}\di\Gamma_{(\alpha)}\right)^{\!\ind{J}} \;&=\; \begin{pmatrix}\big(\tilde{u}_{(\alpha)}^{-\mathrm{T}}\big)^{\;j}_{i} \big(\gamma_{(\alpha)}^{-1}\di\gamma_{(\alpha)}\big)^i \\[0.8em] \big(\tilde{\gamma}_{(\alpha)}^{-1}\di\tilde{\gamma}_{(\alpha)}\big)_i + \tilde{b}_{(\alpha)ij} \big(\gamma_{(\alpha)}^{-1}\di\gamma_{(\alpha)}\big)^i \end{pmatrix}
    \end{aligned}
\end{equation}
Now, on our local bundle $K|_{U_\alpha}\cong U_\alpha\times D$, we can define the following local $\mathfrak{d}$-valued $1$-form
\begin{equation}
    \Gamma_{(\alpha)}^{-1}\di\Gamma_{(\alpha)} + \mathrm{Ad}_{\Gamma_{(\alpha)}^{-1}}\mathcal{A}_{(\alpha)} \;\;\in\,\Omega^1(U_\alpha\times D,\,\mathfrak{d})
\end{equation}
by requiring the identity
\begin{equation}\label{eq:plconn}
\begin{aligned}
    \Gamma_{(\alpha)}^{-1}\di\Gamma_{(\alpha)} + \mathrm{Ad}_{\Gamma_{(\alpha)}^{-1}}\mathcal{A}_{(\alpha)} \;\,&:=\,\; \begin{pmatrix}\xi^i, & \tilde{\xi}_{(\alpha)i}\end{pmatrix} \begin{pmatrix}\big(\tilde{u}_{(\alpha)}^{-\mathrm{T}}\big)^{\;j}_{i} & \tilde{b}_{(\alpha)ij} \\[0.5em] 0 & \delta^{i}_{\;j} \end{pmatrix}
    \begin{pmatrix}e_j \\[0.3em] \tilde{e}^j\end{pmatrix}
    \end{aligned}
\end{equation}
where $\xi^i$ and $\tilde{\xi}_{(\alpha)i}$ are the local connections defined in \eqref{eq:locon}. This $1$-form can be seen as a $D$-connection (not necessarily principal) on each local bundle $U_\alpha\times D$. In terms of generators of the Drinfel'd double we have the $1$-forms
\begin{equation}
\begin{aligned}
    \left(\Gamma_{(\alpha)}^{-1}\di\Gamma_{(\alpha)} + \mathrm{Ad}_{\Gamma_{(\alpha)}^{-1}}\mathcal{A}_{(\alpha)}\right)^{\!\ind{J}} \;&=\; \begin{pmatrix}\big(\tilde{u}_{(\alpha)}^{-\mathrm{T}}\big)^{\;j}_{i} \xi^i \\[0.8em] \tilde{\xi}_i + \tilde{b}_{(\alpha)ij} \,\xi^i \end{pmatrix}
    \end{aligned}
\end{equation}
From equation \eqref{eq:plconn} combined with the identity $\mathrm{Ad}_{\Gamma_{(\alpha)}^{-1}}=\mathrm{Ad}_{\tilde{\gamma}_{(\alpha)}^{-1}}\circ\mathrm{Ad}_{\gamma_{(\alpha)}^{-1}}$ we can get an explicit expression for the local $1$-form $\mathcal{A}_{(\alpha)}=\mathcal{A}_{(\alpha)}^\ind{I}\otimes E_\ind{I}$ where $\{E_\ind{I}\}$ are collectively the generators of $\mathfrak{d}$. We find that the relation between $\mathcal{A}_{(\alpha)}$ and the potential $A_{(\alpha)}^i$ and the component $B^{(1)}_{(\alpha)i}$ of the Kalb-Ramond field is
\begin{equation}\label{eq:dconn}
\begin{aligned}
    \mathcal{A}^i_{(\alpha)} \,=\, A^i - B^{(1)}_{(\alpha)k}(u_{(\alpha)}^{\mathrm{T}})^k_{\;\ell}\, \beta^{\ell n}_{(\alpha)}\, (u_{(\alpha)}^{-1})^i_{\;n} \quad \text{ and } \quad \widetilde{\mathcal{A}}_{(\alpha)i} \,=\, B^{(1)}_{(\alpha)k}(u_{(\alpha)}^{\mathrm{T}})^k_{\;i}
\end{aligned}
\end{equation}
When the Drinfel'd double $D$ is an abelian group we immediately recover the usual abelian $1$-form potentials $\mathcal{A}^I_{(\alpha)}=\big(A^i_{(\alpha)},\,B^{(1)}_{(\alpha)i}\big)\in\Omega^1(U_\alpha,\,\mathbb{R}^{2n})$ of the abelian T-fold.
\vspace{0.25cm}

\noindent We can now combine equation \eqref{eq:d1} with equation \eqref{eq:plconn} to find the relation between the global $1$-form $\Xi^I$, encoding the global connection data of the generalized correspondence space, and the local $D$-bundle connection in \eqref{eq:dconn}. The relation is thus given as it follows:
\begin{equation}
    \Xi^I\; = \; U^{\;\;\;\,I}_{(\alpha)\,\ind{J}}  \left(\Gamma_{(\alpha)}^{-1}\di\Gamma_{(\alpha)} + \mathrm{Ad}_{\Gamma_{(\alpha)}^{-1}}\mathcal{A}_{(\alpha)} \right)^{\! \ind{J}}
\end{equation}
where we defined the following matrix
\begin{equation}
    U^{\;\;\;\,I}_{(\alpha)\,\ind{J}} \;\,:=\,\; \begin{pmatrix} \tilde{u}^{\;\;\;i}_{(\alpha)j} & 0 \\[0.6em] \big(\tilde{b}_{(\alpha)i\ell}-\tilde{u}^{\;\;k}_{(\alpha)i} B_{(\alpha)k\ell}^{(0)}\big)\tilde{u}^{\;\;\ell}_{(\alpha)j} & \delta_{i}^{\; j} \end{pmatrix}
\end{equation}
which generally depends on both the local coordinates $\big(\theta_{(\alpha)}^i,\,\tilde{\theta}_{(\alpha)i}\big)$ of the fibers. Now we must calculate its inverse matrix and find
\begin{equation}
    \left(U_{(\alpha)}^{-1}\right)_{\;J}^{\!\ind{I}} \;\,=\,\; \begin{pmatrix} (\tilde{u}_{(\alpha)}^{-1})_{\;j}^{i} & 0 \\[0.5em] \tilde{u}^{\;\;k}_{(\alpha)i} B_{(\alpha)kj}^{(0)}-\tilde{b}_{(\alpha)ij} & \delta_{i}^{\;j} \end{pmatrix}
\end{equation}
Finally we can define a \v{C}ech cocycle which is given on two-fold overlaps of patches by
\begin{equation}
    N_{(\alpha\beta)}\;:=\; U_{(\alpha)}^{-1}\,U_{(\beta)}
\end{equation}
We can calculate this matrix and find
\begin{equation}
   N_{(\alpha\beta)\,\ind{J}}^{\quad\;\ind{I}} \;\,=\,\; \begin{pmatrix} (\tilde{u}_{(\alpha)}^{-1})_{\;k}^{i} \, \tilde{u}^{\;\;k}_{(\beta)j} & 0 \\[0.8em] \big[ \big(\tilde{u}^{\;\;m}_{(\alpha)i} B_{(\alpha)m\ell}^{(0)}-\tilde{b}_{(\alpha)i\ell}\big) - \big(\tilde{u}^{\;\;m}_{(\beta)i} B_{(\beta)m\ell}^{(0)}-\tilde{b}_{(\beta)i\ell}\big)\big]\tilde{u}^{\;\;\ell}_{(\beta)j} & \delta_{i}^{\; j} \end{pmatrix}
\end{equation}
Thus we can finally write the patching conditions for our local $D$-bundle connections by
\begin{equation}\label{eq:obstru}\boxed{\;
    \left(\Gamma_{(\alpha)}^{-1}\di\Gamma_{(\alpha)} + \mathrm{Ad}_{\Gamma_{(\alpha)}^{-1}}\mathcal{A}_{(\alpha)} \right)^{\! \ind{I}} \;=\; N_{(\alpha\beta)\,\ind{J}}^{\quad\;\ind{I}}\, \left(\Gamma_{(\beta)}^{-1}\di\Gamma_{(\beta)} + \mathrm{Ad}_{\Gamma_{(\beta)}^{-1}}\mathcal{A}_{(\beta)} \right)^{\! \ind{J}} \;}
\end{equation}
Therefore \textit{the cocycle $N_{(\alpha\beta)}$ represents the obstruction of the generalized correspondence space $K$ from being a global $D$-bundle on the base manifold $M_0$}. In physical terms this means that, whenever the cocycle $N_{(\alpha\beta)}$ is non-trivial,\textit{ the Poisson-Lie T-dual spacetime is not a geometric background, but a T-fold}. This is directly analogous to how the abelian T-fold rises from the generalized correspondence space not being a global $T^{2n}$-bundle (see previous section).
\vspace{0.25cm}

\noindent Moreover, if we include the higher form field, we have that the cocycle $N_{(\alpha\beta)}$ is also the obstruction of the bundle gerbe $\mathcal{M}$ from being equivalent to a global $\String(G\bowtie\widetilde{G})$-bundle on $M_0$. This observation is the key to understand how the tensor hierarchy of the Poisson-Lie T-fold is globalized on the base manifold.
\vspace{0.25cm}

\noindent In the next part of the subsection, like we did for the abelian T-fold, we will compare this structure with the one emerging by gauging the algebra of tensor hierarchies we defined in section \ref{s2}. We will briefly show that they do not perfectly match, like in the abelian case.

\paragraph{Tensor hierarchy of a Poisson-Lie T-fold.}
Let us define the $2$-algebra of doubled vectors $\mathscr{D}(D)$ on the Drinfel'd double $D=G\bowtie \widetilde{G}$ by directly generalizing the $2$-algebra $\mathscr{D}(\mathbb{R}^{2n})$ we saw in \eqref{eq:dofd}. We can then consider the $2$-algebra
\begin{equation}
    \mathscr{D}(D) \;:=\; \Big( \Coo(D) \xrightarrow{\;\;\mathfrak{D}\;\;} \mathfrak{X}(D)\Big)
\end{equation}
The manifestly strong constrained version of the $2$-algebra $\mathscr{D}(D)$ will be given by the following
\begin{equation}
    \mathscr{D}_{\mathrm{sc}}(D) \;=\; \Big( \Coo(G) \xrightarrow{\;\;\mathfrak{D}\;\;} \Gamma(G,\,TG\oplus T^\ast G)\Big)
\end{equation}
since there exists an isomorphism $T^\ast G \cong T\widetilde{G}$, which implies $TG\oplus T^\ast G \,\cong\, TG\oplus T\widetilde{G} \,\cong\, TD$. The algebroid $TG\oplus T^\ast G$ is then generally equipped with anti-symmetrized Roytenberg brackets. However notice that, for a frame $\{E_{\ind{I}}\}$ of $D$-left-invariant generalized vectors, the anti-symmetrized Roytenberg bracket are just
\begin{equation}
    [ E_{\ind{I}}, E_\ind{J} ]_{\mathrm{Roy}} \;=\; C^\ind{K}_{\;\;\,\ind{I}\ind{J}\,}E_\ind{K}
\end{equation}
where the $C^\ind{K}_{\;\;\,\ind{I}\ind{J}}$ are the structure constants of the Drinfel'd double algebra $\mathfrak{d}=\mathfrak{g}\bowtie \tilde{\mathfrak{g}}$. This means that for such generalized vectors the anti-symmetrized Roytenberg bracket reduces to the Lie bracket $[ -, - ]_{\mathrm{Roy}} = [ -, - ]_{\mathfrak{d}}$ of the Drinfel'd double algebra $\mathfrak{d}$. \vspace{0.25cm}

\noindent Now let us try to construct the stackification of the prestack $\Omega\big(U,\, \mathscr{D}_{\mathrm{sc}}(D)\big)$ of local tensor hierarchies and let us consider a cocycle 
$\big(\mathcal{A}_{(\alpha)},\,\mathcal{B}_{(\alpha)},\, \lambda_{(\alpha\beta)}, \Xi_{(\alpha\beta)},\,  g_{(\alpha\beta\gamma)})$ where the local differential forms are given as it follows:
\begin{equation}
    \begin{aligned}
        \mathcal{A}_{(\alpha)}&\in\Omega^1(U_\alpha)\otimes \mathfrak{X}(D), & \mathcal{B}_{(\alpha)}&\in\Omega^2(U_\alpha)\otimes \Coo(G), \\
        \lambda_{(\alpha\beta)}&\in\Coo(U_\alpha\cap U_\beta)\otimes \mathfrak{X}(D) & \Xi_{(\alpha\beta)} &\in \Omega^1(U_\alpha\cap U_\beta)\otimes\Coo(G),\\
        && g_{(\alpha\beta\gamma)}&\in \Coo(U_\alpha\cap U_\beta\cap U_\gamma \times G).
    \end{aligned}
\end{equation}
We also impose the conditions that the $1$-form connection is of the form $\mathcal{A}_{(\alpha)}:=\mathcal{A}_{(\alpha)}^\ind{I}\otimes E_\ind{I}$ where $\{E_\ind{I}\}$ is a frame of vertical $D$-left-invariant generalized vectors and that the $2$-form connection satisfies $\mathfrak{D}\mathcal{B}_{(\alpha)}=0$. The patching conditions of the cocycle will then be of the following form:
\begin{equation*}\boxed{\quad
    \begin{aligned}
    \mathcal{F}_{(\alpha)} \;&=\; \di \mathcal{A}_{(\alpha)}  + \big[ \mathcal{A}_{(\alpha)}\,\overset{\wedge}{,}\, \mathcal{A}_{(\alpha)}\big]_{\mathfrak{d}} \\
    \mathcal{H} \;&=\;  \di \mathcal{B}_{(\alpha)} + \frac{1}{2}\big\langle \mathcal{A}_{(\alpha)} \,\overset{\wedge}{,}\, \mathcal{F}_{(\alpha)}\big\rangle + \frac{1}{3!}\big\langle \mathcal{A}_{(\alpha)} \,\overset{\wedge}{,}\,\big[ \mathcal{A}_{(\alpha)}\,\overset{\wedge}{,}\, \mathcal{A}_{(\alpha)}\big]_{\mathfrak{d}} \big\rangle\\[0.8em]
    \mathcal{A}_{(\alpha)} \;&=\; \lambda_{(\alpha\beta)}^{-1} \big(\mathcal{A}_{(\beta)}+\di\big) \lambda_{(\alpha\beta)} + \mathfrak{D}\Xi_{(\alpha\beta)} \\
    \mathcal{B}_{(\alpha)} - \mathcal{B}_{(\beta)} \;&=\; \mathrm{D} \Xi_{(\alpha\beta)} - \langle \log\lambda_{(\alpha\beta)},\,\mathcal{F}_{(\alpha)}\rangle \\[0.8em]
    \lambda_{(\alpha\beta)}\,\lambda_{(\beta\gamma)}\,\lambda_{(\gamma\alpha)} \;&=\; \exp \mathfrak{D} g_{(\alpha\beta\gamma)} \\
    \Xi_{(\alpha\beta)} + \Xi_{(\beta\gamma)} + \Xi_{(\gamma\alpha)} \;&=\; \di g_{(\alpha\beta\gamma)} \\[0.8em]
    g_{(\alpha\beta\gamma)} - g_{(\beta\gamma\delta)} +  g_{(\gamma\delta\alpha)} -  g_{(\delta\alpha\beta)}\;&\in\;2\pi\mathbb{Z}
    \end{aligned}\quad}
\end{equation*}
where we used the map $\log: D \rightarrow \mathfrak{d}$ and where $\mathrm{D}$ is the covariant derivative of the field $A_{(\alpha)}$. \vspace{0.25cm}

\noindent Again, the globalization of tensor hierarchy that we obtain by gauging the local prestack of tensor hierarchies is not the most general globalization we can think of. This is because it does not take into account the obstruction $N_{(\alpha\beta)}$ cocycle, appearing in equation \eqref{eq:obstru}, which we get by dimensional reduction of the bundle gerbe.

\paragraph{Tensor hierarchy from reduction of the gerbe.}
Thus this discussion motivates again the definition of global strong constrained tensor hierarchy by the following dimensional reduction of a bundle gerbe:
\begin{equation}
    \tenhie^{\,G}_{\!\mathrm{sc}}(M_0) \;=\; \left\{\begin{tikzcd}[row sep=7ex, column sep=5ex]
    & \left[G,\mathbf{B}^2U(1)\right]\!/T^n \arrow[d]\\
    M_0 \arrow[ru]\arrow[r, ""] & \mathbf{B}G
    \end{tikzcd}\right\}
\end{equation}

\subsection{Example: semi-abelian T-fold}

In this subsection we will consider (1) a spacetime which is a general $S^3$-fibration $M\twoheadrightarrow M_0$ on some base manifold $M_0$ and (2) a gerbe bundle $\mathcal{M}\twoheadrightarrow M$ with satisfies the simple T-dualization condition $\mathcal{L}_{e_i}\bigsp{B}_{(\alpha)}=0$, where $\bigsp{B}_{(\alpha)}$ is the gerbe connection and $\{e_i\}$ are a basis of $SU(2)$-left invariant vectors on spacetime $M$.\vspace{0.2cm}

\noindent As seen in the previous subsection the dimensional reduction of these gerbe contains a global bundle $K$, the generalized correspondence space, defined by the following diagram:
\begin{equation}
    \begin{tikzcd}[row sep=5ex, column sep=5ex]
    T^3\arrow[r, hook] & K \arrow[d, two heads] \\
    SU(2)\arrow[r, hook] & M \arrow[d, two heads] \\
    & M_0
    \end{tikzcd}
\end{equation}
This case is often called \textit{semi-abelian}, because spacetime is a principal fibration whose algebra has non-zero structure constants $[e_i,\, e_j]_{\mathfrak{su}(2)}=\epsilon^k_{\;ij}\,e_k$, but the dual ones $\tilde{C}^{jk}_{\;\; i}=0$ vanish.\vspace{0.25cm}

\noindent From calculations which are analogous to the ones for the abelian T-fold we find that the moduli of the flux are related to the moduli of the Kalb-Ramond field by $H^{(0)}_{ijk}\,=\, \mathfrak{D}_{[i}B^{(0)}_{(\alpha)jk]} + B^{(0)}_{(\alpha)[i|\ell}\,\epsilon^{\;\;\;\ell}_{|ij]}$  where $\mathfrak{D}_i = \mathcal{L}_{e_i}$. Also notice that the moduli of the Kalb-Ramond field is patched on overlaps of patches by 
$B^{(0)}_{(\beta)ij}-B^{(0)}_{(\alpha)ij} \,=\, \mathfrak{D}_{[i}\Lambda^{(0)}_{(\alpha)j]} + \Lambda^{(0)}_{(\alpha)\ell}\,\epsilon^{\;\;\ell}_{ij}$. This will be useful very soon.
We must now apply all the machinery from the previous subsection to this particular example.

\begin{itemize}
\item We can combine the pullback on $K$ of the global connection of the $SU(2)$-bundle $M\twoheadrightarrow M_0$ and the global connection of the $T^3$-bundle $K\twoheadrightarrow M$ in a single object by
\begin{equation}
    \Xi^I\;\, = \,\;\begin{pmatrix} \xi^i \\[0.8em] \di \tilde{\theta}_{(\alpha)i} + B^{(1)}_{(\alpha)i} + B^{(0)}_{(\alpha)ij}\xi^j \end{pmatrix}\;\;\in \, \Omega^1(K,\,\mathfrak{su}(2)\oplus\mathbb{R}^3) 
\end{equation}

\item Now we can consider a local patch $U_\alpha\subset M_0$ of the base manifold. The total space $K$ restricted on this local patch will be isomorphic to $K|_{U_\alpha} = U_\alpha\times SU(2) \times T^3$. These local bundles can be equipped with local connections
\begin{equation}\label{eq:natdconn2}
    \begin{pmatrix} \xi^i \\[0.8em] \di \tilde{\theta}_{(\alpha)i} + B^{(1)}_{(\alpha)i}  \end{pmatrix} \;\;\in \, \Omega^1\big(U_\alpha\times SU(2)\times T^3,\,\mathfrak{su}(2)\oplus\mathbb{R}^3\big). 
\end{equation}
As derived in \cite[pag.\,61]{Alf19}, these local connections are glued on two-fold overlaps of patches $(U_\alpha\cap U_\beta)\times SU(2) \times T^3$ by a cocycle of $B$-shifts of the form
\begin{equation}\label{eq:natdpat2}
    \begin{pmatrix} \xi^i \\[0.8em] \di \tilde{\theta}_{(\alpha)i} + B^{(1)}_{(\alpha)i}  \end{pmatrix} \;=\; \begin{pmatrix} \delta^i_{\;j} & 0 \\[0.8em] n_{(\alpha\beta)ij}+ \epsilon^{\;\;\,k}_{ij}\tilde{\lambda}_{(\alpha\beta)k} & \delta_i^{\;j} \end{pmatrix}  \begin{pmatrix} \xi^j \\[0.8em] \di \tilde{\theta}_j + B^{(1)}_{(\beta)j}  \end{pmatrix} 
\end{equation}
where we defined the matrix $n_{(\alpha\beta)ij} := \mathfrak{D}_{[i}\tilde{\lambda}_{(\alpha\beta)j]}$. Notice that the T-dualizability condition we imposed on the gerbe implies that $n_{(\alpha\beta)ij}$ is a $\wedge^2\mathbb{Z}^3$-valued \v{C}ech cocycle, similarly to the monodromy matrix cocycle appearing in the abelian T-fold. \vspace{0.1cm}

\noindent As we will explain later, these are the local connection used in most of the non-abelian T-duality literature (before the introduction of Drinfel'd doubles). As noticed by \cite[pag.\,13]{Bug19} they look very similar to an abelian T-fold, but with the monodromy depending on the coordinates via the term $\epsilon^{\;\;\,k}_{ij}\tilde{\lambda}_{(\alpha\beta)k}$. However we will see that it is better to construct and use proper local $D$-connections to make the tensor hierarchy really manifest.

\item Now we must use the fact that the adjoint action of $T^3$ on $D=SU(2)\ltimes T^3$ is specified by setting $\tilde{u}_{(\alpha)j}^{\;\;\;i}=\delta^i_{\;j}$ and $\tilde{b}_{(\alpha)ij}=\epsilon^{\;\;\,k}_{ij}\tilde{\theta}_{(\alpha)k}$. Thus we can construct the local connection for each local $D$-bundle $U_\alpha\times D$ by $\tilde{b}_{(\alpha)}$-twisting the local connection \eqref{eq:natdconn2}, obtaining
\begin{equation}\label{eq:natdpat3}
    \left(\Gamma_{(\alpha)}^{-1}\di\Gamma_{(\alpha)} + \mathrm{Ad}_{\Gamma_{(\alpha)}^{-1}}\mathcal{A}_{(\alpha)} \right)^{\! \ind{I}} \;=\; \begin{pmatrix} \xi^i \\[0.8em] \di \tilde{\theta}_i + B^{(1)}_{(\alpha)i} + \epsilon^{\;\;\,k}_{ij}\tilde{\theta}_{(\alpha)k}\xi^j  \end{pmatrix} \;\; \in\,\Omega(U_\alpha\times D,\, \mathfrak{d})
\end{equation}
Crucially, this new $1$-form is invariant under gauge transformation $\bigsp{B}_{(\alpha)}\mapsto\bigsp{B}_{(\alpha)}+\bigsp{\di}\bigsp{\eta}_{(\alpha)}$ of the original gerbe bundle on $M$, with gauge parameter $\bigsp{\eta}_{(\alpha)}\in\Omega^1_{\mathrm{inv}}(U_\alpha\times G)$. Thus this gives a proper local connection for the internal manifold-fibration rising from the vertical part of the dimensional reduction of the bundle gerbe.
\end{itemize}

\noindent To verify that the $1$-form \eqref{eq:natdpat3} is a proper connection we need to verify that the local potential $\mathcal{A}_{(\alpha)}\in\Omega^1(U_{\alpha},\,\mathfrak{d})$ is actually the pullback of a $1$-form from the base $U_\alpha$. Since for $D=SU(2)\ltimes T^3$ we have $\beta^{ij}=0$, the first component $\mathcal{A}^i=A^i$ is just the local potential of the $SU(2)$-bundle. To check the second component $\widetilde{\mathcal{A}}_{(\alpha)i}$, let us notice that the T-dualizability condition $\mathcal{L}_{e_i}\bigsp{B}_{(\alpha)}=0$ on the gerbe immediately implies $\mathcal{L}_{e_i}(B^{(1)}_{(\alpha)k}\wedge \xi^k)=0$. Now notice that, since the matrix $u^{\mathrm{T}}_{(\alpha)}$ encompasses the adjoint action of the inverse of $\gamma_{(\alpha)}=\exp(\theta^i_{(\alpha)}e_i)$, it must be equal to the exponential of the matrix $\epsilon^{i}_{\;jk}\theta^k_{(\alpha)}$. Therefore we can re-write the $1$-form $B^{(1)}_{(\alpha)k} = \widetilde{\mathcal{A}}_{(\alpha)i}(u^{\mathrm{T}}_{(\alpha)})^i_{\;k}$, where the $\widetilde{\mathcal{A}}_{(\alpha)i}$ depend only on the base $U_\alpha$.
\vspace{0.25cm}

\noindent The field strengths of these principal connections is then given by their covariant derivative
\begin{equation}
    \mathcal{F}_{(\alpha)} \;=\; \di \mathcal{A}_{(\alpha)} + \big[\mathcal{A}_{(\alpha)}\,\overset{\wedge}{,}\,\mathcal{A}_{(\alpha)}\big]_{\mathfrak{d}} \;\;\in\,\Omega^2(U_\alpha,\,\mathfrak{d}).
\end{equation}
In components of the generators of the algebra $\mathfrak{d}$, these assume the following form:
\begin{equation}
    \begin{aligned}
    \mathcal{F}^i_{(\alpha)} \;&=\; \di \mathcal{A}^i_{(\alpha)} + \epsilon^i_{\;jk}\mathcal{A}^j_{(\alpha)}\wedge \mathcal{A}^k_{(\alpha)} \\[0.2em]
    \widetilde{\mathcal{F}}_{(\alpha)i} \;&=\; \di \widetilde{\mathcal{A}}_{(\alpha)i} + \epsilon^k_{\;ij}\mathcal{A}^j_{(\alpha)}\wedge \widetilde{\mathcal{A}}_{(\alpha)k} \\
    \end{aligned}
\end{equation}
What we need to find out now is how these local $D$-bundles are globally glued together.

\noindent We can immediately see that the global connections of the generalized correspondence space encoded in the global $1$-form $\Xi^I$ are related to the local $SU(2)\ltimes T^3$-connections \eqref{eq:natdpat3} by a local patch-wise $B$-shift $U^{\;\;\;\,I}_{(\alpha)\,\ind{J}}$ of the following form:
\begin{equation}\label{eq:natdloctoglob}
    \Xi^I\; = \; \begin{pmatrix} \delta^i_{\;j} & 0 \\[0.8em] -\big(B^{(0)}_{(\alpha)ij}+\epsilon_{\;ij}^{k}\tilde{\theta}_{(\alpha)k}\big) & \delta_i^{\; j} \end{pmatrix} \left(\Gamma_{(\alpha)}^{-1}\di\Gamma_{(\alpha)} + \mathrm{Ad}_{\Gamma_{(\alpha)}^{-1}}\mathcal{A}_{(\alpha)} \right)^{\! \ind{J}}
\end{equation}
which, crucially, depends both on the physical and on the extra coordinates. The geometric flux is here just given by the structure constants $\epsilon_{ij}^{\;\;\,k}$ of $\mathfrak{su}(2)$. Thus from this expression we immediately get that the wanted patching condition on twofold overlaps of patches are
\begin{equation}\label{eq:natdpat3new}\boxed{\;
    \left(\Gamma_{(\alpha)}^{-1}\di\Gamma_{(\alpha)} + \mathrm{Ad}_{\Gamma_{(\alpha)}^{-1}}\mathcal{A}_{(\alpha)} \right)^{\! \ind{I}} \;=\; \big(e^{n_{(\alpha\beta)}}\big)^\ind{I}_{\;\ind{J}} \left(\Gamma_{(\beta)}^{-1}\di\Gamma_{(\beta)} + \mathrm{Ad}_{\Gamma_{(\beta)}^{-1}}\mathcal{A}_{(\beta)} \right)^{\! \ind{J}} \;}
\end{equation}
where, \textit{generalizing the abelian case, the monodromy cocycle $e^{n_{(\alpha\beta)}}\in\mathrm{Aut}(D;\,\mathbb{Z})$ is an integer-valued automorphism of the Drinfel'd double}. The interesting point is that semi-abelian T-folds are still glued by integer $B$-shifts, similarly to the abelian ones, if we consider local $D$-bundles. Consequently the curvatures are glued by
\begin{equation}
\mathcal{F}_{(\alpha)}^\ind{I} \,=\, \big(e^{n_{(\alpha\beta)}}\big)^\ind{I}_{\;\ind{J}} \,\Big(\mathrm{Ad}_{\lambda_{(\alpha\beta)}^{-1}}\mathcal{F}_{(\beta)}\Big)^\ind{J}
\end{equation}
where $\lambda_{(\alpha\beta)}^i$ are the expected transition functions of the $SU(2)$-bundle $M\twoheadrightarrow M_0$.
As usual, if we want to include also the higher form field of which characterizes a tensor hierarchy
\[
\mathcal{H} \;=\;  \di \mathcal{B}_{(\alpha)} + \frac{1}{2}\big\langle \mathcal{A}_{(\alpha)}\,\overset{\wedge}{,}\, \di \mathcal{A}_{(\alpha)}\big\rangle - \frac{1}{3!}\big\langle \mathcal{A}_{(\alpha)}\,\overset{\wedge}{,}\,[\mathcal{A}_{(\alpha)} \,\overset{\wedge}{,}\, \mathcal{A}_{(\alpha)}]_{\mathfrak{d}} \big\rangle
\]
this will be a globally defined, but not closed, $3$-form on the base manifold.

\paragraph{Non-abelian T-duality in the literature.}
We will now explain how the conventional non-abelian T-duality picture we are used in the literature, see for instance \cite{KLMC15} and \cite[pag.\,13]{Bug19}, emerges and is clarified.
Usually, in the literature, (1) we start from the moduli field of the Kalb-Ramond field $B^{(0)}_{(\alpha)ij}\xi^i\wedge\xi^j$ and of the metric $g^{(0)}_{ij}\xi^i\otimes\xi^j$. Then (2) we perform a shift which depends on the dual coordinates $B^{(0)}_{(\alpha)ij}\xi^i\wedge\xi^j \,\mapsto\,  \big(B^{(0)}_{(\alpha)ij}+\epsilon^{\;\;\,k}_{ij}\tilde{\theta}_{(\alpha)k}\big)\xi^i\wedge\xi^j$. Finally (3) we perform the proper inversion of the matrix of the moduli field by
\[
\widetilde{g}^{(0)ij}+\widetilde{B}^{(0)ij}_{(\alpha)} \;:=\; \Big(g_{ij}^{(0)}+B^{(0)}_{(\alpha)ij}\,+\,\epsilon^{\;\;\,k}_{ij}\tilde{\theta}_{(\alpha)k}\Big)^{-1}
\]
to obtain the non-abelian T-dual Kalb-Ramond field
$\widetilde{B}^{(0)ij}_{(\alpha)}\tilde{\xi}_{(\alpha)i}\wedge\tilde{\xi}_{(\alpha)j}$ where the role of the connection is now played by the local $1$-forms $\tilde{\xi}_{(\alpha)i}=\di\tilde{\theta}_{(\alpha)i}+B^{(1)}_{(\alpha)i}$, which are nothing but the last three components of \eqref{eq:natdconn2}. As we know, these connections are not globally defined and hence, as already observed in \cite[pag.\,13]{Bug19}, we obtain a particular T-fold. If we explicit the transition functions $\tilde{\lambda}_{(\alpha\beta)}=\tilde{\theta}_{(\beta)}-\tilde{\theta}_{(\alpha)}$ of $K$ and we consider the simplest case with $n_{(\alpha\beta)}=0$, we indeed notice that the $B$-shifts \eqref{eq:natdpat2} which glue the $\xi^i$ and $\tilde{\xi}_{(\alpha)i}$ reduce to the transformations $\epsilon_{ij}^{\;\;\,k}\big(\tilde{\theta}_{(\beta)}-\tilde{\theta}_{(\alpha)}\big)_k$ which were firstly illustrated by \cite[pag.\,13]{Bug19} for $S^3$.\vspace{0.25cm}

\noindent However these operations and their underlying geometry can be better understood in terms of local $D$-bundles $U_\alpha\times D$. This is because, as seen in equation \eqref{eq:natdloctoglob}, the "effective" moduli field of the Kalb-Ramond field is the sum $B^{(0)}_{(\alpha)ij}+\epsilon^{\;\;\,k}_{ij}\tilde{\theta}_{(\alpha)k}$ and T-duality is nothing but an automorphism of the fiber $D$. As we have shown, working by considering local $D$-bundles leads to the simpler patching conditions \eqref{eq:natdpat3}, than the ones \eqref{eq:natdpat2} that are obtained by analogy with abelian T-folds.

\paragraph{The geometric case: the global $\String(SU(2)\ltimes T^3)$-bundle.}
When our bundle gerbe $\mathcal{M}\twoheadrightarrow M$ is equivariant under the principal $SU(2)$-action of $M$, we get that the flux is just $H^{(0)}_{ijk}=B^{(0)}_{(\alpha)[i|\ell}\,F^{(0)\;\ell}_{\;\;|ij]}$ and the monodromy matrix cocycle $n_{(\alpha\beta)}=0$ is zero. This means that the local $D$-bundles $U_\alpha\times D$ are glued together to form a global $D$-bundle on the base manifold $M_0$. If we include also the higher form field $\mathcal{H}$, we will have a global $\String(D)$-bundle on the base manifold $M_0$. In other words we have the following equivalence of $2$-groupoids
\begin{equation}
    \bigsqcup_{\begin{subarray}{c}M\text{ s.t. }
    M\twoheadrightarrow M_0\\\text{is a }SU(2)\text{-bundle}\end{subarray}} \!\!\!\!SU(2)\text{-}\mathrm{equivGerbes}(M) \;\;\cong\;\; \String(SU(2)\!\ltimes \tilde{T}^3)\text{-}\mathrm{Bundles}(M_0)
\end{equation}
where we called $SU(2)\text{-}\mathrm{equivGerbes}(M)$ the $2$-groupoid of $SU(2)$-equivariant gerbes on $M$ and $ \String(SU(2)\!\ltimes \tilde{T}^3)\text{-}\mathrm{Bundles}(M_0)$ the $2$-groupoid of $\String(SU(2)\!\ltimes \tilde{T}^3)$-bundles on the base $M_0$.\vspace{0.25cm}

\noindent For example, we can look at the case where both our spacetime $M=M_0\times S^3$ is a trivial fibration and our gerbe bundle $\mathcal{M}\twoheadrightarrow M$ is topologically trivial with vanishing Dixmier-Douady class $[H]=0\in H^3(M,\mathbb{Z})$. In this case we obtain a trivially-fibered generalized correspondence space $K=M_0\times D$ with doubled fiber $D=SU(2)\ltimes T^3$. Consequently the global tensor hierarchy rising from the dimensional reduction of this trivial gerbe will be just the connection of the trivial principal $\infty$-bundle $M_0\times \String(D)\twoheadrightarrow M_0$.

\section{Outlook}

We clarified some aspects of the Higher Kaluza-Klein approach to DFT. In particular we defined an atlas for the bundle gerbe which locally matches what we expect from the doubled space of DFT. Moreover we illustrated how (strong constrained) tensor hierarchies can be globalized by starting from the bundle gerbe.

\paragraph{Exceptional Field Theory as geometrized M-theory.}
One of the strengths of Higher Kaluza-Klein geometry is that it can, in principle, be generalized to any bundle $n$-gerbe and more generally to any non-abelian principal $\infty$-bundle. This can overcome the usual difficulty of DFT geometries in directly being generalized to M-theory.
We are intrigued by the prospect that Exceptional Field Theory could be formalized as a Higher Kaluza-Klein Theory on the total space of the (twisted) M2/M5-brane bundle gerbe on the $11d$ super-spacetime, such as the one described by \cite{FSS18}.
\vspace{0.25cm}

\noindent The cases of Heterotic DFT and Exceptional Field Theory will be explored in papers to come.


\section*{Acknowledgement}
I thank my supervisor prof.$\,\,$David Berman for essential and inspiring discussion about the state of the art of Double Field Theory. I thank Christian S\"{a}mann, Urs Schreiber and Francesco Genovese for indispensable discussion. I would like to thank the organizers Vicente Cortés, Liana David and Carlos Shahbazi of the workshop \href{https://www.math.uni-hamburg.de/projekte/gg2020/}{\textit{Generalized Geometry and Applications 2020}} at University of Hamburg.
I would like to thank Emanuel Malek and all the organizers of the \href{https://sites.google.com/view/egseminars}{\textit{Exceptional Geometry Seminar Series}}.
I would also like to thank Vincenzo Marotta, Lukas M\"{u}ller, David Svoboda and Richard Szabo for extremely helpful comments.
The author is grateful to Queen Mary University of London (QMUL) for its partial support.

\medskip
\addcontentsline{toc}{section}{References}
\bibliographystyle{fredrickson}
\bibliography{sample}
\end{document}